\documentclass[12pt, oneside]{article}   	

\usepackage{jheppub}

\usepackage{amsmath}
\usepackage{amssymb}
\usepackage{amsfonts}
\usepackage{amsthm}
\usepackage{amsbsy}
\usepackage{verbatim}
\usepackage{array}
\usepackage[linesnumbered,lined,boxed,commentsnumbered]{algorithm2e}

\usepackage[OT2,OT1]{fontenc}
\newcommand\cyr
{
\renewcommand\rmdefault{wncyr}
\renewcommand\sfdefault{wncyss}
\renewcommand\encodingdefault{OT2}
\normalfont
\selectfont
}
\DeclareTextFontCommand{\textcyr}{\cyr}

\usepackage{mathtools}
\usepackage[usenames,dvipsnames]{xcolor}

\usepackage{subcaption}

\usepackage{dsdshorthand}

\usepackage[vcentermath]{youngtab}

\def\ben{\begin{equation}}
\def\een{\end{equation}}
\def\bea{\begin{eqnarray}}
\def\eea{\end{eqnarray}}
\def\tx{\tilde{x}}

\def\cJ{\cal{J}}

\newcommand{\vardbtilde}[1]{\tilde{\raisebox{0pt}[0.85\height]{$\tilde{#1}$}}}

\newcommand{\AG}[1]{{\color{red}\bf AG: #1}}

\DeclareMathOperator{\sgn}{sgn}
\def\ben{\begin{equation}}
\def\een{\end{equation}}
\def\bne{\begin{equation}}
\def\ene{\end{equation}}

     \let\r=v
 \let\t=\tau

\def\ba{\begin{array}}
\def\ea{\end{array}}
\def\beq{\begin{equation}}
\def\eeq{\end{equation}}

\makeatletter
\def\@fpheader{\ }
\makeatother

\title{Near Conformal Perturbation Theory in SYK Type Models}
\author[1]{Sumit R. Das}
\affiliation[1]{Department of Physics and Astronomy, University of Kentucky, Lexington, KY 40506, USA}
\author[1]{Animik Ghosh}
\author[2]{Antal Jevicki}
\affiliation[2]{Department of Physics, Brown University, 182 Hope Street, Providence, RI 02912, USA}
\author[3]{Kenta Suzuki}
\affiliation[3]{CPHT, CNRS, Ecole Polytechnique, Institut Polytechnique de Paris, Route de Saclay,
91128 PALAISEAU, France}

\emailAdd{das@pa.uky.edu}\emailAdd{animik.ghosh@uky.edu}
\emailAdd{antal\_jevicki@brown.edu}\emailAdd{kenta.suzuki@polytechnique.edu}

\abstract{We present a systematic procedure to extract the dynamics of the low energy soft mode in SYK type models with a single energy scale $J$ and emergent reparametrization symmetry in the IR. This is given in the framework of the perturbative scheme of \url{arXiv:1608.07567}  based on a specific (off-shell) breaking of conformal invariance in the UV,  adjusted to  yield  the exact large-$N$ saddle point. While this breaking term formally vanishes on-shell, it has a non-trivial effect on correlation functions and the effective action. In particular, it leads to the Schwarzian action with a specific coupling to bi-local matter. The method is applied
to the evaluation of $O(1)$ corrections to the correlation function of bi-locals. As a byproduct we confirm precise agreement with the explicit, symmetry breaking procedure. We provide a verification in the large $q$ limit (Liouville theory), where the correlators can be calculated exactly at all length scales. In this case, our scheme illuminates how the enhanced $O(J)$ and the subleading $O(1)$ contributions originate from the Schwarzian dynamics of the soft mode and its interaction with $h=2$ (bi-local) matter.}

\date{}

\begin{document}
\maketitle

\section{Introduction}
\label{sec:int}
In the large $N$ dynamics of models of SYK type \cite{Sachdev:1992fk, Kitaev:2015, Sachdev:2015efa, Polchinski:2016xgd, Maldacena:2016hyu, Jevicki:2016bwu, Jevicki:2016ito,
Davison:2016ngz, Gross:2017hcz, Gross:2017aos, Kitaev:2017awl, Das:2017wae}\footnote{For reviews of the SYK model see \cite{Sarosi:2017ykf,Rosenhaus:2018dtp}.
Some extensions of the SYK models are discussed for example in \cite{Gu:2016oyy,Gross:2016kjj,Berkooz:2016cvq,Fu:2016vas,Nishinaka:2016nxg,Erdmenger:2016jjg,Turiaci:2017zwd,Peng:2017kro,Dartois:2017xoe,Yoon:2017nig,a:2018kvh,Nosaka:2018iat,Peng:2018zap,Jia:2018ccl,Nayak:2019khe,Qi:2020ian,Klebanov:2020kck}.}
(including tensor models \cite{Gurau:2016lzk,Witten:2016iux,Klebanov:2016xxf,Peng:2016mxj,Krishnan:2016bvg,Li:2017hdt,Gurau:2017xhf,Itoyama:2017emp,Krishnan:2017ztz,Narayan:2017qtw,deMelloKoch:2017bvv,Azeyanagi:2017drg,Giombi:2017dtl,Ferrari:2017jgw,Benedetti:2018goh,Krishnan:2018jsp,Delporte:2018iyf,Diaz:2018eik,deMelloKoch:2019lsx,Benedetti:2019sop}),
a central role is played by the emergent Schwarzian mode which is dual to the gravitational mode in the dual theory.
\footnote{Further investigation of this mode in several directions include \cite{Cotler:2016fpe,Stanford:2017thb, Mertens:2017mtv,Raben:2018sjl,Mertens:2018fds,Maldacena:2018lmt,Lam:2018pvp,Blommaert:2018oro,Chen:2019qqe,Belokurov:2018fnn,Belokurov:2018aol,Khveshchenko:2017mvj,Cardella:2019kum,Ghosh:2019rcj,Iliesiu:2020qvm}}
While the origin of this mode can be traced to (time) reparametrization symmetry of the critical theory, its dynamics and its couplings to the matter degrees of freedom emerges in the near-critical region.

Generally, the low energy dynamics of the soft modes arising from the spontaneous breaking of an approximate symmetry in a quantum field theory
is an important problem which appears in many areas of physics.
In quantum field theories with a set of global symmetries in many physical situations,
one is interested in fluctuations around a classical solution which break some of these symmetries (e.g. a soliton solution \cite{Gervais:1975pa,Gervais:1975yg}).
Then a naive perturbation theory around this solution will be infrared divergent because of a zero mode associated with the broken symmetry.
The way to deal with this is well known: one needs to introduce collective coordinates whose dynamics provides the essential low energy physics.

In many cases, however, the symmetry is approximate, and there is a parameter $\lambda$ such that only when $\lambda = 0$ the action has the symmetry. 
This would cause the saddle point to shift and the zero mode will be lifted to an eigenvalue of order $\lambda$.
In an expansion around this shifted saddle, there is no need for a collective coordinate and the answer for e.g. the connected correlation function would be then proportional to $\lambda^{-1}$
which is large for small $\lambda$, hence ``enhanced''.
Nevertheless, we would like to have an understanding of this enhanced contribution in terms of an effective low energy description.

In the present case, the bi-local collective picture offers a framework to study in detail the emergent zero mode dynamics coupled to bi-local matter.
In particular, in \cite{Jevicki:2016ito} a framework was given for both evaluating the leading Schwarzian action,
and also developing a systematic perturbation expansion near criticality.
It relied on understanding the mechanism for (spontaneously) broken conformal symmetry in these nonlinear theories:
which was introduced in \cite{Jevicki:2016bwu} through an off-shell (source) mechanism.  A related scheme was also used in \cite{Kitaev:2017awl}.

In addition to the Schwarzian action, there is the action of fluctuations with conformal dimension $h=2$
(which should not be confused with the infinite tower corresponding to operators with conformal dimensions $h_n > 2$)
which we call ($h=2$) ``matter" and interaction between this fluctuation field and the Schwarzian mode.
The interpretation of these fluctuations in the gravity dual is not clear at this moment. The importance  of a full understanding of the emergence of the Schwarzian mode (and even more of its interaction with matter) 
 lies in the fact that this relates to  the dual description of the gravitational mode in Jackiw-Teitelboim gravity in two dimensions
\cite{Teitelboim:1983ux, Jackiw:1984je, Almheiri:2014cka, Jensen:2016pah, Maldacena:2016upp, Engelsoy:2016xyb}, and its various possible interactions with matter.
\footnote{Several possibilities for the dual gravity theory for the SYK model are investigated for example in
\cite{Mandal:2017thl,Das:2017pif,Taylor:2017dly,Grumiller:2017qao,Das:2017hrt,Nayak:2018qej,Gaikwad:2018dfc,Lala:2018yib,Gonzalez:2018enk,Moitra:2019bub,Moitra:2019xoj,Hirano:2019iwt,Afshar:2019axx,Alkalaev:2019xuv}}
In this paper we discuss further and complete the scheme of  \cite{Jevicki:2016bwu, Jevicki:2016ito} developing fully the Schwarzian-matter coupled description. The key idea is to replace the explicit symmetry breaking term in the SYK action by a regularized $O(1/J)$ source term.
A systematic diagrammatic picture is then established,
which will allow evaluation of perturbative effects. The singular nature of interactions following from \cite{Jevicki:2016bwu} makes the explicit evaluations of contributions nontrivial.
One of the improvements provided by the present work will be a clear, systematic procedure for their evaluation
(where each diagram will reduce to well defined matrix elements of perturbation theory). Specifically, we prove that once the regularized source is chosen such that one gets the correct large $N$ saddle, this diagrammatic procedure yields the correct $1/J$ expansion of the exact two point correlation function.
These results receive a verification as follows.

In the $q \rightarrow \infty$ limit, the SYK model can be solved with large $N$ for any finite $J$.
In this limit, the bi-local theory reduces to Liouville theory.
\footnote{The fact that the Dyson-Schwinger equations reduce to those which follow from a Liouville action is well known. Here we show that the action itself reduces to the Liouville action.
Correlators in the large $q$ limit are discussed in \cite{Maldacena:2016hyu,Gross:2017hcz,Tarnopolsky:2018env,Streicher:2019wek,Choi:2019bmd}.
} 
The situation here is a bit different from the finite $q$ model where an explicit symmetry breaking term in the action for finite $J$ is present. This breaking term plays an essential role in the large $q$ limit. Nevertheless, as is well known, the resulting Liouville theory has an emergent conformal symmetry. The large-$N$ saddle point breaks this symmetry. 
However, the finite $J$ theory comes with a boundary condition -
and the expansion around the saddle point which obeys this boundary condition does not lead to a zero mode since a candidate zero mode does not obey this boundary condition.
Therefore a calculation for correlators of fluctuations around this saddle is well defined.
This calculation was performed in \cite{Maldacena:2016hyu} at finite temperature $T$ and expanded in an expansion in powers of $T/J$.
We perform this calculation at zero temperature in a different way and reproduce the zero temperature limit of the result of \cite{Maldacena:2016hyu}. Instead of expanding around this saddle point one could have expanded around the saddle point appropriate to the large $J$ (or long distance) limit. Now one gets a zero mode and the formalism developed above can be applied. We then compare an expansion of the exact result with this collective coordinate calculation and show a perfect agreement as expected.

\subsection*{Outline:}
In Section \ref{sec:Schwarzian}, we review the bilocal formalism leading to the coupled Schwarzian/(bi-local) matter system and a corresponding  diagrammatic scheme.
For correlation functions we describe techniques for evaluation of contributing Feynman diagrams using various Schwarzian identities. We apply this to the $O(1)$ correction to the enhanced propagator and show agreement with a perturbative expansion around the exact saddle point solution of the theory. In Section {\ref{sec:Liouville}} we consider the explicit example of large $q$ SYK/ Liouville theory and perform explicit computations  implementing the general method. We conclude with some open questions in Section \ref{sec:conclusion}.

\section{Schwarzian Dynamics in SYK}
\label{sec:Schwarzian}

\subsection{The Method}
\label{sec:method}
In this subsection, we will give a brief review of our formalism \cite{Jevicki:2016bwu, Jevicki:2016ito}which was introduced to exhibit the Schwarzian mode and generate a perturbative expansion 
around the conformal point.
The Sachdev-Ye-Kitaev model \cite{Kitaev:2015} is a quantum mechanical many body system with all-to-all interactions on fermionic $N$ sites ($N \gg 1$), represented by the Hamiltonian
	\begin{equation}
		H \, = \, \frac{1}{4!} \sum_{i,j,k,l=1}^N J_{ijkl} \, \chi_i \, \chi_j \, \chi_k \, \chi_l \, ,
	\end{equation}
where $\chi_i$ are Majorana fermions, which satisfy $\{ \chi_i, \chi_j \} = \delta_{ij}$.
The coupling constant $J_{ijkl}$ are random with a Gaussian distribution,
\ben
P(J_{ijkl}) \, \propto \, {\rm{exp}} \left[-\frac{N^3 J_{ijkl}^2}{12 J^2} \right]
\label{xone}
\een
The original model is given by this four-point interaction, with a simple generalization to analogous $q$-point interacting model \cite{Kitaev:2015,Maldacena:2016hyu}.
In this paper, we follow the more general $q$ model, unless otherwise specified.

After the disorder averaging of the random couplings (which corresponds to annealed averaging at large $N$), the effective action is written as
	\begin{equation}
		S_q \, = \, - \, \frac{1}{2} \int d\tau \sum_{i=1}^N \sum_{a=1}^n \chi_i^a \partial_\tau \chi_i^a
		\, - \, \frac{J^2}{2qN^{q-1}} \int d\tau_1 d\tau_2 \sum_{a, b=1}^n \left( \sum_{i=1}^N \chi_i^a(\tau_1) \chi_i^b(\tau_2) \right)^q \, ,
	\end{equation}
where $a, b$ denote the replica indices. Here $\tau$ is treated as an Euclidean time.
We do not expect a spin glass state in this model \cite{Sachdev:2015efa} and we can restrict to replica diagonal subspace \cite{Jevicki:2016bwu}.
Therefore, introducing a (replica diagonal) bi-local collective field:
	\begin{equation}
		\Psi(\tau_1, \tau_2) \, \equiv \, \frac{1}{N} \sum_{i=1}^N \chi_i(\tau_1) \chi_i(\tau_2) \, ,
	\end{equation}
the model is described by a path-integral
	\begin{equation}
		Z \, = \, \int \prod_{\tau_1, \tau_2} \mathcal{D}\Psi(\tau_1, \tau_2) \ \mu(\Psi) \, e^{-S_{\rm col}[\Psi]} \, , 
	\label{eq:collective partition function}
	\end{equation}
with an appropriate $\mathcal{O}(N^0)$ measure $\mu$ and the collective action:
	\begin{equation}
		S_{\rm col}[\Psi] \, = \, \frac{N}{2} \int d\tau \, \Big[ \partial_\tau \Psi(\tau, \tau')\Big]_{\tau' = \tau} \, + \, \frac{N}{2} \, {\rm Tr} \log \Psi
		\, - \, \frac{J^2N}{2q} \int d\tau_1 d\tau_2 \, \Big[ \Psi(\tau_1, \tau_2) \Big]^q \, ,
	\label{S_col}
	\end{equation}
where the trace log term comes from a Jacobian factor due to the change of path-integral variable, and the trace is taken over the bi-local time.
This action being of order $N$ gives a systematic $1/N$ expansion, while the measure $\mu$ found as in \cite{Jevicki:2014mfa} begins to contribute at one loop level (in $1/N$).

At zero temperature, one can redefine the time $\tau$ to get rid of the energy scale $J$. 
This also shows that in the IR the theory is strongly coupled and the first term linear in the bi-local field can be dropped. At finite temperature $T$ this redefinition rescales the thermal circle to have dimensionless size $J/T$, which then becomes the coupling. In this paper we will deal with the zero temperature limit.

In the IR with the strong coupling  $J|\tau| \gg 1$, the collective action reduces to the critical action 
	\begin{equation}
		S_{\rm c}[\Psi] \, = \, \frac{N}{2} \, {\rm Tr} \log \Psi \, - \, \frac{J^2N}{2q} \int d\tau_1 d\tau_2 \, \Big[ \Psi(\tau_1, \tau_2) \Big]^q \, ,
	\label{S_c}
	\end{equation}
which exhibits an emergent conformal reparametrization symmetry $\tau \to f(\tau)$ with
	\begin{equation}
		\Psi(\tau_1, \tau_2) \, \to \, \Psi_f(\tau_1, \tau_2) \, \equiv \, \Big| f'(\tau_1) f'(\tau_2) \Big|^{\frac{1}{q}} \, \Psi(f(\tau_1), f(\tau_2)) \, .
	\label{reparametrization}
	\end{equation}
The first term in (\ref{S_col}) explicitly breaks this symmetry. 
The saddle point solution of the action (\ref{S_c}) gives the critical classical solution,
	\begin{equation}
		\Psi^{(0)}(\tau_1, \tau_2) \, = \, b \, \frac{{\rm sgn}(\tau_{12})}{|\tau_{12}|^{\frac{2}{q}}} \, , \qquad
		\Psi^{(0)}_f(\tau_1, \tau_2) \, = \, b \left( \frac{|f'(\tau_1)f'(\tau_2)|}{|f(\tau_1) -f(\tau_2)|^2} \right)^{\frac{1}{q}} \, ,
	\label{Psi^0}
	\end{equation}
where $\tau_{ij}\equiv \tau_i - \tau_j$ and $b$ is given by 
\ben
b^q \, = \, \left( \frac{1}{2} - \frac{1}{q} \right) \frac{\tan(\pi/q)}{\pi J^2} \, .
\label{eq:b1}
\een
In general, for a bi-local primary field $\phi_h(\tau_1, \tau_2)$ with the conformal dimension $h$, the transformation is given by
	\begin{equation}
		\phi_h(\tau_1, \tau_2) \, \to \, \phi_{h,f}(\tau_1, \tau_2) \, \equiv \, |f'(\tau_1) f'(\tau_2)|^h \phi_h(f(\tau_1), f(\tau_2)) \, .
	\end{equation}
For an infinitesimal transformation $f(\tau)=\tau+\varepsilon(\tau)$, we get the variation of the field
	\begin{equation}
		\delta \phi_h(\tau_1, \tau_2) \, = \, \int d\tau \, \varepsilon(\tau) \hat{d}_{h, \tau}(\tau_1, \tau_2) \phi_h(\tau_1, \tau_2)
		\, + \, \frac{1}{2} \int d\tau d\tau' \, \varepsilon(\tau) \varepsilon(\tau') \, \hat{d}^{\, (2)}_{h, \tau, \tau'} \phi_h(\tau_1, \tau_2) \, + \, \cdots \, ,
	\label{delta phi_h}
	\end{equation}
where 
	\begin{equation}
		\hat{d}_{h, \tau}(\tau_1, \tau_2) \, = \, h \Big( \delta'(\tau_1 - \tau) + \delta'(\tau_2 - \tau) \Big) + \delta(\tau_1 - \tau) \partial_{\tau_1}+ \delta(\tau_2 - \tau) \partial_{\tau_2} \, ,
	\label{dhat}	
	\end{equation}
and
	\begin{align}
		\hat{d}^{\, (2)}_{h, \tau, \tau'}(\tau_1, \tau_2) \, &= \, 2 h^2 \delta'(\tau_1 - \tau) \delta'(\tau_2 - \tau') \nonumber\\
		&\ \ + h(h-1) \Big( \delta'(\tau_1 - \tau) \delta'(\tau_1 - \tau') + \delta'(\tau_2 - \tau) \delta'(\tau_2 - \tau') \Big) \nonumber\\
		&\ \ + 2h \Big( \delta'(\tau_1 - \tau) + \delta'(\tau_2 - \tau) \Big) \Big( \delta(\tau_1 - \tau') \partial_{\tau_1}+ \delta(\tau_2 - \tau') \partial_{\tau_2} \Big) \nonumber\\
		&\ \ + \delta(\tau_1 - \tau) \delta(\tau_1 - \tau') \partial_{\tau_1}^2 + \delta(\tau_2 - \tau) \delta(\tau_2 - \tau') \partial_{\tau_2}^2  \nonumber\\
		&\ \ + 2 \delta(\tau_1 - \tau) \delta(\tau_2 - \tau') \partial_{\tau_1} \partial_{\tau_2} \, ,	
	\end{align}
The critical saddle point spontaneously breaks the conformal reparametrization symmetry, leading to the appearance of zero modes in the strict IR critical theory.
This problem was addressed in \cite{Jevicki:2016bwu} in analogy with the quantization of extended systems with symmetry modes \cite{Gervais:1975pa}.
The above symmetry mode representing time reparametrization can be elevated to a dynamical variable introduced according to \cite{Gervais:1975yg}
through the Faddeev-Popov method which we summarize below. We insert into the partition function (\ref{eq:collective partition function}), the functional identity:
	\begin{equation}
		\int \prod_{\tau} \mathcal{D}f(\tau) \ \prod_{\tau}\delta\big( \smallint u \cdot \Psi_f \big) \left| \frac{\delta \left(\int u \cdot \Psi_f \right)}{\delta f} \right| \, = \, 1 \, ,
	\end{equation}
where $\delta\big( \smallint u \cdot \Psi_f \big)$ projects out the zero mode. After an inverse change of the integration variable, it results in a combined representation 
	\begin{equation}
		Z \, = \, \int \prod_{\tau} \mathcal{D}f(\tau) \prod_{\tau_1, \tau_2} \mathcal{D}\Psi(\tau_1, \tau_2) \ \mu(\Psi_f) \, \delta\big( \smallint  u \cdot \Psi \big) \, e^{-S_{\rm col}[\Psi_f]} \ ,
	\label{eq:Z}
	\end{equation}
with an appropriate Jacobian. Here
\ben
S_{\text{col}}[\Psi_f] = S_c[\Psi] + \frac{N}{2} \int d\tau \left[ \partial_\tau \Psi_f (\tau,\tau^\prime) \right]_{\tau = \tau^\prime}
\label{scol2}
\een
where we have used the invariance of the critical action. 

The symmetry breaking term implies a modification of the critical, conformal theory, which we would like to calculate perturbatively (in $1/J$) in a long distance expansion.
However since the breaking term is singular and non-vanishing only at {\em short distances}, the corrected solution cannot be obtained by treating it as an ordinary  perturbation. To see this more clearly, let the exact classical solution be 
\ben
\Psi_{\rm cl}=\Psi^{(0)}+\Psi^{(1)} + \cdots
\een 
Here $\Psi^{(1)}$ is the lowest order shift which should satisfy the linearized equation
\begin{equation}
		\int d\tau_3 d\tau_4 \, \mathcal{K}^{(0)}(\tau_1, \tau_2; \tau_3, \tau_4) \Psi^{(1)}(\tau_3, \tau_4) \, = \, \partial_1 \delta(\tau_{12}) \, ,
	\label{Psi_1-eq}
	\end{equation}
where the kernel is given by
	\begin{align}
		\mathcal{K}^{(0)}(\tau_1, \tau_2; \tau_3, \tau_4) \, &= \, \frac{\delta^2 S_c[\Psi_{(0)}]}{\delta\Psi_{(0)}(\tau_1, \tau_2) \delta\Psi_{(0)}(\tau_3, \tau_4)} 	\label{K^0} \\
		&= \, \big[ \Psi_{(0)}\big]_\star^{-1}(\tau_1, \tau_3) \big[ \Psi_{(0)}\big]_\star^{-1}(\tau_2, \tau_4)
		\, + \, (q-1) J^2 \, \delta(\tau_{13}) \delta(\tau_{24}) \, \Psi_{(0)}^{q-2}(\tau_1, \tau_2) \, , \nonumber
	\end{align}
where the inverse $\big[\Psi_{(0)}\big]_\star^{-1}$ is defined in the sense of the star product (i.e. matrix product):
$\int d\tau' A(\tau_1, \tau') \big[A\big]_\star^{-1}(\tau', \tau_2) = \delta(\tau_{12})$ and explicitly given by
	\begin{equation}
		\big[ \Psi_{(0)}\big]_\star^{-1}(\tau_1, \tau_2) \, = \, - \, J^2 b^{q-1} \ \frac{{\rm sgn}(\tau_{12})}{|\tau_{12}|^{2-\frac{2}{q}}} \, .
	\label{Psi_0^-1}
	\end{equation}
Since the kernel transforms under scaling with a dimension $1-\frac{1}{q}$, i.e. 
$\mathcal{K}^{(0)} \sim |\tau|^{-4+4/q}$, one might expect $\Psi^{(1)}$ to have the form
$\Psi^{(1)}(\tau_1, \tau_2) \propto {\rm sgn}(\tau_{12}) |\tau_{12}|^{-4/q}$
which, however is in disagreement with  the desired $\delta'$-source of $\eqref{Psi_1-eq}$. Indeed, one sees that
the exact $\delta'$-source could only  be matched  at the non-perturbative level, where  the $1/J$ corrections are all summed up. 

Nevertheless, it was shown in \cite{Jevicki:2016ito} that a consistent $1/J$ perturbation theory around the critical solution is possible.
The key idea is to {\em replace} the explicit symmetry breaking term in (\ref{S_col}) by a regularized source term which is determined as follows.
One considers an off-shell extension (of $\Psi^{(1)}$):
	\begin{equation}
		\Psi_s^{(1)}(\tau_1, \tau_2) \, = \, B_1 \ \frac{\, {\rm sgn}(\tau_{12}) \, }{|\tau_{12}|^{\frac{2}{q}+2s}} \, ,
	\label{Psi_1}
	\end{equation}
with a parameter $s>0$, because the dimension of $\Psi^{(1)}$ needs to be less than the scaling dimension of $\Psi^{(0)}$.
At the end of the calculation we will take the on-shell limit $s \rightarrow 1/2$. With this ansatz, we 
have
	\begin{equation}
		\int d\tau_3 d\tau_4 \, \mathcal{K}^{(0)}(\tau_1, \tau_2; \tau_3, \tau_4) \Psi_s^{(1)}(\tau_3, \tau_4)
		\, = \, (q-1) B_1 b^{q-2} \gamma(s, q) \, \frac{{\rm sgn}(\tau_{12})}{|\tau_{12}|^{2-\frac{2}{q}+2s}} \, .
	\label{regularized source eq}
	\end{equation}
The coefficient $\gamma$ is uniquely specified by the integral with $\gamma(1/2, q)=0$ \cite{Jevicki:2016ito},
so that the ansatz (\ref{Psi_1}) reduces to the homogeneous (on-shell) equation in the limit $s \to 1/2$.
The overall coefficient $B_1$ obviously does not follow from the above, but must be taken from the numerical result found in \cite{Maldacena:2016hyu}.
In turn, this overall coefficient can be left arbitrary and its value is only fixed after summing the expansion.

The RHS in Eq.(\ref{regularized source eq}) defines an off-shell regularized source term, which takes the form 
	\begin{align}
		Q_s(\tau_1, \tau_2) \, &\equiv \, \int d\tau_3 d\tau_4 \, \mathcal{K}^{(0)}(\tau_1, \tau_2; \tau_3, \tau_4) \Psi_s^{(1)}(\tau_3, \tau_4) \nonumber\\
		&\propto \, (s- \tfrac{1}{2}) \ \frac{{\rm sgn}(\tau_{12})}{|\tau_{12}|^{2-\frac{2}{q}+2s}} \, + \, \mathcal{O}\big( (s-\tfrac{1}{2})^2 \big) \, ,
	\label{eq:Q_s}
	\end{align}
and replaces the non-perturbative source in  (\ref{scol2})  through
	\begin{equation}
		\int \, \big[ \Psi_{ f} \big]_s \, \equiv \, - \, \lim_{s \to \frac{1}{2}} \int d\t_1 d\t_2 \, \Psi_{ f}(\t_1, \t_2) \, Q_s(\t_1, \t_2) \, .
	\label{[Psi_f]_s}
	\end{equation}
We stress that on-shell (with the limit $s \rightarrow 1/2$) so that $\Psi_{1/2}^{(1)}(\tau_1, \tau_2)$ this vanishes.
It is a highly nontrivial feature (of this breaking procedure) that it leads to systematic nonzero effects. In particular, separating the critical classical solution
from the bi-local field: $\Psi = \Psi^{(0)}+ {\bar{\Psi}}$ we get
\begin{equation}
		S_{\rm col}[\Psi, f] \, = \, S[f] \, + \, \lim_{s \to \frac{1}{2}} \int \overline{\Psi}_f \cdot Q_s \, + \, S_{\rm c}[\Psi] \, ,
	\label{S_col[Psi, f]}
	\end{equation}
where $S[f]$ is the action of the collective coordinate
	\begin{equation}
		S[f] \, = \, \lim_{s \to \frac{1}{2}} \int \, \Psi^{(0)}_ f \cdot Q_s \, .
	\label{sf1}
	\end{equation}
This can be evaluated by using the $Q_s$ given above. Taking $s \rightarrow 1/2$ at the end of the calculation one gets the Schwarzian action \cite{Jevicki:2016bwu, Jevicki:2016ito}.
	\begin{equation}
		S[f] \, = \, - \, \frac{\alpha N}{J} \int d\tau \, \left[ \, \frac{f'''(\tau)}{f'(\tau)} \, - \, \frac{3}{2} \, \left( \frac{f''(\tau)}{f'(\tau)} \right)^2 \, \right] \, ,
	\label{S[f]}
	\end{equation}
where $\alpha$ is a dimensionless coefficient \cite{Jevicki:2016ito}. The limit $ s \rightarrow 1/2$ is non-trivial because the evaluation of the integral (\ref{sf1}) leads to a simple pole at $s = 1/2$. This cancels the overall $s-1/2$ in $Q_s$ to yield a finite result. In this sense we can view the source $Q_s$ as an off-shell extension.

The above strategy can alternatively be described as follows.
Suppose we knew the correction to the saddle point solution by e.g. solving the equation of motion following from the SYK action exactly and then performing a long distance expansion.
 Then we can {\em define} a source $Q_s$ using (\ref{eq:Q_s}),
 and replace the symmetry breaking term in the SYK action by the source term $\int d\tau_1 d\tau_2\ Q_s (\tau_1,\tau_2) \Psi (\tau_1,\tau_2)$. 
 In the rest of this section we will show that once this is done, a well defined perturbation expansion around the critical solution can be developed.
 The non-trivial aspect of this scheme is that a specification of this off-shell source is all that is needed to obtain a full near-conformal perturbation expansion.
 We will show for the example of sub-leading corrections for the two point correlation function that the method  produces results which agree with a $1/J$ expansion of the exact answer.

\subsection{Vertices}
\label{sec:vertices}
From (\ref{eq:Z}) and (\ref{S_col[Psi, f]}), near the critical theory, we have the partition function
	\begin{equation}
		Z \, = \, \int \mathcal{D}f \int  \mathcal{D}\Psi \, \mu(f,\Psi) \, \delta\big( \smallint  u \cdot \Psi_f \big) \,
		e^{- \big( S[f] \, + \, \lim_{s\rightarrow \frac{1}{2}} \int \overline{\Psi}_f \cdot Q_s \, + \, S_{\rm c}[\Psi] \big)} \ ,
	\end{equation}
where $S[f]$ is given by the Schwarzian action (\ref{S[f]}).
This leads to the interaction between the Schwarzian mode and the bi-local matter $\Psi$.

First taking an infinitesimal transformation $f(\tau)=\tau+\varepsilon(\tau)$, we expand the Schwarzian action (\ref{S[f]}) to obtain
	\begin{align}
		S[\varepsilon] \, = \, \frac{\alpha N}{2J} \int d\tau \Big[ (\varepsilon'')^2 \, + \, (\varepsilon')^2 \varepsilon''' \, + \, \mathcal{O}(\varepsilon^4) \Big] \, .
	\end{align}
which gives the Schwarzian propagator
	\begin{equation}
		\big\langle \varepsilon(\tau_1) \varepsilon(\tau_2) \big\rangle \, = \, \frac{J}{12\alpha N} \, |\tau_{12}|^3 \, , \qquad
		\big\langle \varepsilon(\omega) \varepsilon(-\omega) \big\rangle \, = \, \frac{J}{\alpha N} \frac{1}{\omega^4} \, ,
	\end{equation}
and the cubic vertex
	\begin{equation}
		V^{(3)}(\tau_1, \tau_2, \tau_3) \, = \, \frac{\alpha N}{6J} \Big[ \partial_1^3 \partial_2 \partial_3 \, + \, (\rm permutations) \Big] \, .
	\end{equation}
Because of the projector $\delta\big( \smallint  u \cdot \Psi_f \big)$ which projects out the zero mode,
the matter bi-local propagator is given by the non-zero mode propagator $\mathcal{D}_c$ (\ref{D_c}).

\begin{figure}[t!]
	\begin{center}
		\scalebox{1.2}{\hspace{0pt} \includegraphics{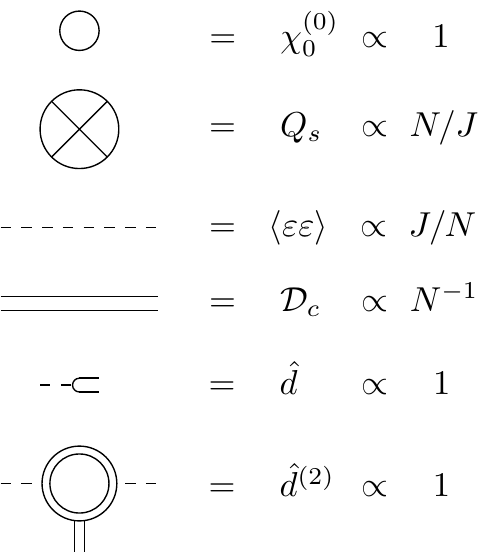}} 
	\end{center}
	\vspace{-10pt}
	\caption{Diagrammatic expressions for the Feynman rules.}
	\label{fig:rules}
\end{figure}

Finally the interaction between the Schwarzian mode and the bi-local matter is given by the interaction term $\int \overline{\Psi}_f \cdot Q_s$.
Taking an infinitesimal transformation $f(\tau)=\tau+\varepsilon(\tau)$ with (\ref{delta phi_h}), we have the interaction term 
	\begin{align}
		\lim_{s \to \frac{1}{2}} \int d\tau_1 d\tau_2 \, Q_s(\tau_{12}) &\bigg[ 1 \, + \, \int d\tau \, \varepsilon(\tau) \hat{d}_{\tau}(\tau_1, \tau_2) \nonumber\\
		&\quad + \, \frac{1}{2} \int d\tau d\tau' \, \varepsilon(\tau) \varepsilon(\tau') \, \hat{d}_{\tau,\tau'}^{\, (2)} \, + \, \cdots \bigg]  \overline{\Psi}(\tau_1, \tau_2) \, .
	\end{align}
where we have used a shorthand notation $\hat{d}_{\tau} \equiv \hat{d}_{\frac{1}{q},\tau}$.
This leads to the $\epsilon(\tau) \overline{\Psi}(\tau_1, \tau_2)$ vertex:
	\begin{align}
		Q_s(\tau_{12})  \hat{d}_{\tau}(\tau_1, \tau_2) \, ,
	\end{align}
the $\epsilon(\tau) \epsilon(\tau') \overline{\Psi}(\tau_1, \tau_2)$ vertex:
	\begin{align}
		\frac{1}{2} \, Q_s(\tau_{12}) \hat{d}_{\tau,\tau'}^{\, (2)}(\tau_1, \tau_2) \, ,
	\end{align}
and so on.
The diagrammatic expressions of these Feynman rules are shown in Figure \ref{fig:rules}.

\subsection{Critical SYK eigenvalue problem}
\label{sec:eigenvalue}
Now we focus on the eigenvalue problem of the critical theory.
We have the conformal symmetry: $\tau \to f(\tau)$ in the critical theory (i.e. strict large $J$):
	\begin{equation}
		\frac{\delta S_c[\Psi^{(0)}]}{\delta f(\tau)} \, = \, 0 \, ,
	\end{equation}
where $S[\Psi^{(0)} + J^{-1} \Psi^{(1)} + \cdots] = S_c[\Psi^{(0)}] \, + \, \cdots$.
This leads to a zero mode in the strict large $J$ limit:
	\begin{equation}
		\chi^{(0)}_{0,\omega}(\tau_1, \tau_2) \, = \, \int d\tau \, e^{i \omega \tau} \chi^{(0)}_{0,\tau}(\tau_1, \tau_2) \, , \ \quad
		\chi^{(0)}_{0,\tau}(\tau_1, \tau_2) \, = \, \big( N^{(0)}_0 \big)^{-\frac{1}{2}} \, \frac{\delta \Psi^{(0)}_f(\tau_1, \tau_2)}{\delta f(\tau)} \bigg|_{f(\tau) = \tau} \, , 
	\label{zero mode}
	\end{equation}
with $\lambda^{(0)}_{0, \omega} = 0$.
The normalization factor $N^{(0)}_0$ for the second equation is fixed by the orthonormality condition of the eigenfunction $\chi^{(0)}$ (\ref{orthonormality}).
This zero mode leads to the enhanced contribution of the Green's function $G^{(-1)}$,
which will be discussed in Section \ref{sec:D^(-1)} as well as in Appendix \ref{app:Perturbation}.
Also for the notation we will use below for the perturbation theory, readers are referred to Appendix \ref{app:Perturbation}.

For general modes, the second term in (\ref{K^0}) is already diagonal, but contains an extra factor $\Psi_{(0)}^{q-2}$.
Because of this extra factor, the eigenvalue problem of this kernel must be like 
	\begin{equation}
		\int d\tau_3 d\tau_4 \, \mathcal{K}^{(0)}(\tau_1, \tau_2; \tau_3, \tau_4) \, \chi^{(0)}_{n, \omega}(\tau_3, \tau_4)
		\, = \, \lambda^{(0)}_{n,\omega} \, \Psi_{(0)}^{q-2}(\tau_1, \tau_2) \, \chi^{(0)}_{n, \omega}(\tau_1, \tau_2) \, .
	\label{eigenvalue0}
	\end{equation}
The extra factor appearing in the RHS can be compensated by defining new eigenfunctions by $\chi^{(0)} = \Psi_{(0)}^{1-q/2} \, \tilde{\chi}^{(0)}$ 
and a new kernel \cite{Polchinski:2016xgd,Maldacena:2016hyu,Jevicki:2016bwu} by
	\begin{equation}
		\widetilde{\mathcal{K}}(\tau_1, \tau_2; \tau_3, \tau_4)
		\, = \, \Psi_{(0)}^{1-\frac{q}{2}}(\tau_1, \tau_2) \, \mathcal{K}(\tau_1, \tau_2; \tau_3, \tau_4) \, \Psi_{(0)}^{1-\frac{q}{2}}(\tau_3, \tau_4) \, ,
	\label{Ktilde}
	\end{equation}
so that the new kernel and eigenfunctions obey the standard eigenvalue problem
	\begin{equation}
		\int d\tau_3 d\tau_4 \, \widetilde{\mathcal{K}}^{(0)}(\tau_1, \tau_2; \tau_3, \tau_4) \, \tilde{\chi}^{(0)}_{n, \omega}(\tau_3, \tau_4)
		\, = \, \lambda^{(0)}_{n,\omega} \, \tilde{\chi}^{(0)}_{n, \omega}(\tau_1, \tau_2) \, .
	\end{equation}
We note that the eigenvalue does not change under this new definition.
Since the new eigenfunctions $\tilde{\chi}^{(0)}$ obey the standard eigenvalue problem, we impose the orthonormality
	\begin{equation}
		\int d\tau_1 d\tau_2 \, \tilde{\chi}^{(0)}_{n, \omega}(\tau_1, \tau_2) \tilde{\chi}^{(0)}_{n', \omega'}(\tau_1, \tau_2)
		\, = \, \delta(n-n') \delta(\omega - \omega') \, , 
	\label{orthonormality}
	\end{equation}
and the completeness 
	\begin{equation}
		\sum_{n, \omega} \, \tilde{\chi}^{(0)}_{n, \omega}(\tau_1, \tau_2) \tilde{\chi}^{(0)}_{n, \omega}(\tau_3, \tau_4)
		\, = \, \delta(\tau_{13}) \delta(\tau_{24}) \, .
	\label{completeness}
	\end{equation}

For the eigenvalue problem (\ref{eigenvalue0}), we denote that $\chi^{(0)}_{n, \omega} = \chi^{(0)}_{h, \omega}$ and $\lambda^{(0)}_{n,\omega} = \lambda^{(0)}_q(h)$
by using the fact that the complete eigenfunctions are spanned by the conformal dimension $h$
containing the continuous modes ($h=1/2+i r$, $0<r<\infty$) and the discrete modes ($h=2n$, $n=0,1,2, \cdots$)
with $\omega$ ($-\infty < \omega < \infty$) \cite{Polchinski:2016xgd, Maldacena:2016hyu}.
The eigenvalue is known to be independent of $\omega$ and we can parametrize it as
	\begin{equation}
		\lambda^{(0)}_q(h) \, = \, (q-1) J^2 \big( 1 \, - \, \widetilde{g}(q; h) \big) \, ,
	\end{equation}
which excludes the trivially diagonalized second term in (\ref{K^0}).
Therefore, the non-trivial eigenvalue problem becomes
	\begin{align}
		&\quad \int d\tau_3 d\tau_4 \, \big[\Psi_{(0)}\big]_\star^{-1}(\tau_1, \tau_3) \big[\Psi_{(0)}\big]_\star^{-1}(\tau_2, \tau_4) \, \chi^{(0)}_{h, \omega}(\tau_3, \tau_4) \nonumber\\
		&= \, - \, (q-1) J^2 \, \widetilde{g}(q; h) \, \Psi_{(0)}^{q-2}(\tau_1, \tau_2) \, \chi^{(0)}_{h, \omega}(\tau_1, \tau_2) \, ,
	\label{critical eigenvalue}
	\end{align}
where the explicit form of the inverse is given in (\ref{Psi_0^-1}).

Next we note that the eigenvalue problem of the following form was solved in \cite{Maldacena:2016hyu}:
	\begin{equation}
		- \frac{1}{\alpha_0(q)} \int d\tau_3 d\tau_4 \, \frac{{\rm sgn}(\tau_{13}) {\rm sgn}(\tau_{24})}{|\tau_{13}|^{\frac{2}{q}} |\tau_{24}|^{\frac{2}{q}} |\tau_{34}|^{2-\frac{4}{q}}} \ 
		\vardbtilde{\chi}^{(0)}_{h, \omega}(\tau_3, \tau_4) \, = \, k_c(q; h) \, \vardbtilde{\chi}^{(0)}_{h, \omega}(\tau_1, \tau_2) \, ,
	\label{MS-eigenvalue}	
	\end{equation}
where the eigenfunctions are independent of $q$ with
	\begin{equation}
		\alpha_0(q) \, = \, \frac{1}{(q-1)J^2 b^q} \, , \qquad 
		\label{eq:b}
	\end{equation}
and the eigenvalue
	\begin{equation} \label{eq:kc(h,q)}
		k_c(q; h) \, = \, - (q-1) \,
		\frac{\Gamma(\frac{3}{2}-\frac{1}{q})\Gamma(1-\frac{1}{q})\Gamma(\frac{1}{q}+\frac{h}{2})\Gamma(\frac{1}{2}+\frac{1}{q}-\frac{h}{2})}
		{\Gamma(\frac{1}{2}+\frac{1}{q})\Gamma(\frac{1}{q})\Gamma(\frac{3}{2}-\frac{1}{q}-\frac{h}{2})\Gamma(1-\frac{1}{q}+\frac{h}{2})} \, .
	\end{equation}
We note that this eigenvalue has a symmetry
	\begin{equation}
		k_c\big( \tfrac{q}{q-1}; h \big) \, = \, \frac{1}{k_c(q; h)} \, .
	\end{equation}
Therefore, from (\ref{MS-eigenvalue}) transforming $q\to q/(q-1)$, we obtain
	\begin{equation}
		- \frac{1}{\alpha_0(\tfrac{q}{q-1})} \int d\tau_3 d\tau_4 \,
		\frac{{\rm sgn}(\tau_{13}) {\rm sgn}(\tau_{24})}{|\tau_{13}|^{2-\frac{2}{q}} |\tau_{24}|^{2-\frac{2}{q}} |\tau_{34}|^{\frac{4}{q}-2}} \ 
		\vardbtilde{\chi}^{(0)}_{h, \omega}(\tau_3, \tau_4) \, = \, \frac{1}{k_c(q; h)} \, \vardbtilde{\chi}^{(0)}_{h, \omega}(\tau_1, \tau_2) \, .
	\end{equation}
This is related to our eigenvalue problem (\ref{critical eigenvalue}) by $\vardbtilde{\chi}^{(0)} = \Psi_{(0)}^{q-2} \, \chi^{(0)}$.
Therefore comparing these expressions, we find 
	\begin{equation}
		\widetilde{g}(q; h) \, = \, \frac{ \alpha_0(\tfrac{q}{q-1})}{(q-1)^2\alpha_0(q)} \frac{1}{k_c(q; h)} \, = \, \frac{1}{k_c(q; h)} \, .
	\label{gtilde}
	\end{equation}

After determining the eigenfunctions \cite{Polchinski:2016xgd, Maldacena:2016hyu, Jevicki:2016bwu}, one can obtain the non-zero mode bi-local propagator $\mathcal{D}_c$.
This propagator contains three types of contributions:
	\begin{equation}
		\mathcal{D}_c \, = \, \mathcal{D}' \, + \, \mathcal{D}'' \, + \, \sum_{m=1}^{\infty} \mathcal{D}_m \, .
	\label{D_c}	
	\end{equation}
The $h=2$ contributions has a single-pole $\mathcal{D}'$ and a double-pole $\mathcal{D}''$ contributions.
Detailed evaluation of these are given in Section \ref{sec:bilocal propagator} and Appendix \ref{app:nuint}.
The former is given by
	\begin{equation}
		\mathcal{D}'(t,z;t',z') \, = \, \frac{|zz'|^{\frac{1}{2}}}{N \tilde{g}'(\frac{3}{2})} \int_{-\infty}^{\infty} \frac{d\omega}{2\pi} \, e^{- i \omega(t-t')} \,
		\frac{3\pi}{2} \, J_{-\frac{3}{2}}(|\omega z^>|)J_{\frac{3}{2}}(|\omega| z^<) \, ,
	\end{equation}
while the latter contribution is given by
	\begin{equation}
		\mathcal{D}''(t,z;t',z') \, = \, - \, \frac{|zz'|^{\frac{1}{2}}}{N \tilde{g}'(\frac{3}{2})} \int \frac{d\omega}{2\pi} \, e^{-i \omega (t-t')}
		 \left( 3 \partial_{\nu} + 2 - \frac{3\tilde{g}''(\frac{3}{2})}{2\tilde{g}'(\frac{3}{2})} \right) \Big( J_{\nu}(|\omega z|) J_{\nu}(|\omega z'|) \Big)_{\nu=\frac{3}{2}} \, , \nonumber
	\label{eq:D''}
	\end{equation}
where we used the change of variables in (\ref{eq:t and z}).
The other contributions are given by single-poles located at $h=p_m+1/2>2$. \cite{Jevicki:2016bwu}
	\begin{equation}
		\mathcal{D}_m(t,z;t',z') \, = \, - \, \frac{|zz'|^{\frac{1}{2}}}{N \tilde{g}'(p_m)} \int \frac{d\omega}{2\pi} \, e^{- i \omega(t-t')} \,
		\frac{\pi p_m}{\sin(\pi p_m)} \, Z_{-p_m}(|\omega z^>|)J_{p_m}(|\omega| z^<) \, .
	\end{equation}

\subsection{Enhanced contribution: $G^{(-1)}$}
\label{sec:D^(-1)}
In this and next subsections, we consider the bi-local two-point function:
	\begin{equation}
		\Big\langle \Psi(\tau_1, \tau_2) \Psi(\tau_3, \tau_4) \Big\rangle \, ,
	\end{equation}
where the expectation value is evaluated by the path integral (\ref{eq:collective partition function}).
After the Faddeev-Popov procedure and changing the integration variable as we discussed in Section \ref{sec:int}, this two-point function becomes
	\begin{equation}
		\Big\langle \Psi_f(\tau_1, \tau_2) \Psi_f(\tau_3, \tau_4) \Big\rangle \, ,
	\end{equation}
where now the expectation value is evaluated by the gauged path integral (\ref{eq:Z}).
We expand the bi-local field around the critical saddle-point solution $\Psi = \Psi^{(0)} + \overline{\Psi}$, so that the two-point function is now
	\begin{equation}
		\Big\langle \Psi_f(\tau_1, \tau_2) \Psi_f(\tau_3, \tau_4) \Big\rangle
		\, = \, \Big\langle \big( \Psi^{(0)}_f + \overline{\Psi}_f \big)(\tau_1, \tau_2) \big( \Psi^{(0)}_f + \overline{\Psi}_f \big)(\tau_3, \tau_4) \Big\rangle\, .
	\end{equation}
Taking an infinitesimal transformation $f(\tau)=\tau+\varepsilon(\tau)$ with (\ref{delta phi_h}), we find the order $\mathcal{O}(J)$ contribution
	\begin{equation}
		G^{(-1)}(\tau_1, \tau_2; \tau_3, \tau_4) \, = \, \int \frac{d\omega}{2\pi} \frac{d\omega'}{2\pi} \, \big( N^{(0)}_{0,\omega} N^{(0)}_{0,\omega'} \big)^{\frac{1}{2}} \, 
		\chi_{0, \omega}^{(0)}(\tau_1, \tau_2) \, \big\langle \epsilon(\omega) \epsilon(\omega') \big\rangle \, \chi_{0, \omega'}^{(0)}(\tau_3, \tau_4) \, ,
	\label{diagram D^-1}
	\end{equation}
where we used the zero mode expression (\ref{zero mode}). $N^{(0)}_{0,\omega}$ is the normalization of the Fourier transform of the zero mode.
With the explicit form of the Schwarzian propagator this leading contribution is of order $\mathcal{O}(J)$ .

Following the standard perturbation theory, (which is described in detail in Appendix \ref{app:Perturbation}), the exact Green's function can be written as
	\begin{equation}
		\widetilde{G}_{(\rm ex)} \, = \, \sum_{n} \, \frac{\tilde{\chi}_n^{(\rm ex)}\tilde{\chi}_n^{(\rm ex)}}{\lambda_n^{(\rm ex)}}
		\,\equiv \, \Psi_{\rm cl}^{\frac{q}{2}-1} \, G_{(\rm ex)} \, \Psi_{\rm cl}^{\frac{q}{2}-1} \, .
	\end{equation}
In the following, we will be interested in the contribution coming from the zero mode of the lowest order kernel, $n=0$. Since $\lambda_0^{(0)}=0$, we have
\begin{equation} \label{eq:D^(-1)}
\begin{split}	
G^{(-1)} =\frac{\chi_0^{(0)} \chi_0^{(0)}} {\lambda_0^{(1)}}
\end{split}
\end{equation}
\begin{equation} \label{eq:D^(0)}
\begin{split}
G^{(0)} =\frac{\chi_0^{(0)} \chi_0^{(1)}}{\lambda_0^{(1)}} +  \frac{\chi_0^{(1)} \chi_0^{(0)}}{\lambda_0^{(1)}} -\frac{\lambda_0^{(2)} \chi_0^{(0)} \chi_0^{(0)}}{(\lambda_0^{(1)})^2} + \mathcal{D}_c, \qquad \mathcal{D}_c = \sum_{n \neq 0}\frac{\chi_n^{(0)} \chi_n^{(0)}} {\lambda_n^{(0)}}  .
\end{split}
\end{equation}
Here we suppressed all $\tau$ (and $\omega$) dependence since they don't play any crucial role here.
The expression of the perturbative kernels are obtained as (see Appendix \ref{app:Perturbation})
	\begin{align}
		\mathcal{K}^{(0)}(\tau_1, \tau_2; \tau_3, \tau_4)
		\, &= \, S_c^{(2)}(\tau_{1, 2}; \tau_{3,4}) \, , \label{eq:K^0-app} \\
		\mathcal{K}^{(1)}(\tau_1, \tau_2; \tau_3, \tau_4) \, &= \, \int d\tau_5 d\tau_6 \, S_c^{(3)}(\tau_{1,2}; \tau_{3,4}; \tau_{5,6}) \, \Psi^{(1)}(\tau_{56}) \, , \label{eq:K^1-app} \\
		\mathcal{K}^{(2)}(\tau_1, \tau_2; \tau_3, \tau_4) \, &= \, \frac{1}{2} \int d\tau_5 d\tau_6 d\tau_7 d\tau_8 \,
		S_c^{(4)}(\tau_{1,2}; \tau_{3,4}; \tau_{5,6}; \tau_{7,8}) \,  \Psi^{(1)}(\tau_{56}) \Psi^{(1)}(\tau_{78}) \nonumber\\
		&\quad + \, \int d\tau_5 d\tau_6 \, S_c^{(3)}(\tau_{1,2}; \tau_{3,4}; \tau_{5,6}) \, \Psi^{(2)}(\tau_{56}) \, , \label{eq:K^2-app}
	\end{align}
where we used the short-hand notation for $S_c^{(n)}$ defined in (\ref{S_c^n}).
The first and second order eigenvalue and eigenfunction corrections are given by
	\begin{align}
		\lambda_0^{(1)} \, &= \, \int \chi_0^{(0)} \cdot \mathcal{K}^{(1)} \cdot \chi_0^{(0)} \, , \label{l01} \\
		\chi_0^{(1)} \, &= \, - \, \sum_{k \ne 0} \frac{\chi_k^{(0)}}{\lambda_k^{(0)}} \, \int \chi_k^{(0)} \cdot \mathcal{K}^{(1)} \cdot \chi_0^{(0)} \, , \ \label{chi01a} \\
		\lambda_0^{(2)} \, &= \, - \, \sum_{k \ne 0} \frac{1}{\lambda_k^{(0)}} \left| \int \chi_k^{(0)} \cdot \mathcal{K}^{(1)} \cdot \chi_0^{(0)} \right|^2
		\, + \, \int \chi_0^{(0)} \cdot \mathcal{K}^{(2)} \cdot \chi_0^{(0)} \, \label{l02} ,
	\end{align}
where for the first order shift of the zero-mode eigenfunction $\chi^{(1)}_0$ we used (\ref{eq:chi^1}). In the rest of this subsection, we relate the diagrammatic expression of $G^{(-1)}$ in (\ref{diagram D^-1}) to the eigenvalue perturbation expression (\ref{eq:D^(-1)}).

Now we note that the Schwarzian action is given by 
	\ben
		S_{\rm Sch}[f] \,  =  \, - \, \lim_{s \rightarrow 1/2} \int d\tau_1 d\tau_2 \, Q_s(\tau_1, \tau_2) \, \Psi^{(0)}_f(\tau_1, \tau_2) \,  
	\label{S_Sch}
	\end{equation}
Expanding $f(\tau) = \tau + \epsilon (\tau)$ we know that in the $ s \rightarrow 1/2$ limit, the first nonzero term in an expansion in $\epsilon(\tau)$ is the term which is quadratic in $\epsilon(\tau)$ and given by \\
\ben
S^{(2)}_{\rm Sch}[\epsilon] = \int d\tau_1 d\tau_2\ Q_s (\tau_1,\tau_2) \int \frac{d\omega}{2\pi} \, \epsilon(\omega) \epsilon(-\omega) {\hat{d}}^{\, (2)}_{\omega,-\omega} \Psi^{(0)} \, .
\label{quadsch}
\een
For the zero mode (\ref{zero mode}) with $\lambda^{(0)}_{0,\omega}=0$, the eigenvalue problem is written as
	\begin{equation}
		0 \, = \, \int d\tau_3 d\tau_4 \, \frac{\delta^2 S_c[\Psi_f^{(0)}]}{\delta \Psi_f^{(0)}(\tau_1, \tau_2) \delta \Psi_f^{(0)}(\tau_3, \tau_4)} \,
		\frac{\delta \Psi_f^{(0)}(\tau_3, \tau_4)}{\delta f(\tau)} \, .
	\label{zero mode eigenvalue}
	\end{equation}
Taking one more variation respect to $f(\tau')$, we obtain
	\begin{align}
		&\quad \int d\tau_3 d\tau_4 d\tau_5 d\tau_6 \,
		\frac{\delta^3 S_c[\Psi^{(0)}]}{\delta \Psi^{(0)}(\tau_1, \tau_2) \delta \Psi^{(0)}(\tau_3, \tau_4) \delta \Psi^{(0)}(\tau_5, \tau_6)} \, 
		\chi^{(0)}_{0, \omega}(\tau_3, \tau_4) \chi^{(0)}_{0, \omega'}(\tau_5, \tau_6) \nonumber\\
		&= \, - \, \big( N^{(0)}_{0,\omega} N^{(0)}_{0,\omega'} \big)^{-\frac{1}{2}} \int d\tau_3 d\tau_4 \,
		\mathcal{K}^{(0)}(\tau_1, \tau_2; \tau_3, \tau_4) {\hat{d}}^{\, (2)}_{\omega,\omega'} \Psi^{(0)}(\tau_3, \tau_4) \, ,
	\end{align}
where after the variation we set $f(\bullet) = \bullet$ and then Fourier transformed $\tau$ to $\omega$ (and $\tau'$ to $\omega'$).
Multiplying $\Psi^{(1)}(\tau_1, \tau_2)$ and integrating, the first term becomes $\mathcal{K}^{(1)}$ (\ref{eq:K^1-app}).
Therefore, now we find
	\begin{equation}
		\frac{1}{J} \, \int \chi^{(0)}_{0, \omega} \cdot \mathcal{K}^{(1)} \cdot \chi^{(0)}_{0, \omega'} \, = \, - \, \big( N^{(0)}_{0,\omega} N^{(0)}_{0,\omega'} \big)^{-\frac{1}{2}}
		\int Q_s \cdot \Big({\hat{d}}^{\, (2)}_{\omega,\omega'} \Psi^{(0)} \Big) \, .
	\label{zero mode identity}
	\end{equation}
Then, the first order zero-mode eigenvalue correction $\lambda^{(1)}_{0,\omega}$ (\ref{l01}) is written as
\begin{equation}
		\frac{\lambda^{(1)}_{0,\omega}}{J} \, \delta(\omega + \omega^\prime) \, = \, - \, \big( N^{(0)}_{0,\omega} \big)^{-1} \int d\tau_1 d\tau_2 \, Q_s(\tau_1, \tau_2)
		\Big( {\hat{d}}^{\, (2)}_{\omega,-\omega} \Psi^{(0)}\Big)(\tau_1, \tau_2) \, .
	\end{equation}
Now this equation immediately shows that 
\begin{equation}
		\frac{J}{\lambda^{(1)}_{0,\omega}} \ \delta(\omega + \omega') \, = \, \frac{N^{(0)}_{0,\omega}}{2\pi} \, \big\langle \epsilon(\omega) \epsilon(\omega') \big\rangle \, .
	\label{<epseps>}
	\end{equation} 
Therefore, $\lambda^{(1)}_0$  is the quadratic kernel of the Schwarzian action.
	
Hence, using this relationship, we relate the diagrammatic expression (\ref{diagram D^-1}) to the perturbation expression (\ref{eq:D^(-1)}) as
	\begin{equation}
		G^{(-1)}(\tau_1, \tau_2; \tau_3, \tau_4) \, = \, J \int \frac{d\omega}{2\pi} \, \frac{\chi_{0, \omega}^{(0)}(\tau_1, \tau_2) \chi_{0, -\omega}^{(0)}(\tau_3, \tau_4)}
		{\lambda_{0,\omega}^{(1)}} \, .
	\end{equation}

\subsection{Diagrams for $G^{(0)}$}
\label{sec:D^(0)}

\begin{figure}[t!]
	\begin{center}
		\scalebox{1.24}{\includegraphics{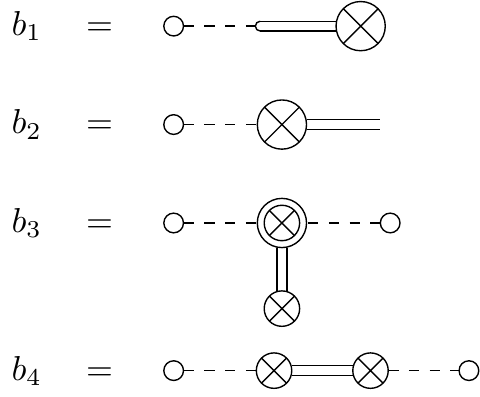}}
	\end{center}
	\caption{Diagrammatic expressions for the $\mathcal{O}(J^0)$ contributions of the Green's function.}
	\label{fig:diagrams}
\end{figure}

Next we consider the subleading  $O( J^0)$ contribution $G^{(0)}$.
Besides the non-zero mode contribution $\mathcal{D}_c$, the contributions to $G^{(0)}$  given by four diagrams (in Figure \ref{fig:diagrams})  that  arise from interactions between the Schwarzian mode and the bilocal field,
and also the Schwarzian redefinition of the field itself:
	\begin{equation}
		G^{(0)} \, = \, b_1 + b_2 + b_3 + b_4 + \mathcal{D}_c \, .
	\end{equation}
The corresponding Feynman diagrams are shown in Figure \ref{fig:diagrams} and they are given by
\footnote{
In this subsection, we always omit the normalization factor $N^{(0)}_{0,\omega}$ introduced through (\ref{zero mode}) in order to simplify the notation.
This factor does not play any crucial role here and the final expressions are independent of this factor.
}
	\begin{align}
		b_1 \, &\equiv \, \bigg\langle \int \frac{d\omega}{2\pi} \, \varepsilon(\omega) \chi^{(0)}_{0, \omega}(\tau_1, \tau_2) \int \frac{d\omega'}{2\pi} \, \varepsilon(\omega')
		\hat{d}_{\frac{1}{q}, \omega'}(\tau_3, \tau_4) \int d\tau_5 d\tau_6 \, \mathcal{D}_c(\tau_3, \tau_4; \tau_5, \tau_6) \, Q_s(\tau_{56}) \bigg\rangle \, , \nonumber\\[4pt]
		b_2 \, &\equiv \, \bigg\langle \int \frac{d\omega}{2\pi} \, \varepsilon(\omega) \chi^{(0)}_{0, \omega}(\tau_1, \tau_2) \int d\tau_5 d\tau_6 \, Q_s(\tau_{56})
		\int \frac{d\omega'}{2\pi} \, \varepsilon(\omega') \hat{d}_{\frac{1}{q}, \omega'}(\tau_5, \tau_6) \mathcal{D}_c(\tau_3, \tau_4; \tau_5, \tau_6) \bigg\rangle \, , \nonumber\\[4pt]
		b_3 \, &\equiv \, \frac{1}{2} \, \bigg\langle \int \frac{d\omega_a}{2\pi} \, \varepsilon(\omega_a) \chi^{(0)}_{0, \omega_a}(\tau_1, \tau_2) 
		\int \frac{d\omega_b}{2\pi} \frac{d\omega_c}{2\pi} \, \varepsilon(\omega_b) \varepsilon(\omega_c) \int d\tau_5 d\tau_6 d\tau_7 d\tau_8 \, Q_s(\tau_{56}) \nonumber\\[4pt]
		&\qquad \quad \times \hat{d}^{\,(2)}_{\frac{1}{q}, \omega_b, \omega_c}(\tau_5, \tau_6) \mathcal{D}_c(\tau_5, \tau_6; \tau_7, \tau_8) \, Q_s(\tau_{78}) \int \frac{d\omega_d}{2\pi} \,
		\varepsilon(\omega_d) \chi^{(0)}_{0, \omega_d}(\tau_3, \tau_4) \bigg\rangle \, , \\[4pt]
		b_4 \, &\equiv \, \bigg\langle \int \frac{d\omega_a}{2\pi} \, \varepsilon(\omega_a) \chi^{(0)}_{0, \omega_a}(\tau_1, \tau_2) \int d\tau_5 d\tau_6 \, Q_s(\tau_{56})
		\int \frac{d\omega_b}{2\pi} \, \varepsilon(\omega_b) \hat{d}_{\frac{1}{q}, \omega_b}(\tau_5, \tau_6) \bigg\rangle \nonumber\\[4pt]
		&\ \times \bigg\langle \int \frac{d\omega_c}{2\pi} \, \varepsilon(\omega_c) \chi^{(0)}_{0, \omega_c}(\tau_3, \tau_4) \int d\tau_7 d\tau_8 \, Q_s(\tau_{78})
		\int \frac{d\omega_d}{2\pi} \, \varepsilon(\omega_d) \hat{d}_{\frac{1}{q}, \omega_d}(\tau_7, \tau_8) \bigg\rangle 
		\mathcal{D}_c(\tau_5, \tau_6; \tau_7, \tau_8) \, , \nonumber
	\end{align}
where we defined $\hat{d}_{h, \omega}$ as the Fourier transform of $\hat{d}_{h, \tau}$ defined in Eq.(\ref{dhat}).
For $b_1$ and $b_2$, there are also ($\tau_{1,2} \leftrightarrow \tau_{3,4}$) contributions we did not write explicitly above.
We note that the integration by parts of this operator is given by a shadow transform:
	\begin{align}
		\int d\tau_1 d\tau_2 \, A(\tau_1, \tau_2) \hat{d}_{h, \tau}(\tau_1, \tau_2) B(\tau_1, \tau_2)
		\, = \, - \, \int d\tau_1 d\tau_2 \, B(\tau_1, \tau_2) \hat{d}_{1-h, \tau}(\tau_1, \tau_2) A(\tau_1, \tau_2) \, .
	\label{shadow}
	\end{align}
From the definition of the operator (\ref{dhat}), the action of this operator onto a product can also be written as
	\begin{align}
		&\quad \, \hat{d}_{h_A+h_B, \tau}(\tau_1, \tau_2) \Big( A(\tau_1, \tau_2) B(\tau_1, \tau_2) \Big) \nonumber\\
		&= \, B(\tau_1, \tau_2) \hat{d}_{h_A, \tau}(\tau_1, \tau_2) A(\tau_1, \tau_2) \, + \, A(\tau_1, \tau_2) \hat{d}_{h_B, \tau}(\tau_1, \tau_2) B(\tau_1, \tau_2) \, .
	\label{product rule}
	\end{align}
Now we evaluate the diagrams.
For $b_1$, using (\ref{eq:Q_s}) and (\ref{<epseps>}), we find
	\begin{align}
		b_1 \, &= \, \int \frac{d\omega}{2\pi} \frac{d\omega'}{2\pi} \, \chi^{(0)}_{0, \omega}(\tau_1, \tau_2) \, \big\langle \varepsilon(\omega) \varepsilon(\omega') \big\rangle \,
		\hat{d}_{\frac{1}{q}, \omega'}(\tau_3, \tau_4) \Psi^{(1)}(\tau_3, \tau_4) \nonumber\\ 
		&= \, J \int \frac{d\omega}{2\pi} \, \frac{\chi^{(0)}_{0, \omega}(\tau_1, \tau_2)}{\lambda^{(1)}_{0,\omega}} \, \big( \hat{d}_{\frac{1}{q}, -\omega} \Psi^{(1)} \big)(\tau_3, \tau_4) \, .
	\end{align}
For $b_2$, first using (\ref{eq:D^(0)}) and (\ref{<epseps>}) we have
	\begin{equation}
		b_2 \, = \, J \int \frac{d\omega}{2\pi} \, \frac{\chi^{(0)}_{0, \omega}(\tau_1, \tau_2)}{\lambda^{(1)}_{0,\omega}} \int \frac{d\omega'}{2\pi} \sum_{k \ne 0}
		\frac{\chi^{(0)}_{k, -\omega'}(\tau_3, \tau_4)}{\lambda^{(0)}_{k,\omega'}}
		\int d\tau_5 d\tau_6 \, Q_s(\tau_{56}) \big( \hat{d}_{\frac{1}{q}, -\omega} \chi^{(0)}_{k, \omega'} \big)(\tau_5, \tau_6) \, .
	\label{b2}
	\end{equation}
Now following the same manipulation as in (\ref{zero mode eigenvalue}) - (\ref{zero mode identity}), but now for non-zero mode, we find
	\begin{equation}
		\frac{1}{J} \int \chi^{(0)}_{0, \omega} \cdot \mathcal{K}^{(1)} \cdot \chi^{(0)}_{k, \omega'} \, + \, \int Q_s \cdot \Big({\hat{d}}_{\frac{1}{q}, \omega} \chi^{(0)}_{k, \omega'} \Big)
		\, = \, \lambda_{k,\omega'}^{(0)} \, \int \Psi^{(1)} \cdot \hat{d}_{1-\frac{1}{q}, \omega} \Big( \Psi_{(0)}^{q-2} \chi^{(0)}_{k, \omega'} \Big) \, .
	\label{non-zero mode identity}
	\end{equation}
For the RHS, we can move $\hat{d}$ onto $\Psi_{(1)}$ by the shadow transform (\ref{shadow}).
Using this relation, $b_2$ is now written as
	\begin{align}
		b_2 \, &= \, - \, J \int \frac{d\omega}{2\pi} \, \frac{\chi^{(0)}_{0, \omega}(\tau_1, \tau_2)}{\lambda^{(1)}_{0,\omega}} \,
		\big( \hat{d}_{\frac{1}{q}, -\omega} \Psi_{(1)} \big)(\tau_3, \tau_4) \nonumber\\
		&\quad - \, \int \frac{d\omega}{2\pi} \, \frac{\chi^{(0)}_{0, \omega}(\tau_1, \tau_2)}{\lambda^{(1)}_{0,\omega}} \int \frac{d\omega'}{2\pi} \sum_{k \ne 0}
		\frac{\chi^{(0)}_{k, -\omega'}(\tau_3, \tau_4)}{\lambda^{(0)}_{k,\omega'}} \int \chi^{(0)}_{k, \omega'} \cdot \mathcal{K}^{(1)} \cdot \chi^{(0)}_{0, -\omega} \, ,
	\end{align}
where for the first term, we used the completeness of $\tilde{\chi}^{(0)}$ (\ref{completeness}).
The first term cancels with $b_1$, while the second term combines into the first order shift of the eigenfunctions (\ref{chi01a}).
Therefore, we find
	\begin{align}
		b_1 \, + \, b_2 \, &= \, \int \frac{d\omega}{2\pi} \, \frac{\chi^{(0)}_{0, \omega}(\tau_1, \tau_2) \chi^{(1)}_{0, -\omega}(\tau_3, \tau_4)}{\lambda^{(1)}_{0,\omega}}
		\, + \, (\tau_{1,2} \leftrightarrow \tau_{3,4}) \, ,
	\label{b1+b2}
	\end{align}
where we implemented the explicit symmetrization of the external legs. 
\\For $b_3$, using (\ref{eq:Q_s}) and (\ref{<epseps>}), we find
	\begin{equation}
		b_3 \, = \, \frac{J^2}{2} \int \frac{d\omega}{2\pi} \frac{d\omega'}{2\pi} \, \frac{\chi^{(0)}_{0, \omega}(\tau_1, \tau_2) \chi^{(0)}_{0, \omega'}(\tau_3, \tau_4)}
		{\lambda^{(1)}_{0,\omega} \, \lambda^{(1)}_{0,\omega'}} \, S_c^{(2)}\Big[\Psi^{(1)}; \hat{d}^{\, (2)}_{h, -\omega, -\omega'} \Psi^{(1)} \Big] \, .
	\end{equation}
Now we consider the equation of motion of $\Psi^{(1)}$ (\ref{eq:Q_s}) with the on-shell limit ($s\to 1/2$):
$\int \mathcal{K}_f^{(0)} \cdot \Psi_f^{(1)} = 0$, where we transformed $\Psi \to \Psi_f$.
Taking variations respect to $f(\tau)$ and $f(\tau')$, then multiplying another $\Psi^{(1)}$, we find
	\begin{align}
		\int \Psi^{(1)} \cdot \mathcal{K}^{(0)} \cdot \Big( \hat{d}^{\, (2)}_{\frac{1}{q}, \tau, \tau'} \Psi^{(1)} \Big)
		\, = \, & - \, \frac{1}{J} \int \chi_{0, \tau}^{(0)} \cdot \mathcal{K}^{(1)} \cdot \Big( \hat{d}_{\frac{1}{q}, \tau'} \Psi^{(1)} \Big) \, - \, (\tau \leftrightarrow \tau') \nonumber\\
		& - \, \frac{2}{J^2} \int \chi_{0, \tau}^{(0)} \cdot \mathcal{K}^{(2)} \cdot \chi_{0, \tau'}^{(0)} \, .
	\end{align}
Therefore, we can write $b_3$ as
	\begin{align}
		b_3 \, &= \, - \int \frac{d\omega}{2\pi} \frac{d\omega'}{2\pi} \, \frac{\chi^{(0)}_{0, \omega}(\tau_1, \tau_2) \chi^{(0)}_{0, \omega'}(\tau_3, \tau_4)}
		{\lambda^{(1)}_{0,\omega} \, \lambda^{(1)}_{0,\omega'}} \nonumber\\
		&\qquad \times \Bigg[ \, J \int \chi_{0, -\omega}^{(0)} \cdot \mathcal{K}^{(1)} \cdot \Big( \hat{d}_{\frac{1}{q}, -\omega'} \Psi^{(1)} \Big)
		\, + \, \int \chi_{0, -\omega}^{(0)} \cdot \mathcal{K}^{(2)} \cdot \chi_{0, -\omega'}^{(0)} \Bigg] \, .
	\end{align}
For $b_4$, first using (\ref{eq:D^(-1)}) and (\ref{<epseps>}) we have
	\begin{align}
		b_4 \, &= \, J^2 \int \frac{d\omega}{2\pi} \frac{d\omega'}{2\pi} \, \frac{\chi^{(0)}_{0, \omega}(\tau_1, \tau_2) \chi^{(0)}_{0, \omega'}(\tau_3, \tau_4)}
		{\lambda^{(1)}_{0,\omega} \, \lambda^{(1)}_{0,\omega'}} \int d\tau_5 d\tau_6 d\tau_7 d\tau_8 \, Q_s(\tau_{56}) Q_s(\tau_{78}) \nonumber\\
		&\hspace{60pt} \times \int \frac{d\omega''}{2\pi} \sum_{k \ne 0} \frac{1}{\lambda^{(0)}_{k,\omega''}}
		\Big( \hat{d}_{h, -\omega} \chi^{(0)}_{k, \omega''}\Big) (\tau_5, \tau_6)
		\Big( \hat{d}_{h, -\omega'} \chi^{(0)}_{k, -\omega''}\Big) (\tau_7, \tau_8) \, .
	\end{align}
Again using the relation (\ref{non-zero mode identity}), this is written as
	\begin{align}
		b_4 \, &= \, \int \frac{d\omega}{2\pi} \frac{d\omega'}{2\pi} \, \frac{\chi^{(0)}_{0, \omega}(\tau_1, \tau_2) \chi^{(0)}_{0, \omega'}(\tau_3, \tau_4)}
		{\lambda^{(1)}_{0,\omega} \, \lambda^{(1)}_{0,\omega'}} \int \frac{d\omega''}{2\pi} \sum_{k \ne 0} \frac{1}{\lambda^{(0)}_{k,\omega''}} \nonumber\\
		&\quad \times \Bigg[ J \lambda^{(0)}_{k,\omega''} \int \Psi_{(1)} \cdot \hat{d}_{1-\frac{1}{q}, -\omega} \Big( \Psi_{(0)}^{q-2} \, \chi^{(0)}_{k, \omega''} \Big)
		\, - \, \int \chi^{(0)}_{0, -\omega} \cdot \mathcal{K}^{(1)} \cdot \chi^{(0)}_{k, \omega''} \Bigg] \nonumber\\
		&\quad \times \Bigg[ \omega \to \omega'\, , \, \omega'' \to - \omega'' \Bigg] \, .
	\end{align}
For the first term, we use the shadow transform (\ref{shadow}) to move $\hat{d}$ onto $\Psi_{(1)}$.
After using the completeness of $\tilde{\chi}^{(0)}$ (\ref{completeness}), we obtain
	\begin{align}
		b_4 \, &= \, \int \frac{d\omega}{2\pi} \frac{d\omega'}{2\pi} \, \frac{\chi^{(0)}_{0, \omega}(\tau_1, \tau_2) \chi^{(0)}_{0, \omega'}(\tau_3, \tau_4)}
		{\lambda^{(1)}_{0,\omega} \, \lambda^{(1)}_{0,\omega'}} \\
		&\ \times \Bigg[ J \int \chi_{0, -\omega}^{(0)} \cdot \mathcal{K}^{(1)} \cdot \Big( \hat{d}_{\frac{1}{q}, -\omega'} \Psi^{(1)} \Big)
		\, + \, \int \frac{d\omega''}{2\pi} \sum_{k \ne 0} \, \frac{1}{\lambda_{k,\omega''}^{(0)}}
		\left| \int \chi_k^{(0)} \cdot \mathcal{K}^{(1)} \cdot \chi_0^{(0)} \right|^2 \Bigg] \, . \nonumber
	\end{align}
The first term precisely cancels with $b_3$, while the second term combines with the other term in $b_3$ to give the second order eigenvalue shift.
Therefore, we finally obtain
	\begin{align}
		b_3 \, + \, b_4 \, &= \, - \, \int \frac{d\omega}{2\pi} \, \frac{\lambda_{0, \omega}^{(2)} \, \chi^{(0)}_{0, \omega}(\tau_1, \tau_2) \chi^{(0)}_{0, -\omega}(\tau_3, \tau_4)}
		{\lambda^{(1)}_{0,\omega} \, \lambda^{(1)}_{0,-\omega}} \, .
	\label{b3+b4}
	\end{align}
The first and second order eigenvalue and eigenfunction corrections which will be needed below are given by (\ref{l01}) - (\ref{l02}).
\\We now found that the diagrammatic expressions denoted by $b_1$ - $b_4$ shown in Figure \ref{fig:diagrams} are given by the matrix elements (\ref{b1+b2}) and (\ref{b3+b4}). These are in agreement with the standard perturbative evaluations summarized in Appendix \ref{app:Perturbation}.
In the next section, we will explicitly evaluate these matrix elements in the large $q$ SYK model as an example.
\\The total $O(J^0)$ contribution to the correlation function is given by the following
	\begin{equation}
		G^{(0)} \, = \, b_1 + b_2 + b_3 + b_4 + \mathcal{D}_c \, ,
	\end{equation}
It is seen that the contribution of diagrams $b_3+b_4$ is precisely of the same form, and cancels the double pole term in the propagator $\mathcal{D}^{\prime\prime}$.
The result is therefore given by the single pole propagator $\mathcal{D}'$ and the result from diagrams $b_1+b_2$
which as we have shown above involves the first order correction to the zero mode eigenfunction. We evaluate this (and the other matrix elements) in the large $q$ case in the following section.
In this limit (with the higher massive modes decoupling) the result reads:
	\begin{align} 
		\langle \eta(t,z) \eta(t^\prime,z^\prime)\rangle \, \simeq \, \frac{(zz^\prime)^{\frac{1}{2}}}{2} &\int_{-\infty}^{\infty} d\omega\ e^{i\omega(t-t^\prime)}
		\Bigg[J_{-\frac{3}{2}}(\vert \omega \vert z^>) J_{\frac{3}{2}}(\vert \omega \vert z^<) \\
		&\qquad +\frac{1}{\vert\omega\vert} \left( \frac{1}{2z}+\frac{1}{2z^\prime} +\partial_z + \partial_{z'} \right)
		J_{\frac{3}{2}}(\vert \omega \vert z) J_{\frac{3}{2}}(\vert \omega \vert z^\prime) \Bigg] \nonumber \, .
	\end{align}
This result consists of two terms: the first one due to Schwarzian interaction with the bi-local and the second representing the contribution of the $h=2$ matter.
For general $q$ one would have the further contribution of massive mode propagators.
We also mention that a similar evaluation in \cite{Kitaev:2017awl} gave a different result. We are not sure why the method of \cite{Kitaev:2017awl} disagrees at $O(J^0)$, however we have compared 
our result in the large $q$ limit, where a finite $J$ evaluation is possible, and observed disagreement (this will be seen in the next Section). We also note that the contributions of diagrams $b_3+b_4$  represents the $O(J^0)$  correction to the $O(J)$ Schwarzian action (at the quadratic level). This again differes from the correction found in\cite{Kitaev:2017awl} .

\section{Large $q$: Liouville Theory } 
\label{sec:Liouville}
The scheme described in the previous section ensures that once we have an expression for the first correction to the critical saddle point, $\Psi^{(1)}$, an unambiguous perturbation theory (which is really a derivative expansion) can be developed by replacing the actual symmetry breaking source term by a new source term which would give rise to this $\Psi^{(1)}$. This procedure is necessary since the symmetry breaking term is singular and nonzero only at short distances, which implies that this cannot be used as a perturbation in a long distance expansion. This means one has to make an ansatz for $\Psi^{(1)}$ given in  (\ref{Psi_1}). 

In the large $q$ limit, the SYK model simplifies considerably and at the leading order the action is that of Liouville theory on the bilocal space. At large $N$ this model can be now solved exactly at all scales, i.e. for any finite $J$. Even though the symmetry breaking term of the SYK model needs to be included to obtain this large $q$ limit, the resulting Liouville action acquires an emergent reparametrization symmetry. The large $N$ saddle point breaks this symmetry. However, this does {\em not} lead to a zero mode, and a calculation of the exact bilocal function proceeds without any obstruction. In particular, in this exact calculation there is no emergent Schwarzian dynamics.

Nevertheless, we would like to understand the long distance derivative expansion in this limit, and extract the dynamics of the soft mode, since the Schwarzian action is the most direct link to a dual description in terms of JT gravity. In this section we develop this expansion using the scheme of the previous section. However, now we can simply use the leading correction to the critical saddle point to determine the regularized source $Q_s$. The result of the previous section then ensures that the scheme using this source reproduces the large $J$ expansion of the exact answer for correlation functions.


\subsection{Bilocal theory at large $q$} \label{subsec:deraction}
Let us begin with the collective action for SYK model $\eqref{S_col}$
	\begin{equation} \label{eq:collaction}
		S_{\rm col}[\Psi] \, = \, - \, \frac{N}{2} \int d\tau \, \Big[ \partial_\tau \Psi(\tau, \tau')\Big]_{\tau' \to \tau} \, + \, \frac{N}{2} {\rm Tr} \log \Psi
		\, - \, \frac{2^{q-2}\mathcal{J}^2 N}{q^2} \, \int d\tau_1 d\tau_2 \, \big[ \Psi(\tau_1, \tau_2) \big]^q\, ,
	\end{equation}
where we have defined	
	\begin{equation}
		\mathcal{J}^2 \, \equiv \, \frac{q}{2^{q-1}} \, J^2 \, .
	\label{eq:mathcalJ}
	\end{equation}
This $\mathcal{J}$ is kept fixed as $q \to \infty$. At large $q$, we can do the following field redefinition 
	\begin{equation}
		\Psi(\tau_1, \tau_2) \, = \, \frac{{\rm sgn}(\tau_{12})}{2} \, \left[ 1 \, + \, \frac{\Phi(\tau_1, \tau_2)}{q}  \right] \, .
	\label{eq:Phi-first}
	\end{equation}
Using $\eqref{eq:Phi-first}$ in $\eqref{eq:collaction}$, and performing a $1/q$ expansion we get a field independent $O(1)$ term and the next contribution is $O(1/q^2)$, given by the Liouville action
	\begin{align}
		S_{\rm L}[\Phi] \, = \, - \, \frac{N}{16q^2} \int d\tau_1 d\tau_2 \, &\bigg[ \partial_1 \big( {\rm sgn}(\tau_{12}) \Phi(\tau_1, \tau_2) \big)
		\partial_2 \big( {\rm sgn}(\tau_{21}) \Phi(\tau_2, \tau_1) \big)  \nonumber\\
		&\quad \, + \, 4 \mathcal{J}^2 \, e^{\Phi(\tau_1, \tau_2)} \bigg] \, + \, O(q^{-3})\, .
	\label{eq:S_L}
	\end{align}
The details of the derivation of (\ref{eq:S_L}) is given in Appendix \ref{app:SLderiv}. 
Note that there is no $O(1/q)$ term in this expansion. The kinetic term of SYK - the first term of (\ref{eq:collaction})- provides a $1/q$ piece which cancels with a $1/q$ piece coming from the second term. The inclusion of the symmetry breaking term is crucial.

Nevertheless the action $S_{\rm L}[\Phi]$ has an emergent reparametrization symmetry for $\tau_i \rightarrow f(\tau_i)$,
\bea
 \Phi(\tau_1,\tau_2) & \rightarrow & \Phi(f(\tau_1),f(\tau_2)) + \log| f^\prime (\tau_1) f^\prime (\tau_2) | \, .
\label{aone}
\eea
At finite temperature and finite $\mathcal{J}$, we need to impose the physical requirement that the expectation value of the bilocal field $\langle\Psi (\tau_1,\tau_2) \rangle$ should be equal to the free fermion two point function $\frac{1}{2} {\rm sgn} (\tau_{12})$ in the short distance limit. This means we need to impose a boundary condition
\ben
\Phi (\tau,\tau) \, = \, 0 \, .
\label{atwo}
\een
At zero temperature the expansion is really in $\mathcal{J} |\tau_{12}|$: this means that we cannot really access the point $\tau_1 = \tau_2$. However the zero temperature theory should be really thought of as a limit of the finite temperature theory. Accordingly we should impose the condition (\ref{atwo}) even at zero temperature. 
In fact, as is well known, the Liouville action on an infinite plane has a symmetry which has two copies of Virasoro, i.e. $\tau_1$ and $\tau_2$ can be reparametrized by different functions. The restriction to the same function as in (\ref{aone}) comes from this boundary condition (\ref{atwo}).

The above derivation is in a $1/q$ expansion. The fact that the resulting action is a standard two derivative action signifies that to leading order of this expansion there is a single pole of the two point correlation function of the bi-local field. This appears to be in conflict with the well known fact that even for $q=\infty$  there are an infinite number of poles in the conformal limit, i.e. an infinite number of solutions of $\tilde{g}(q;h)=1$ where $\tilde{g}(q;h)$ is defined in $\eqref{gtilde}$. In fact at large $q$ the solutions to this equation are given by $h=2$ and the tower \cite{Gross:2017hcz} 
\beq
h_n = 2n+1+\frac{2}{q}\frac{2n^2+n+1}{2n^2+n-1} +O(1/q^2),\qquad\qquad n=1,2,\ldots
\eeq
However the residues of these poles all vanish as $q\rightarrow \infty$ except for $h=2$ \cite{Das:2017hrt}. The $1/q$ expansion of the function $\tilde{g}(q;h)$ is given by
\begin{align}
\tilde{g}(q;h) \, = \, \frac{h(h-1)}{2} & \Bigg[1+\frac{1}{q} \bigg(\frac{2}{h(h-1)}+3  \nonumber\\
&\quad -2\left[2\gamma+\log 4+\psi\left(\frac{1}{2}-\frac{h}{2}\right)+\psi\left(\frac{h}{2}\right)\right]\bigg) +O(q^{-2})\Bigg] \, , 
\end{align}
where $\psi(x)$ denotes the digamma function and $\gamma$ is the Euler-Mascheroni constant. Since $\psi(-n)$ for integer $n$ has a pole, the coefficient of the $1/q$ term is singular for $h = 2n + 1$. This is the signature of the infinite number of solutions of the spectral equation in a $1/q$ expansion. In the following, however, we will restrict our attention to values of $h$ close to 2. Therefore these other solutions will not be relevant for us. Note, however, on-shell modes corresponding to these infinite tower of solutions do have non-trivial higher point correlation functions \cite{Gross:2017hcz}.

\subsection{Quantum fluctuations at leading order} \label{subsec:qfluc}
The equation of motion which follows from the action (\ref{eq:S_L}) is given by
	\begin{equation}
		\partial_1 \partial_2 \, \Big( {\rm sgn}(\tau_{12}) \Phi(\tau_1, \tau_2) \Big) \, =- \, 2 \mathcal{J}^2 \, {\rm sgn}(\tau_{12}) \, e^{\Phi(\tau_1, \tau_2)} \, .
	\label{eq:L-EOM}
	\end{equation}
We should really view the zero temperature theory as a limit of the finite temperature theory. At finite temperature, periodicity in both $\tau_1$ and $\tau_2$, together with the boundary condition (\ref{atwo}) determines the solution uniquely \cite{Maldacena:2016hyu} and its zero temperature limit is given by
	\begin{equation}
		\Phi_{\text{cl}}(\tau_{12}) \, = \, - \, 2 \log\Big( \mathcal{J} |\tau_{12}| + 1 \Big) \, ,
	\label{eq:phi_0}
	\end{equation}
where $\tau_{12} \equiv \tau_1-\tau_2$.
Let us now consider quantum fluctuations around the saddle point solution 
$\eqref{eq:phi_0}$ by defining 
	\begin{equation}
		\Phi(\tau_1, \tau_2) \, = \, \Phi_{\text{cl}}(\tau_{12}) \, + \, \sqrt{\frac{2}{N}} \ \eta(\tau_1, \tau_2) \, ,
	\label{eta}
	\end{equation}
where $\eta$ denotes the quantum fluctuations. The fluctuation must also obey the boundary condition
\ben
\eta (\tau,\tau) \, = \, 0 \, .
\label{asix}
\een
Substituting $\eqref{eta}$ in $\eqref{eq:S_L}$, we get the quadratic action for quantum fluctuations as
	\begin{equation}
		S_{(2)}[\eta]
		\, = \, - \, \frac{1}{8q^2} \int d\tau_1 d\tau_2 \ \eta(\tau_1, \tau_2) \left[ \partial_1 \partial_2 \, + \, 2 \mathcal{J}^2 \, e^{\Phi_{\text{cl}}(\tau_{12})} \right] \eta(\tau_1, \tau_2) \, .
	\label{Seta}
	\end{equation}
Using the bilocal map
\begin{align}
		t \, \equiv \, \frac{\tau_1+\tau_2}{2} \, , \qquad z \, \equiv \, \frac{\tau_1-\tau_2}{2} \, ,
	\label{eq:t and z}
	\end{align}
we can write, using $\eqref{eq:phi_0}$
\ben
S^{(2)}[\eta]=\frac{1}{16q^2}\int_{-\infty}^\infty dt\int_0^\infty dz\ \eta(t,z) \left[-\partial_t^2+\partial_z^2-\frac{8\mathcal{J}^2}{(2\mathcal{J}z+1)^2}\right]\eta(t,z) \, .
 \label{SLiouville}
\een
The operator which appears in the square bracket is the expression for ${\cal K}_{{\rm {ex}}}$ which is defined in (\ref{K_ex}). 
\\In the following it will be convenient to define
\ben
\tilde{z} \, \equiv \, z+\frac{1}{2\mathcal{J}} \, .
\een
It is convenient to define a conformally covariant operator ${\widetilde{\cal{K}}_L}$ (similar to (\ref{Ktilde}))
\ben
{\widetilde{\cal{K}}}_{\text{L}} \, = \, {\tilde{z}}\left[-\partial_t^2+\partial_z^2-\frac{2}{\tilde{z}^2}\right]  {\tilde{z}}\, ,
\label{fone}
\een
and redefine the field
\ben
\tilde{\eta} (t,z) \, \equiv \, \frac{\eta (t,z)}{\tilde{z}} \, .
\label{fthree}
\een
The eigenvalue problem we need to solve is
\ben
{\widetilde{\cal{K}}}_{\text{L}} \, \tilde{\chi}^{\text{(ex)}}_{\nu}(t,z) \, = \, \lambda_{\nu}^{(\rm{ex})} \tilde{\chi}^{\text{(ex)}}_{\nu}(t,z) \, .
\label{ffive}
\een
These eigenfunctions are given by
\ben
\tilde{\chi}^{\text{(ex)}}_{\nu,\omega}(t,z) \, = \, \frac{e^{i\omega t} \tilde{z}^{-\frac{1}{2}}}{\sqrt{2\pi \hat{N}_{\nu,\omega}}}\hat{Z}_\nu (\vert \omega \vert \tilde{z}) \, ,
\label{ffour}
\een
where 
	\begin{equation} \label{exacteigenfn}
		\hat{Z}_\nu (\vert \omega \vert \tilde{z})=J_\nu (\vert \omega \vert \tilde{z}) +\xi_\omega(\nu) J_{-\nu} (\vert \omega \vert \tilde{z}) \, ,
	\end{equation}
\vspace{0pt}
	\begin{equation}
		\xi_{\omega}(\nu)= -\frac{\Gamma\left(\frac{1}{4}-\frac{\nu}{2}-\frac{\omega}{2\pi\mathcal{J}}\right) \Gamma\left(\frac{3}{4}+\frac{\nu}{2}+\frac{\omega}{2\pi\mathcal{J}}\right)}{\Gamma\left(\frac{1}{4}+\frac{\nu}{2}-\frac{\omega}{2\pi\mathcal{J}}\right) \Gamma\left(\frac{3}{4}-\frac{\nu}{2}+\frac{\omega}{2\pi\mathcal{J}}\right)} \, ,
	\end{equation}
and the eigenvalue $\lambda_{\nu}^{(\rm{ex})}$ is given by
\ben
\lambda^{(\text{ex})}_\nu \, \equiv \, \left[\nu^2 - \frac{9}{4} \right] \, .
\label{fseven}
\een
The eigenfunctions satisfy the following orthonormality condition which follows from self-adjointness of the operator ${\widetilde{\cal{K}}}_{\text{L}}$ in the interval $[0,\infty)$ with the boundary condition ${\tilde{\chi}}^{\rm{ex}}_{\nu,\omega}(t,0)=0$
\ben
\int_{-\infty}^\infty dt \int_0^\infty dz\  \tilde{\chi}^{\text{(ex)}}_{\nu,\omega}(t,z) \tilde{\chi}^{\text{(ex)}}_{\nu^\prime,\omega^\prime}(t,z)  =  \delta (\nu - \nu^\prime) \delta (\omega + \omega^\prime) \, .
\label{ffour1}
\een
Here $\hat{N}_{\nu,\omega}$ appears in the orthonormality relation

\begin{equation} \label{exactorthonormality}
\int_{0}^\infty \frac{dz}{\tilde{z}}\ \hat{Z}_{\nu_1} (\vert \omega \vert \tilde{z})\hat{Z}_{\nu_2} (\vert \omega \vert \tilde{z}) \, = \, \hat{N}_{\nu_1,\omega}\ \delta(\nu_1-\nu_2) \, .
\end{equation}
We do not have an analytic expression for $\hat{N}_{\nu,\omega}$, but we can determine it perturbatively in $\omega/ \mathcal{J}$ to get for the discrete modes
\begin{equation} \label{exactnorm}
\hat{N}_{\nu} \, = \, \frac{1}{2\nu} \, + \, \mathcal{O}\left(\frac{\vert\omega\vert}{2\mathcal{J}}\right)^3 \, . 
\end{equation}
For the continuous modes, we will not need to know the $\omega/ \mathcal{J}$ corrections since there is no enhancement. So, we use
\begin{equation} \label{exactnormcont}
\hat{N}_{\nu} \, = \, \frac{2\sin\pi \nu}{\nu} \, + \, \mathcal{O}\left(\frac{\vert\omega\vert}{2\mathcal{J}}\right) \, .
\end{equation}
We now perform a mode expansion of the fluctuation field in terms of these exact eigenfunctions as follows \footnote{The operator ${\cal{K}}_{\rm{ex}}$ is the zero temperature version of the operator considered in \cite{Maldacena:2016hyu}, where the finite temperature exact eigenfunctions were determined.  So our exact eigenfunctions should arise as the zero temperature limit of those eigenfunctions. We discuss details of how to take this zero temperature limit and the derivation of various properties of the exact eigenfunctions in Appendix \ref{app:exacteigfn}. }
\begin{equation} \label{modedecomposition}
\begin{split}
\eta(t,z) \, & = \, \tilde{z} \int_{-\infty}^\infty \frac{d\omega}{2\pi}\int d\nu\ \widetilde{\Phi}_{\nu,\omega} \, \tilde{\chi}^{\text{(ex)}}_{\nu,\omega}(t,z)\\
\, & = \, \int_{-\infty}^\infty \frac{d\omega}{2\pi}\int d\nu\ \widetilde{\Phi}_{\nu,\omega} \, \chi^{\text{(ex)}}_{\nu,\omega}(t,z) \, .
\end{split}
\end{equation}
The boundary condition (\ref{asix}), i.e. $\eta (t,z=0) =0$ determines the allowed values $\nu$. These are of the form
\beq \label{aseven}
\begin{split}
 \nu  & =  ir + \mathcal{O}\left(\frac{\omega}{\cal{J}}\right), \qquad\qquad r  \in {\rm Real} \\
\nu & =  2n + \frac{3}{2} + a_1\left(\frac{\omega}{\cal{J}}\right) + a_2 \left(\frac{\omega}{\mathcal{J}}\right)^2 + \ldots \qquad n=0,1,2, \ldots
\end{split}
\eeq
In the following we will  not need the correction to the continuous series. The coefficients $a_1$ and $a_2$ for the $n=0$ case are given by $1/ \pi$ and zero respectively (see Appendix \ref{app:exacteigfn} for details). 
Using $\eqref{exactorthonormality}-\eqref{modedecomposition}$ in $\eqref{SLiouville}$, we get 
\begin{equation} \label{Sdiag}
\begin{split}
S^{(2)} & = \frac{1}{16q^2}  \int_{-\infty}^\infty \frac{d\omega}{2\pi} \int d\nu\ \widetilde{\Phi}_{\nu,\omega} \lambda^{(\text{ex})}_\nu
\widetilde{\Phi}_{\nu,-\omega} \, .
\end{split}
\end{equation}
Note that the eigenvalue  $ \lambda^{(\text{ex})}_\nu$ arises from the large $q$ limit of the eigenvalue of the finite $q$  bilocal kernel (see $\eqref{eq:kc(h,q)}, \eqref{gtilde} $), with $h=\nu+1/2$
\begin{equation}
\begin{split} \label{Lgtilde}
\widetilde{g}(\nu, \infty) \, = \, \frac{1}{2}\left(\nu^2-\frac{1}{4}\right) \, .
\end{split}
\end{equation}
From $\eqref{Sdiag}$, we can read off the momentum space bilocal correlator
\begin{equation} \label{momprop}
\langle \widetilde{\Phi}_{\nu_1,\omega_1}\widetilde{\Phi}_{\nu_2,\omega_2}\rangle \, \simeq \, \frac{8q^2}{\lambda^{(\text{ex})}_{\nu_1}}\ \delta(\nu_1-\nu_2) \delta(\omega_1+\omega_2) \, .
\end{equation}
Since the solution $\eqref{eq:phi_0}$ breaks the symmetry (\ref{aone}), one might expect that there is a zero mode given by its variation $\delta \Phi_{\text{cl}}$. This would be of the form 
\ben
\eta_{(0)} (t,z) \sim \tilde{z}^{1/2} \int d\omega\ e^{i\omega t} \epsilon (\omega) \left[
\cos\big(\tfrac{|\omega|}{2\mathcal{J}} \big) J_{\frac{3}{2}} (\vert\omega\vert \tilde{z}) + \sin\big(\tfrac{|\omega|}{2\mathcal{J}} \big) J_{-\frac{3}{2}} (\vert\omega\vert \tilde{z}) \right] \, .
\label{athree}
\een
As expected this solves the equation of motion which follows from (\ref{SLiouville}), but does not satisfy the boundary condition (\ref{asix}) at any finite $J$. Indeed, 
(\ref{aseven}) shows that there is no eigenfunction with $\nu = 3/2$. 
Since there is no zero mode, a calculation of the bilocal two point function proceeds in a straightforward fashion.
\\Using $\eqref{modedecomposition}$ and $\eqref{momprop}$, we can write down the exact bilocal two point correlator in position space as
\begin{equation} \label{posprop}
\begin{split}
\langle \eta(t,z) \eta(t^\prime, z^\prime) \rangle 
\, & = \, 8q^2 
\int_{-\infty}^\infty \frac{d\omega}{2\pi}\int d\nu\ \frac{\chi^{\text{(ex)}}_{\nu,\omega}(t,z)\chi^{\text{(ex)}}_{\nu,-\omega}(t^\prime,z^\prime)}{\lambda^{(\text{ex})}_\nu} \\
\, &= \, 4q^2 (\tilde{z}\tilde{z}^\prime)^{\frac{1}{2}} \int_{-\infty}^\infty \frac{d\omega}{2\pi}\ e^{i\omega(t-t^\prime)} \int\frac{d\nu}{\hat{N}_\nu}\ \frac{\hat{Z}^*_\nu(\vert \omega \vert \tilde{z}) \hat{Z}_\nu(\vert \omega \vert \tilde{z}^\prime)}{\widetilde{g}(\nu,\infty)-1} \, .
\end{split}
\end{equation}



\subsection{Perturbative expansion}
In this subsection, we calculate the various ingredients needed to obtain a perturbative expansion of the bi-local propagator, which are  the matrix elements given by $\eqref{l01}-\eqref{l02}$ for Liouville theory.
The full (or ``exact") kernel  $\widetilde{\mathcal{K}}_{\text{L}}$ can be expanded in powers of $( |\omega| \mathcal{J}^{-1})$ as follows. This is the analog of the expansion (\ref{Kexpansion}). 
\beq
\widetilde{\mathcal{K}}_{\text{L}} = \widetilde{\mathcal{K}}^{(0)} + \frac{1}{\mathcal{J}} \, \widetilde{\mathcal{K}}^{(1)} + \frac{1}{\mathcal{J}^2} \, \widetilde{\mathcal{K}}^{(2)} + \ldots
\eeq
where
\beq \label{LK0}
\widetilde{\mathcal{K}}^{(0)}  = z^2 (-\partial_t^2 + \partial_z^2) + 2z\partial_z -2 \, ,
\eeq
\beq \label{LK1}
\widetilde{\mathcal{K}}^{(1)}  = z(-\partial_t^2 + \partial_z^2) + \partial_z \, ,
\eeq
\beq \label{LK2}
\widetilde{\mathcal{K}}^{(2)}  = \frac{-\partial_t^2 + \partial_z^2}{4} \, .
\eeq
The kernel $\widetilde{\mathcal{K}}^{(0)}$ now has a zero mode,
\beq \label{eig0}
\lambda_0^{(0)} = 0 \, ,
\eeq
\beq\label{ef0}
\tilde{\chi}_{0,\omega}^{(0)}(t,z) = \sqrt{\frac{3}{2\pi}}z^{-\frac{1}{2}} e^{i\omega t} J_{\frac{3}{2}}(|\omega|z) \, .
\eeq
The eigenvalue and eigenfunction for this mode will be corrected by the perturbations. The corresponding exact eigenvalue is $\lambda_{0,\omega}$ which has the expansion
\ben
\lambda_{0,\omega} = \lambda_0^{(0)}+\frac{1}{\mathcal{J}}
\lambda_0^{(1)} + \left( \frac{1}{\mathcal{J}} \right)^2 
\lambda_0^{(2)} + \cdots \, ,
\label{lambdaexpansion}
\een
analogous to (\ref{lambda_ex}). 
The corrections to the eigenvalues can be read off from (\ref{aseven}) and (\ref{fseven}),
\beq\label{eig1}
\lambda_0^{(1)} \, = \, \frac{3|\omega|}{\pi} \, .
\eeq
Similarly, for the first order correction to the eigenfunction, we get 
	\begin{equation}
		\tilde{\chi}_{0,\omega}^{(1)}(t,z) \, = \, \sqrt{\frac{3}{2\pi}}\frac{\vert \omega \vert}{2 \mathcal{J}} \, z^{-\frac{1}{2}} e^{i\omega t}
		\left( \frac{1}{|\omega|} \, \partial_z - \frac{1}{2|\omega| z} + \frac{2}{3\pi} + \frac{2}{\pi} \, \partial_{\nu} \, \right) J_\nu(|\omega|z)\bigg|_{\nu=\frac{3}{2}} \, , 
	\label{ef1}
	\end{equation}
and
\beq\label{eig2}
\lambda_0^{(2)} \, = \, \frac{\omega^2}{\pi^2} \, .
\eeq
While we have written these down using the expansion of the known exact eigenvalue and eigenfunction, one can of course calculate these directly in perturbation theory. The first order correction follows easily from $\eqref{LK1}$ and $\eqref{ef0}$ in  $\eqref{l01}$. For the first order correction to the eigenfunction, we were unable to perform the sum analytically in $\eqref{l02}$. The expression (\ref{ef1}) can be used to verify that standard perturbation theory indeed leads to the correct second order correction to the eigenvalue. The details of this calculation have been included in Appendix \ref{app:lambda2}. 
These calculations are in agreement with the results of \cite{Maldacena:2016hyu}.


\subsection{Evaluation of bilocal two point function}
\label{sec:bilocal propagator}
We now have all the necessary ingredients to evaluate the bilocal two point function $\eqref{posprop}$ perturbatively in $\left(|\omega |\mathcal{J}^{-1}\right)$ using the formalism of Section \ref{sec:Schwarzian} and Appendix \ref{app:Perturbation}. 
First, we note that the values of $\nu$ to be integrated over are given by $\eqref{aseven}$. The imaginary and discrete values give rise to a continuous and discrete contribution respectively. The discrete sum receives an enhancement from the zero mode so it needs to be treated separately. We separate it from the non zero modes in order to write
\beq
\langle \eta(t,z) \eta(t^\prime, z^\prime) \rangle \, = \, G^{(-1)}(t,z ; t^\prime, z^\prime) \, + \, G^{(0)}(t,z ; t^\prime, z^\prime) \, ,
\eeq
where $G^{(-1)}$ and $G^{(0)}$ are given by equations $\eqref{eq:D^(-1)}$ and $\eqref{eq:D^(0)}$ respectively. The perturbative corrections to the zero mode eigenvalue and eigenfunction can be substituted in these expressions to write down the  bilocal propagator explicitly.
Using $\eqref{eig1}$ and $\eqref{ef0}$ in $\eqref{eq:D^(-1)}$, we get the enhanced propagator 
\beq
\begin{split}
G^{(-1)}(t,z ; t^\prime, z^\prime) \, &=\, 8q^2\mathcal{J} \int_{-\infty}^\infty \frac{d\omega}{2\pi |\omega|}\ \frac{\chi_{0,\omega}^{(0)}(t,z) \chi_{0,-\omega}^{(0)}(t^\prime,z^\prime)}{\lambda_0^{(1)}}\\
\, &= \, 4q^2\mathcal{J} (zz^\prime)^{\frac{1}{2}} \int_{-\infty}^\infty \frac{d\omega}{\vert \omega \vert}\ e^{i\omega(t-t^\prime)} J_{\frac{3}{2}}(|\omega|z) J_{\frac{3}{2}}(|\omega|z^\prime) \, .
\end{split}
\eeq
Next, let us consider the $\mathcal{O}(1)$ contribution $G^{(0)}$. It receives a contribution from $\mathcal{D}_c$, which for the Liouville case is given by
\beq \label{Dc}
\mathcal{D}_c (t,z;t^\prime,z^\prime) \, = \, 8q^2 \int_{-\infty}^\infty \frac{d\omega}{2\pi}\left[ \sum_{n=1}^\infty  \frac{\chi^{(0)}_{\nu_n,\omega}(t,z)\chi^{(0)}_{\nu_n,\omega}(t^\prime,z^\prime)}{\lambda_{\nu_n}^{\text{(0)}}} + \int d\nu\ \frac{\chi^{(0)*}_{\nu,\omega}(t,z) \chi^{(0)}_{\nu,\omega}(t^\prime,z^\prime)}{\lambda_{\nu}^{\text{(0)}}} \right] \, .
\eeq
The details of evaluation of $\mathcal{D}_c$ is given in Appendix  \ref{app:nuint}.  The result is
	\begin{align}
	\label{Dcfin}
		\mathcal{D}_c(t,z;t^\prime,z^\prime) \, = \, 4q^2 (z z^\prime)^{\frac{1}{2}} &\int_{-\infty}^\infty d\omega\, e^{i\omega(t-t^\prime)}
		\Bigg[\frac{J_{-\frac{3}{2}}(\vert \omega \vert z^>)J_{\frac{3}{2}}(\vert \omega \vert z^<)}{2} \nonumber\\
		&\qquad - \, \frac{1}{\pi} \left( \frac{d}{d\nu} \, + \, \frac{1}{3} \right) J_\nu(\vert \omega \vert z) J_\nu(\vert \omega \vert z^\prime) \bigg|_{\nu=\frac{3}{2}} \Bigg] \, .
	\end{align}
We identify the first term in $\eqref{Dcfin}$ 
with the single pole term $\mathcal{D}^\prime$ and the second term with the double pole term $\mathcal{D}^{\prime\prime}$ in $\eqref{D_c}$ respectively. The remaining terms in $\eqref{eq:D^(0)}$ can also be evaluated using $\eqref{eig0}-\eqref{ef1}$ to get
	\begin{align}
		&8q^2\int_{-\infty}^\infty \frac{d\omega}{2\pi}\left[\frac{\chi_{0,\omega}^{(0)}(t,z) \chi_{0,-\omega}^{(1)}(t^\prime,z^\prime)}{\lambda_0^{(1)}}
		\, + \, \frac{\chi_{0,\omega}^{(1)}(t,z) \chi_{0,-\omega}^{(0)}(t^\prime,z^\prime)}{\lambda_0^{(1)}}
		\, - \, \frac{\lambda_0^{(2)} \chi_{0,\omega}^{(0)}(t,z) \chi_{0,-\omega}^{(0)}(t^\prime,z^\prime)}{(\lambda_0^{(1)})^2}\right] \nonumber\\
		&=4q^2(zz^\prime)^{\frac{1}{2}} \int d\omega \, e^{i\omega(t-t^\prime)} \Bigg[ \frac{1}{\pi} \left( \frac{d}{d\nu} \, + \, \frac{1}{3} \right)  + \frac{1}{2|\omega|}
		\left( \partial_z + \partial_{z'} + \frac{1}{2z} + \frac{1}{2z'} \right) \Bigg]
		J_{\nu}(\vert \omega \vert z) J_{\nu}(\vert \omega \vert z^\prime) \bigg|_{\nu=\frac{3}{2}} \, . 
	\label{D0T1}
	\end{align}
From $\eqref{Dcfin}$ and $\eqref{D0T1}$, we see that the double pole  precisely cancels. This is an explicit illustration of the claim made in Section \ref{sec:Schwarzian} that the final answer for the bilocal propagator receives contribution only from the simple pole in $\mathcal{D}_c$ (which corresponds to $h=2$ matter) and from the $b_1+b_2$ diagrams. Collecting all the terms, we get the bilocal propagator for Liouville theory up to $\mathcal{O}(1)$ to be

	\begin{align} \label{finalans}
		\langle \eta(t,z) \eta(t^\prime,z^\prime)\rangle \, = \, 2q^2 (zz^\prime)^{\frac{1}{2}} &\int_{-\infty}^{\infty} d\omega\ e^{i\omega(t-t^\prime)}
		\Bigg[\frac{2\mathcal{J}}{\vert\omega\vert}\, J_{\frac{3}{2}}(\vert \omega \vert z)J_{\frac{3}{2}}(\vert \omega \vert z^\prime)
		+ J_{-\frac{3}{2}}(\vert \omega \vert z^>) J_{\frac{3}{2}}(\vert \omega \vert z^<) \nonumber\\
		&\ + \, \frac{1}{\vert\omega\vert}\left( \partial_z + \partial_{z'} + \frac{1}{2z}+\frac{1}{2z^\prime}\right)
		J_{\frac{3}{2}}(\vert \omega \vert z) J_{\frac{3}{2}}(\vert \omega \vert z^\prime) \Bigg] \, .
	\end{align}


\subsection{Comparison with zero temperature limit of four-point function} \label{subsec:zerotemplim}
We now show that the zero temperature limit of the finite temperature four point function calculated in \cite{Maldacena:2016hyu} agrees with $\eqref{finalans}$. The SYK four point function at finite temperature up to $\mathcal{O}(1)$ is given by \footnote{Note that the explicit calculation of the $O(1)$ contribution in \cite{Maldacena:2016hyu} is obtained by writing the propagator in terms of eigenfunctions of $-\partial_{\tx}^2+\frac{1}{2\sin^2\frac{\tx}{2}}$ rather than $ - \frac{\sin^2\frac{\tx}{2}}{v^2}\partial_y^2 + 4 \sin^2\frac{\tx}{2}\partial_{\tx}^2 + \frac{1}{4}$ which is the finite temperature version of the Bessel operator we consider. They, of course, lead to the same result since at finite temperature the periodicity conditions ensure a unique Green's function.}
\begin{align} \label{MSprop}
\mathcal{F}(x,y;x^\prime,y^\prime) \, &= \, \Big[\beta\mathcal{J}-2\left[-1+\left(y-\frac{\pi}{2}\right)\partial_y+(x-\pi)\partial_x+(x^\prime-\pi)\partial_{x^\prime}\right]\Big] \nonumber\\
&\qquad \times \sum_{\vert n \vert\geq 2} e^{-in(y-y^\prime)}\frac{f_n(x)f_n(x^\prime)}{\pi^2 n^2 (n^2-1)}
\end{align}
Let us define
\begin{equation}
y \, = \,\frac{2\pi t}{\beta} \,  \qquad\quad x \, = \, \frac{4\pi z}{\beta} \, , \qquad\quad \omega \, = \, \frac{2\pi n}{\beta}
\end{equation}
In the zero temperature limit the $h=2$ eigenfunctions become Bessel functions
\begin{equation} \label{fzerotemp}
\begin{split}
f_n(x) \, &= \, \frac{\sin\frac{nx}{2}}{\tan\frac{x}{2}}-n\cos\frac{nx}{2}\\ &= \, \frac{\beta\omega}{2\pi} \sqrt{\frac{\pi \omega z}{2}} \, J_{\frac{3}{2}}(\omega z) \, .
\end{split}
\end{equation}
Using $\eqref{fzerotemp}$ in $\eqref{MSprop}$  and replacing the sum over the discrete Fourier index $n$ by a continuous integral over $\omega$, we get 
	\begin{align}
		\mathcal{F}(t,z;t^\prime,z^\prime) \, = \, \frac{1}{2}  (zz^\prime)^{\frac{1}{2}} &\int_{-\infty}^\infty d\omega \ e^{-i\omega (t-t^\prime)}
		\Bigg[\frac{2\mathcal{J}}{\vert\omega\vert} \, J_{\frac{3}{2}}(\vert \omega \vert z)J_{\frac{3}{2}}(\vert \omega \vert z^\prime) \nonumber\\
		&\ + \, \frac{1}{\vert\omega\vert}\left( \partial_z + \partial_{z'} + \frac{1}{2z}+\frac{1}{2z^\prime}\right)
		J_{\frac{3}{2}}(\vert \omega \vert z) J_{\frac{3}{2}}(\vert \omega \vert z^\prime) \Bigg] \nonumber\\
		&- (zz^\prime)^{\frac{1}{2}} \int_0^\infty d\omega\ \sin\omega(t-t^\prime)\ J_{\frac{3}{2}}(\omega z) J_{\frac{3}{2}}(\omega z^\prime) \, .
	\label{MSzerotemp}
	\end{align}
Now, we use the following integrals expressed in terms of the quantity $\xi \equiv \frac{-(t-t^\prime)^2+z^2+z^{\prime 2}}{2zz^\prime}$

\[
  \ (zz^\prime)^{\frac{1}{2}} \int_{-\infty}^\infty d\omega\ e^{-i\omega(t-t^\prime)} J_{-\frac{3}{2}}(\vert \omega \vert z) J_{\frac{3}{2}}(\vert \omega \vert z^\prime) \, = \,
  \begin{cases}
    0, & \text{for } |\xi|> 1 \\
    P_1(-\xi), & \text{for }  |\xi| < 1  \end{cases}
\]
and

\[
  (zz^\prime)^{\frac{1}{2}}\int_0^\infty d\omega\ \sin\omega(t-t^\prime)\ J_{\frac{3}{2}}(\omega z) J_{\frac{3}{2}}(\omega z^\prime) \, = \,
  \begin{cases}
    0, & \text{for } |\xi|> 1 \\
    \frac{1}{2} P_1(\xi), & \text{for }  |\xi| < 1  \end{cases}
\]
The details of the derivation of these integrals are given in Appendix \ref{app:Bessel product}. 
Using these integrals, we see from $\eqref{finalans}$ and $\eqref{MSzerotemp}$, 
\begin{equation}
\langle\eta(t,z)\eta(t^\prime,z^\prime)\rangle \, = \, 4q^2 \, \mathcal{F}(t,z;t^\prime,z^\prime)  \, .
\end{equation}
The proportionality factor of $4q^2$ arises as a consequence of the field redefinition in $\eqref{eq:Phi-first}$.

\subsection{Expansion around the critical saddle point}
In the ${\cal{J}}|\tau_{12}| \gg 1$ limit $\eqref{eq:phi_0}$ becomes 
\ben
\Phi^{(0)} \, = \, - 2 \log ({\cal{J}}|\tau_{12}|) \, .
\label{pone}
\een
We will call this the ``critical saddle point".
This is in fact what follows from the large $q$ limit of $\Psi^{(0)}$ in  (\ref{Psi^0}) and plays its role. There is, however, an important difference. For finite $q$ the critical solution  ${\cal{J}}|\tau_{12}|$ is not a solution of the full theory because of the term which breaks the reparametrization symmetry. In this case (\ref{pone}) is also a solution of the classical equations of motion. $\Phi^{(0)}$, however, does not satisfy the boundary condition (\ref{atwo}). In fact, in an expansion around the critical limit, there is no reason to impose this boundary condition. Keeping this point in mind we will call $\Phi_{\text{cl}}$ the ``exact" solution and develop a perturbative expansion around the critical solution following the same steps as in the finite $q$ SYK model.

An expansion around this solution would in fact lead to a normalizable zero mode given by $\eqref{ef0}$ which needs to be treated properly in precisely the same way as the finite $q$ SYK model. 
 In order to deal with this, we introduce a source term given by \footnote{The fact that this source term goes as $\mathcal{J}^{-1}$ is reminiscent of the conformal breaking term in finite $q$ SYK. It should be noted that if we add other source terms which are suppressed by higher powers of $\mathcal{J}$, that would be inconsistent with the finite $q$ picture.  }
\ben
\frac{N}{{8q^2\cal{J}}}
\int d\tau_1 d\tau_2\ Q_s^{\text{L}}(\tau_1,\tau_2) \Phi (\tau_1,\tau_2) \, ,
\label{etwo}
\een
where the source is again a regularized version of (\ref{eq:Q_s}). In this case, the kernel $\mathcal{K}^{(0)}_{\text{L}}$ can be read off from the first line of (\ref{SLiouville}) as
\ben
\mathcal{K}^{(0)}_{\text{L}}(t,z) \, = \, - \partial_t^2 + \partial_z^2 - \frac{2}{z^2} \, .
\label{ethree}
\een
The source is then related to the $O(1/\cJ)$ correction to the classical solution by
\ben
Q_s^{\textrm{L}} (t,z) \, = \, \mathcal{K}^{(0)}_{\text{L}} \Phi^{(1)}_s(t,z) \, .
\label{sone}
\een 
The $\mathcal{O}(1/\cJ)$ correction to the classical solution is given by \footnote{We have stripped off the power of $\mathcal{J}$ in the expression for $\Psi^{(1)}$ in order to maintain consistency in notation with Section \ref{sec:Schwarzian}. Also, since the domain of integration of the $z$ coordinate is always positive, we drop the absolute sign.}
\ben
\Phi_{\text{cl}}-  \Phi_{0} \, \equiv \, \Phi^{(1)}(z) \, = \, - \frac{1}{z} \, .
\een
As expected, the source
in (\ref{eq:Q_s}) with this non-regularized  $\Phi^{(1)}(z)$ vanishes. Following the treatment in finite $q$ SYK, we therefore introduce a regulator $s$ to define
\ben
\Phi^{(1)}_s (z) \, \equiv \, - \lim_{s\rightarrow\frac{1}{2}}\frac{1}{ z^{2s}} \, .
\een
This results in a regularized version of the source $Q_s^{\text{L}}$  given by
\ben
Q_s^{\text{L}} (\tau_1,\tau_2) \, \simeq \, -\lim_{s\rightarrow \frac{1}{2}}\left(s-\frac{1}{2}\right)\frac{s+1}{| \tau_{12} |^{2s+2}} \, .
\label{eeight}
\een
Since the exact classical solution is now known, the coefficient of the resulting Schwarzian action for the soft mode can be now determined unambiguously. The Schwarzian action
is given by
\ben \label{eq:Lsch}
S_{\text{Sch}}^\text{L}[f] \, = \, -\frac{N}{8q^2\mathcal{J}} \lim_{s \rightarrow \frac{1}{2}}\int d\tau_1 d\tau_2\ Q_s^\text{L} (\tau_1,\tau_2) \Phi_{0}^f(\tau_1,\tau_2) \, ,
\een
where $\Phi^f_{0}$  corresponds to the transformed classical solution
\ben
\Phi^f_{0} (\tau_1,\tau_2) \, = \, \log \left[ \frac{f^\prime(\tau_1) f^\prime(\tau_2)}{\vert f(\tau_1)-f(\tau_2)\vert^2} \right] \, .
\label{esix}
\een
To determine the coefficient one could proceed by writing $f(\tau) = \tau + \varepsilon(\tau)$, evaluate $\eqref{eq:Lsch}$ in an expansion in $\varepsilon(t)$, perform the limit $s \rightarrow 1/2$ and resum to obtain the full
expression. Since we know that the answer should be proportional to the Schwarzian
it is adequate to expand to second order in  $\varepsilon(t)$. We have performed this calculation, but will omit the details here.

Alternatively we can calculate the propagator of the Fourier transform of  $\varepsilon(\tau)$, which we call $\varepsilon(\omega)$ given in $\eqref{<epseps>}$. For the Liouville theory this is given by
\beq \label{<epseps>2}
\frac{\mathcal{J}}{\lambda_{0,\omega}^{(1)}} \, \delta(\omega+\omega^\prime) \, = \, \frac{N_{0,\omega}^{(0)}}{2\pi}\ \langle \varepsilon(\omega) \varepsilon(\omega^\prime)\rangle \, .
\eeq
The main ingredient is contained in the set of conformal Ward identities which are described for the arbitrary $q$ model in equations (\ref{zero mode eigenvalue})-(\ref{zero mode identity}). These equations can be explicitly verified in our case with the expressions for 
${\cal K}_{\textrm{L}}^{(0)}$ given in $\eqref{ethree}$ and ${\cal K}_{\textrm{L}}^{(1)}$ given by 
\bea
{\cal K}_{\textrm{L}}^{(1)} & = & -\frac{2}{z^2} \Phi^{(1)} \, .
\eea
For our case $\lambda_0^{(1)}$ is given in $\eqref{eig1}$. To determine the normalization $N_{0,\omega}^{(0)}$, we need 
\ben
\chi_{\textrm{zero},\omega}(\tau_1,\tau_2) = \int d\tau\ e^{i\omega \tau}\left[ \frac{\delta \Phi^{(0)}_f (\tau)}{\delta f(\tau^\prime)}\right]_{f(\tau)=\tau} \, .
\label{stwo}
\een
It is straightforward to see that 
\ben
\chi_{\textrm{zero},\omega}(\tau_1,\tau_2) \, = \, -\sqrt{2\pi}\ i \omega^{\frac{3}{2}} z^{\frac{1}{2}} J_{\frac{3}{2}}(|\omega|z) \, .
\label{sthree}
\een
Comparing with $\eqref{ef0}$, we get (from the definition)
\beq \label{sfour}
N_{0,\omega}^{(0)} \, = \, \frac{4\pi^2|\omega|^3}{3} \, .
\eeq
To apply $\eqref{<epseps>}$ we need to note that while the source term for finite $q$ was defined by $\eqref{eq:Q_s}$ while in the Liouville theory we defined the source term by $\eqref{etwo}$. Thus in Liouville theory we have
\beq \label{sfive}
\langle \varepsilon(\omega) \varepsilon(\omega^\prime) \rangle \, = \, \frac{8q^2\mathcal{J}}{N} \frac{2\pi \delta(\omega+\omega^\prime)}{\lambda_{0,\omega}^{(1)}N_{0,\omega}^{(0)}}
\, = \, \frac{4q^2 \mathcal{J}}{N\omega^4} \, .
\eeq
where we have used $\eqref{sfour}$ and  $\eqref{eig1}$. This propagator can be easily seen to follow from the Schwarzian action
\beq
S_{\textrm{Sch}}^{\textrm{L}}[f] \, = \, -\frac{N}{4q^2 \mathcal{J}} \int d \tau\ \{f(\tau),\tau\} \, .
\eeq
The coefficient is in precise agreement with the large $q$ limit of the action given in \cite{Maldacena:2016hyu}. The calculation of the bilocal two point function now follows the diagrammatic technique
of Section \ref{sec:Schwarzian}.

\section{Conclusion}
\label{sec:conclusion}
In this work, we concentrated on the development of a complete understanding of systematic near conformal perturbation expansion in SYK type models.
It develops further the initial work of \cite{Jevicki:2016bwu, Jevicki:2016ito} where the soft mode with Schwarzian dynamics is extracted
from the bi-local field in a systematic fashion and arises as an emergent degree of freedom.
This mode interacts with the remaining ``matter" degrees of freedom which include a component dual to an operator with conformal dimension $h \sim 2$,
in a manner which is completely determined.
The non-triviality of exhibiting this interacting representation lies in the fact that, as we explain,
the symmetry breaking effects responsible both for the Schwarzian and Schwarzian-matter interactions comes from a very subtle off-shell regularization
which produces non-zero effects when removed in the limiting procedure.
This leaves a series of interacting vertices that are determined explicitly. 
Representing these corrections in a diagrammatic picture provides a complete and transparent scheme.
This allows for a concrete perturbation calculation of corrections to leading (conformal) correlation functions and other physical quantities.
We do this for the bi-local two point functions by considering the evaluation of the first correction (in a low energy expansion) to the leading (enhanced) answer. 
The evaluation of these is facilitated by a series of conformal identities.  

We applied this formalism to the large $q$ limit, where the model is reduced to the Liouville theory which is exactly solvable.
In this case, expansion around the correct saddle-point does not result in a zero mode and the correlators can be calculated exactly.
We can nevertheless expand around a saddle point appropriate for a long distance expansion and apply the perturbation scheme described above.
In addition, it is instructive to see the workings of the method in this example of a prototype conformal field theory. We note that it is of interest to consider the complex SYK model already, 
since issues related to applicability of perturbation theory have been observed recently \cite{Gu:2019jub}. Indeed we believe that the applicability of the scheme goes beyond the SYK model, applying generally to perturbations in conformal field theory and more generally quantum field theory.

Returning to the case of SYK theories, the present systematic reformulation of the model as a bi-local matter coupled Schwarzian theory might offer the needed insight into the outstanding question of its gravity dual. The Liouville theory in particular provides the simplest limiting case. In this limit the higher $h$ modes decouple and we find an interacting picture of the soft mode with bi-local $h=2$ matter.
We emphasize this fact since most studies of the gravity dual focus on the dilaton gravity sector of the dual theory.
In \cite{Almheiri:2014cka, Jensen:2016pah, Maldacena:2016upp, Engelsoy:2016xyb}, the enhanced part of the bi-local two point function has been reproduced
in a dual theory which contains bulk fields which can be thought to be dual to the SYK fermions.
Our preliminary results indicate that some parts of the subleading correction may be obtainable by considering the effect of a coupling to the non-trivial dilaton background  \cite{Das2020:wip}.

However, additional ingredients, related to the contribution of $h=2$ bi-local matter, are probably necessary to understand the low energy sector completely.
We hope to return to this problem in future studies.

\acknowledgments
We thank Gautam Mandal and Pranjal Nayak for discussions during the early stages of this work, and Gustavo J. Turiaci for comments on the draft. 
S.R.D. and A.J. would like to thank the organizers of Fourth Mandelstam Theoretical Physics School and Workshop for hospitality. S.R.D. thanks the Tata Institute of Fundamental Research, Mumbai and Yukawa Insttiute for Theoretical Physics, Kyoto for hospitality during various stages of this work. AG acknowledges support from the KITP, University of California, Santa Barbara and the Dean's Arts and Sciences fellowship at the University of Kentucky while part of this work was in progress. The work of S.R.D. was partially supported by a National Science Foundation grant NSF/PHY-181878 and by a Distinguished Professorship award from the College of Arts and Sciences, University of Kentucky. The work of A.J. was partially supported by the U.S. Department of Energy under contract DE-SC0010010 and DE-SC0019480. The work of KS is supported by the European Research Council (ERC) under the European Union's Horizon 2020 research and innovation program (grant agreement No758759).

\appendix
\section{Perturbative Expansions}
\label{app:Perturbation}
In this appendix, we summarize several relations for the expansions (\ref{Psi_cl}) - (\ref{lambda_ex}) and introduce some short-hand notations.
Formally we suppose to have the exact action $S$ and the exact classical solution $\Psi_{\rm cl}$.
Expanding the action around the exact solution, we have
	\begin{equation}
		S\big[\Psi_{\rm cl} + N^{-1/2} \eta \big] \, = \, S[\Psi_{\rm cl}] \, + \, \frac{1}{2} \int \eta \cdot \mathcal{K}_{(\rm ex)} \cdot \eta \, + \, \cdots \, , 
	\end{equation}
where 
	\begin{equation}
		\mathcal{K}_{(\rm ex)}(\tau_1, \tau_2; \tau_3, \tau_4) \, = \, \frac{\delta^2 S[\Psi_{\rm cl}]}{\delta \Psi_{\rm cl}(\tau_1, \tau_2) \delta \Psi_{\rm cl}(\tau_3, \tau_4)} \, .
	\label{K_ex}
	\end{equation}
Then, introducing $\widetilde{\mathcal{K}}_{(\rm ex)}$ from $\mathcal{K}_{(\rm ex)}$
using the analogous definition as in (\ref{Ktilde}) where we replace all $\Psi_{(0)}$ by $\Psi_{\rm cl}$,
formally we can consider the exact Green's function $\widetilde{G}_{(\rm ex)}$ which is determined by the Green's equation 
	\begin{equation}
		\int d\tau_3 d\tau_4 \, \widetilde{\mathcal{K}}_{(\rm ex)}(\tau_1, \tau_2; \tau_3, \tau_4) \widetilde{G}_{(\rm ex)}(\tau_3, \tau_4; \tau_5, \tau_6)
		\, = \, \delta(\tau_{15}) \delta(\tau_{26}) \, .
	\end{equation}
In order to invert the kernel $\widetilde{\mathcal{K}}_{(\rm ex)}$ in this Green's equation, we consider the eigenvalue problem of the kernel $\widetilde{\mathcal{K}}_{(\rm ex)}$:
	\begin{equation}
		\int d\tau_3 d\tau_4 \, \widetilde{\mathcal{K}}_{(\rm ex)}(\tau_1, \tau_2; \tau_3, \tau_4) \, \tilde{\chi}^{(\rm ex)}_{n, \omega}(\tau_3, \tau_4)
		\, = \, \lambda^{(\rm ex)}_{n,\omega} \, \tilde{\chi}^{(\rm ex)}_{n, \omega}(\tau_1, \tau_2) \, ,
	\label{eigenvalue}
	\end{equation}
with the eigenfunctions $\tilde{\chi}^{(\rm ex)}_{n, \omega}$ and the eigenvalues $\lambda^{(\rm ex)}_{n,\omega}$, where $n, \omega$ are some quantum numbers.
We normalize the eigenfunctions by requiring 
	\begin{equation}
		\int d\tau_1 d\tau_2 \, \tilde{\chi}^{(\rm ex)}_{n, \omega}(\tau_1, \tau_2) \tilde{\chi}^{(\rm ex)}_{n', \omega'}(\tau_1, \tau_2) \, = \, \delta_{n, n'} \delta(\omega-\omega') \, .
	\end{equation}

Now we consider a perturbative expansion in $1/J$ of the above eigenvalue problem.
We expand each of the quantities of interest as
	\begin{align}
		\Psi_{\rm cl} \, &= \, \Psi^{(0)} \, + \, \frac{1}{J} \, \Psi^{(1)} \, + \, \frac{1}{J^2} \, \Psi^{(2)} \, + \, \cdots \, , \label{Psi_cl} \\
		\mathcal{K}_{(\rm ex)} \, &= \, \mathcal{K}^{(0)} \, + \, \frac{1}{J} \, \mathcal{K}^{(1)} \, + \, \frac{1}{J^2} \, \mathcal{K}^{(2)} \, + \, \cdots \, , \label{Kexpansion}\\
		G_{(\rm ex)} \, &= \, J \, G^{(-1)} \, + \, G^{(0)} \, + \, \cdots \, , \label{D_ex}\\
		\chi^{(\rm ex)} \, &= \, \chi^{(0)} \, + \, \frac{1}{J} \, \chi^{(1)} \, + \, \cdots \, , \\
		\lambda^{(\rm ex)} \, &= \, \lambda^{(0)} \, + \, \frac{1}{J} \, \lambda^{(1)} \, + \, \frac{1}{J^2} \, \lambda^{(2)} \, + \, \cdots \, , \label{lambda_ex}
	\end{align}
and similarly for the redefined kernel $\eqref{Ktilde}$, and the bi-local propagator and eigenfunctions corresponding to it.
In this paper, the superscript in a round bracket denotes the order of $1/J$ expansion, while the subscript denotes quantum numbers.

The exact Green's function can be then written as
	\begin{equation}
		\widetilde{G}_{(\rm ex)} \, = \, \sum_{n} \, \frac{\tilde{\chi}_n^{(\rm ex)}\tilde{\chi}_n^{(\rm ex)}}{\lambda_n^{(\rm ex)}}
		\,\equiv \, \Psi_{\rm cl}^{\frac{q}{2}-1} \, G_{(\rm ex)} \, \Psi_{\rm cl}^{\frac{q}{2}-1} \, .
	\end{equation}
In the following, we will be interested in the contribution coming from the zero mode of the lowest order kernel, $n=0$. Since $\lambda_0^{(0)}=0$, we have
\begin{equation}
\begin{split}	
G^{(-1)} =\frac{\chi_0^{(0)} \chi_0^{(0)}} {\lambda_0^{(1)}}
\end{split}
\end{equation}
\begin{equation}
\begin{split}
G^{(0)} =\frac{\chi_0^{(0)} \chi_0^{(1)}}{\lambda_0^{(1)}} +  \frac{\chi_0^{(1)} \chi_0^{(0)}}{\lambda_0^{(1)}} -\frac{\lambda_0^{(2)} \chi_0^{(0)} \chi_0^{(0)}}{(\lambda_0^{(1)})^2} + \mathcal{D}_c, \qquad \mathcal{D}_c = \sum_{n \neq 0}\frac{\chi_n^{(0)} \chi_n^{(0)}} {\lambda_n^{(0)}}  .
\end{split}
\end{equation}
Here we suppressed all $\tau$ (and $\omega$) dependence since they don't play any crucial role here.

The expression of the perturbative kernels are also found by expanding $\mathcal{K}_{\rm ex}$ (\ref{K_ex})
	\begin{align}
		\mathcal{K}^{(0)}(\tau_1, \tau_2; \tau_3, \tau_4)
		\, &= \, S_c^{(2)}(\tau_{1, 2}; \tau_{3,4}) \, , \\
		\mathcal{K}^{(1)}(\tau_1, \tau_2; \tau_3, \tau_4) \, &= \, \int d\tau_5 d\tau_6 \, S_c^{(3)}(\tau_{1,2}; \tau_{3,4}; \tau_{5,6}) \, \Psi^{(1)}(\tau_{56}) \, , \\
		\mathcal{K}^{(2)}(\tau_1, \tau_2; \tau_3, \tau_4) \, &= \, \frac{1}{2} \int d\tau_5 d\tau_6 d\tau_7 d\tau_8 \,
		S_c^{(4)}(\tau_{1,2}; \tau_{3,4}; \tau_{5,6}; \tau_{7,8}) \,  \Psi^{(1)}(\tau_{56}) \Psi^{(1)}(\tau_{78}) \nonumber\\
		&\quad + \, \int d\tau_5 d\tau_6 \, S_c^{(3)}(\tau_{1,2}; \tau_{3,4}; \tau_{5,6}) \, \Psi^{(2)}(\tau_{56}) \, .
	\end{align}
where we used a short-hand notation
	\begin{align}
		S_c^{(n)}(\tau_{1,2}; \cdots; \tau_{2n-1,2n}) \, \equiv \, \frac{\delta^n S_c[\Psi^{(0)}]}{\delta \Psi^{(0)}(\tau_1, \tau_2) \cdots \delta \Psi^{(0)}(\tau_{2n-1}, \tau_{2n})} \, .
	\label{S_c^n}
	\end{align}
We will also use the following simplified notation for the contractions: 
	\begin{align}
		S_c^{(n)}\big[ A_1; A_2; \cdots; A_n \big]
		\, &\equiv \, \int d\tau_1 d\tau_2 \cdots d\tau_{2n-1} d\tau_{2n} \, S_c^{(n)}(\tau_{1,2}; \cdots; \tau_{2n-1,2n}) \nonumber\\
		&\hspace{80pt} \times A_1(\tau_1, \tau_2) \cdots A_n(\tau_{2n-1}, \tau_{2n}) \, .
	\end{align}

Finally expanding the eigenvalue equation (\ref{eigenvalue}), we can fix the perturbative eigenfunctions and eigenvalues
	\begin{align}
		\lambda_n^{(0)} \, &= \, \int \tilde{\chi}_n^{(0)} \cdot \widetilde{\mathcal{K}}^{(0)} \cdot \tilde{\chi}_n^{(0)} \, , \hspace{60pt}
		\lambda_n^{(1)} \, = \, \int \tilde{\chi}_n^{(0)} \cdot \widetilde{\mathcal{K}}^{(1)} \cdot \tilde{\chi}_n^{(0)} \, , \\
		\tilde{\chi}_n^{(1)} \, &= \, \sum_{k \ne n} \frac{\tilde{\chi}_k^{(0)}}{\lambda_n^{(0)} - \lambda_k^{(0)}} \,
		\int \tilde{\chi}_k^{(0)} \cdot \widetilde{\mathcal{K}}^{(1)} \cdot \tilde{\chi}_n^{(0)} \, , \\
		\lambda_n^{(2)} \, &= \, \sum_{k \ne n} \frac{1}{\lambda_n^{(0)} - \lambda_k^{(0)}} \left| \int \tilde{\chi}_k^{(0)} \cdot \widetilde{\mathcal{K}}^{(1)} \cdot \tilde{\chi}_n^{(0)} \right|^2
		\, + \, \int \tilde{\chi}_n^{(0)} \cdot \widetilde{\mathcal{K}}^{(2)} \cdot \tilde{\chi}_n^{(0)} \, . 
	\end{align}
where we used the normalization condition $\int \tilde{\chi}^{(0)}\cdot \tilde{\chi}^{(1)} = 0$.
We note that the zero-mode ($n=0$) first order eigenfunctions without tilde are obtained as
	\begin{align}
		\chi_0^{(1)} \, &= \, - \sum_{k \ne 0} \frac{\chi_k^{(0)}}{\lambda_k^{(0)}} \, \int \tilde{\chi}_k^{(0)} \cdot \widetilde{\mathcal{K}}^{(1)} \cdot \tilde{\chi}_0^{(0)} 
		\, + \, \left( 1 - \frac{q}{2} \right) \Psi_{(0)}^{-1} \Psi_{(1)} \chi_0^{(0)} \nonumber\\
		&= \, - \sum_{k \ne 0} \frac{\chi_k^{(0)}}{\lambda_k^{(0)}} \, \int \chi_k^{(0)} \cdot \mathcal{K}^{(1)} \cdot \chi_0^{(0)}  \, ,
	\label{eq:chi^1}
	\end{align}
where the second term in the first line comes by expanding $\chi^{(\rm ex)} = \Psi_{\rm cl}^{1-q/2} \, \tilde{\chi}^{(\rm ex)}$ and for the second line we used 
	\begin{align}
		\widetilde{\mathcal{K}}^{(1)}(\tau_1, \tau_2; \tau_3, \tau_4) \, &= \, \int d\tau_5 d\tau_6 \, S_c^{(3)}(\tau_{1,2}; \tau_{3,4}; \tau_{5,6}) \, 
		\Psi_{(0)}^{1-\frac{q}{2}}(\tau_{12}) \Psi_{(0)}^{1-\frac{q}{2}}(\tau_{34}) \Psi_{(1)}(\tau_{56}) \\
		&\ - \, \left(\frac{q}{2}-1 \right) \, S_c^{(2)}(\tau_{1,2}; \tau_{3,4}) \, \left[ \Psi_{(0)}^{-\frac{q}{2}}(\tau_{12}) \Psi_{(1)}(\tau_{12}) \Psi_{(0)}^{1-\frac{q}{2}}(\tau_{34})
		+ (\tau_{12} \leftrightarrow \tau_{34}) \right] \, . \nonumber
	\end{align}

\section{Details of Derivation of Liouville from Bilocal}
\label{app:SLderiv}
In this Appendix, we give a detail derivation of the action (\ref{eq:S_L}) from (\ref{eq:collaction}).

For the logarithm term, we use the following fact that 
	\begin{equation}
		\frac{{\rm sgn}(\tau_{12})}{2} \, = \, \theta(\tau_{12}) - \frac{1}{2} \, , \qquad \ \bigg[ \frac{{\rm sgn}}{2} \, \bigg]_{\star}^{-1}(\tau_{12}) \, = \, \partial_1 \delta(\tau_{12}) \, ,
	\end{equation}
where the inverse $[\bullet]_{\star}^{-1}$ is defined in the sense of the star product (i.e. matrix product) $\int d\tau' A(\tau_1, \tau') [A]_{\star}^{-1}(\tau', \tau_2) = \delta(\tau_{12})$.
By using this, one can rewrite the term as
	\begin{align}
		{\rm Tr} \log \Psi \, &= \, {\rm Tr} \left[ \log \left( \frac{{\rm sgn}}{2} \right) \, + \, \log\left( 1 + \frac{\partial({\rm sgn} \times \Phi)}{2q} \right) \right] \nonumber\\
		&= \, {\rm Tr} \log \left( \frac{{\rm sgn}}{2} \right) \, - \, \sum_{n=1}^{\infty} \frac{1}{n} \left( \frac{-1}{2q} \right)^n {\rm Tr} \Big[ \partial ({\rm sgn} \times \Phi) \Big]^n_{\star} \, , 
	\end{align}
where the last term is defined for example for $n=2$ as
	\begin{equation}
		{\rm Tr} \Big[ \partial ({\rm sgn} \times \Phi) \Big]^2_{\star}
		\, \equiv \, \int d\tau_1 d\tau_2 \, \partial_1 \Big({\rm sgn}(\tau_{12}) \Phi(\tau_1, \tau_2) \Big) \partial_2 \Big({\rm sgn}(\tau_{21}) \Phi(\tau_2, \tau_1) \Big) \, .
	\end{equation}
For the interaction term we have
	\begin{equation}
		\Big[ \Psi(\tau_1, \tau_2) \Big]^q \, = \, \frac{1}{2^q} \, \left[ 1 \, + \, \frac{\Phi(\tau_1, \tau_2)}{q} \right]^q \, .
	\end{equation}
We consider only $q=$ even integer case and the sign function disappeared here.
We want to rewrite the RHS as the following form
	\begin{equation}
		\left[ 1 \, + \, \frac{\Phi(\tau_1, \tau_2)}{q} \right]^q
		\, = \, e^{\Phi(\tau_1, \tau_2)} \left[ 1 \, + \, \frac{c_1(\tau_1, \tau_2)}{q} \, + \, \frac{c_2(\tau_1, \tau_2)}{q^2} \, + \, \mathcal{O}(q^{-3}) \right] \, ,
	\end{equation}
where function $c_1$ and $c_2$ are to be determined.
Taking logarithm of both-hand sides and expanding the logarithm, one finds
	\begin{align}
		\log({\rm LHS}) \, &= \, \Phi(\tau_1, \tau_2) \, - \, \frac{\Phi^2(\tau_1, \tau_2)}{2q} \, + \, \frac{\Phi^3(\tau_1, \tau_2)}{3q^2} \, + \, \mathcal{O}(q^{-3}) \, , \nonumber\\
		\log({\rm RHS}) \, &= \, \Phi(\tau_1, \tau_2) \, + \, \frac{c_1(\tau_1, \tau_2)}{q} \, + \, \frac{c_2(\tau_1, \tau_2)}{q^2} \, - \, \frac{c_1^2(\tau_1, \tau_2)}{2q^2}
		\, + \, \mathcal{O}(q^{-3}) \, .
	\end{align}
Comparing the both sides, the unfixed functions are determined as
	\begin{align}
		c_1(\tau_1, \tau_2) \, &= \, - \, \frac{1}{2} \, \Phi^2(\tau_1, \tau_2) \, , \nonumber\\
		c_2(\tau_1, \tau_2) \, &= \, \frac{1}{3} \, \Phi^3(\tau_1, \tau_2) \, + \, \frac{1}{8} \, \Phi^4(\tau_1, \tau_2) \, .
	\end{align}

Combining everything, the collective action is now written as
	\begin{align}
		S_{\rm col}[\Phi] \, &= \, - \, \frac{N}{2} \sum_{n=2}^{\infty} \, \frac{1}{n}\left( \frac{-1}{2q} \right)^n {\rm Tr}\left[ \partial \big( {\rm sgn} \times \Phi \big) \right]_{\star}^n \nonumber\\
		&\quad - \, \frac{\mathcal{J}^2 N}{4q^2} \int d\tau_1 d\tau_2 \, e^{\Phi(\tau_1, \tau_2)} \left[ 1 + \frac{c_1(\tau_1, \tau_2)}{q} + \frac{c_2(\tau_1, \tau_2)}{q^2} + \cdots \right] \, .
	\end{align}
The $\mathcal{O}(q^0)$ and $\mathcal{O}(q^{-1})$ order terms are identically vanish due to the free field equation of motion.
Therefore, the first non-trivial order is $\mathcal{O}(q^{-2})$.
Lower order terms can be explicitly written down as
	\begin{align}
		S_{\rm col}[\Phi] \ &= \, - \, \frac{N}{16q^2} \int d\tau_1 d\tau_2 \, \partial_1 \big( {\rm sgn}(\tau_{12}) \Phi(\tau_1, \tau_2) \big)
		\partial_2 \big( {\rm sgn}(\tau_{21}) \Phi(\tau_2, \tau_1) \big) \, - \, \frac{\mathcal{J}^2 N}{4q^2} \int d\tau_1 d\tau_2 \, e^{\Phi(\tau_1, \tau_2)} \nonumber\\
		&\quad + \, \frac{N}{48q^3} \int d\tau_1 d\tau_2 d\tau_3 \, \partial_1 \big( {\rm sgn}(\tau_{12}) \Phi(\tau_{12}) \big) \partial_2 \big( {\rm sgn}(\tau_{23}) \Phi(\tau_{23}) \big)
		\partial_3 \big( {\rm sgn}(\tau_{31}) \Phi(\tau_{31}) \big) \nonumber\\
		&\quad + \, \frac{\mathcal{J}^2 N}{8q^3} \int d\tau_1 d\tau_2 \, \Phi^2(\tau_1, \tau_2) \, e^{\Phi(\tau_1, \tau_2)} \, + \, \mathcal{O}(q^{-4}) \, .
	\end{align}

\section{Details of Exact Eigenfunctions}
\label{app:exacteigfn}

In this Appendix, we elaborate on various properties of the exact eigenfunctions $\eqref{exacteigenfn}$.
\subsection{Zero temperature limit} \label{sapp:zerotemplim}
Consider the exact eigenfunctions at finite temperature at large $q$  \cite{Maldacena:2016hyu} which can be written as
\begin{equation}
\chi^{(\textrm{ex})}_{n,\nu}(x) \sim \left(\sin \frac{\tilde{x}}{2}\right)^{1/2} \left[P^{-\nu}_{\tilde{n}-\frac{1}{2}}\left(\cos\frac{\tilde{x}}{2}\right)+\kappa^{\text{even/odd}}_{\tilde{n},\nu} P^{\nu}_{\tilde{n}-\frac{1}{2}}\left(\cos\frac{\tilde{x}}{2}\right)\right]
\end{equation}
where
\begin{equation}
\kappa^{\text{even}}_{\tilde{n},\nu} =-2^{-2\nu} \frac{\Gamma\left(\frac{1}{4}-\frac{\tilde{n}}{2}-\frac{\nu}{2}\right) \Gamma\left(\frac{1}{4}+\frac{\tilde{n}}{2}-\frac{\nu}{2}\right)}{\Gamma\left(\frac{1}{4}-\frac{\tilde{n}}{2}+\frac{\nu}{2}\right)\Gamma\left(\frac{1}{4}+\frac{\tilde{n}}{2}+\frac{\nu}{2}\right)} \qquad \kappa^{\text{odd}}_{\tilde{n},\nu} =-2^{-2\nu} \frac{\Gamma\left(\frac{3}{4}-\frac{\tilde{n}}{2}-\frac{\nu}{2}\right) \Gamma\left(\frac{3}{4}+\frac{\tilde{n}}{2}-\frac{\nu}{2}\right)}{\Gamma\left(\frac{3}{4}-\frac{\tilde{n}}{2}+\frac{\nu}{2}\right)\Gamma\left(\frac{3}{4}+\frac{\tilde{n}}{2}+\frac{\nu}{2}\right)}
\end{equation}
and
\begin{equation}
\begin{split}
&\tilde{x}=vx+(1-v)\pi\\
&\tilde{n}=\frac{n}{v}\\
&v=1-\frac{2}{\beta \mathcal{J}}+\mathcal{O}\left(\frac{1}{\beta^2 \mathcal{J}^2}\right)\\
&\omega=\frac{2\pi n}{\beta}
\end{split}
\end{equation}
Let us define
\begin{equation}
x=\frac{4\pi z}{\beta}
\end{equation}
This allows us to write 
\begin{equation}
\tilde{x}=\frac{\omega}{n \mathcal{J}}\ (2 \mathcal{J}z+1)+\mathcal{O}\left[\frac{1}{ (n \mathcal{J})^2}\right]
\end{equation}
Taking the zero temperature limit is equivalent to taking $n \rightarrow \infty$. In this limit, 
\begin{equation}
\sin \frac{\tilde{x}}{2} \simeq  \frac{\tilde{x}}{2}
\end{equation}
and
\begin{equation}
\lim_{n\rightarrow \infty}P^{\pm\nu}_{\tilde{n}-\frac{1}{2}}\left(\cos\frac{\tilde{x}}{2}\right) \sim n^{\pm\nu} J_{\mp\nu}\left[\frac{\omega\ (2\mathcal{J}z+1)}{2\mathcal{J}}\right]
\end{equation}
On doing so, we can write the exact eigenfunction in the zero temperature but finite coupling limit as
\begin{equation} \label{eigenexact}
\chi^{(\textrm{ex})}_{\omega,\nu} (z) \sim \left[\omega z+\frac{\omega}{2\mathcal{J}}\right]^{1/2} \left[J_\nu\left(\omega z+\frac{\omega}{2\mathcal{J}}\right)+\xi_\omega(\nu)\ J_{-\nu}\left(\omega z+\frac{\omega}{2\mathcal{J}}\right)\right]
\end{equation}
where
\begin{equation} \label{exactxi}
\begin{split}
\xi_\omega(\nu) &\equiv \lim_{n\rightarrow\infty} n^{2\nu}\kappa_{\tilde{n},\nu} \\&= -\lim_{n\rightarrow\infty} \left(\frac{n}{2}\right)^{2\nu} \frac{\Gamma\left(\frac{1}{4}-\frac{n}{2}-\frac{\nu}{2}-\frac{\omega}{2\pi\mathcal{J}}\right) \Gamma\left(\frac{1}{4}+\frac{n}{2}-\frac{\nu}{2}+\frac{\omega}{2\pi\mathcal{J}}\right)}{\Gamma\left(\frac{1}{4}-\frac{n}{2}+\frac{\nu}{2}-\frac{\omega}{2\pi\mathcal{J}}\right)\Gamma\left(\frac{1}{4}+\frac{n}{2}+\frac{\nu}{2}+\frac{\omega}{2\pi\mathcal{J}}\right)}\\
&=-\frac{\Gamma\left(\frac{1}{4}-\frac{\nu}{2}-\frac{\omega}{2\pi\mathcal{J}}\right) \Gamma\left(\frac{3}{4}+\frac{\nu}{2}+\frac{\omega}{2\pi\mathcal{J}}\right)}{\Gamma\left(\frac{1}{4}+\frac{\nu}{2}-\frac{\omega}{2\pi\mathcal{J}}\right) \Gamma\left(\frac{3}{4}-\frac{\nu}{2}+\frac{\omega}{2\pi\mathcal{J}}\right)}
\end{split}
\end{equation}
Equation $\eqref{exactxi}$ can be proved as follows.
\begin{equation} \label{xidef}
\begin{split}
\xi_\omega(\nu) &= -\lim_{n\rightarrow\infty} \left(\frac{n}{2}\right)^{2\nu} \frac{\Gamma\left(\frac{1}{4}-\frac{n}{2}-\frac{\nu}{2}-\frac{\omega}{2\pi\mathcal{J}}\right) \Gamma\left(\frac{1}{4}+\frac{n}{2}-\frac{\nu}{2}+\frac{\omega}{2\pi\mathcal{J}}\right)}{\Gamma\left(\frac{1}{4}-\frac{n}{2}+\frac{\nu}{2}-\frac{\omega}{2\pi\mathcal{J}}\right)\Gamma\left(\frac{1}{4}+\frac{n}{2}+\frac{\nu}{2}+\frac{\omega}{2\pi\mathcal{J}}\right)}
\end{split}
\end{equation}
Now, we can use the identity
\begin{equation} \label{gammaid}
\lim_{n\rightarrow\infty} \left(\frac{n}{2}\right)^{\nu}
\frac{\Gamma\left(\frac{1}{4}+\frac{n}{2}-\frac{\nu}{2}+\frac{\omega}{2\pi\mathcal{J}}\right)}{\Gamma\left(\frac{1}{4}+\frac{n}{2}+\frac{\nu}{2}+\frac{\omega}{2\pi\mathcal{J}}\right)}=1
\end{equation}
Using $\eqref{gammaid}$ in $\eqref{xidef}$, we have
\begin{equation}
\begin{split}
\xi_\omega(\nu) &= -\lim_{n\rightarrow\infty} \left(\frac{n}{2}\right)^{\nu} \frac{\Gamma\left(\frac{1}{4}-\frac{n}{2}-\frac{\nu}{2}-\frac{\omega}{2\pi\mathcal{J}}\right) }{\Gamma\left(\frac{1}{4}-\frac{n}{2}+\frac{\nu}{2}-\frac{\omega}{2\pi\mathcal{J}}\right)}\\
&=-\frac{\Gamma\left(\frac{1}{4}-\frac{\nu}{2}-\frac{\omega}{2\pi\mathcal{J}}\right) }{\Gamma\left(\frac{1}{4}+\frac{\nu}{2}-\frac{\omega}{2\pi\mathcal{J}}\right)}\lim_{n\rightarrow\infty} \left(\frac{n}{2}\right)^{\nu} \prod_{p=1}^{n/2} \frac{\Gamma\left(\frac{1}{4}+\frac{\nu}{2}-\frac{\omega}{2\pi\mathcal{J}}-p\right) }{\Gamma\left(\frac{1}{4}-\frac{\nu}{2}-\frac{\omega}{2\pi\mathcal{J}}-p\right)}\\
&=-\frac{\Gamma\left(\frac{1}{4}-\frac{\nu}{2}-\frac{\omega}{2\pi\mathcal{J}}\right) \Gamma\left(\frac{3}{4}+\frac{\nu}{2}+\frac{\omega}{2\pi\mathcal{J}}\right)}{\Gamma\left(\frac{1}{4}+\frac{\nu}{2}-\frac{\omega}{2\pi\mathcal{J}}\right) \Gamma\left(\frac{3}{4}-\frac{\nu}{2}+\frac{\omega}{2\pi\mathcal{J}}\right)}
\end{split}
\end{equation}
as required. We can expand $\xi_\omega(\nu)$ in a power series in $\frac{\omega}{2\mathcal{J}}$ to get
\begin{equation}
\xi_\omega(\nu)=\frac{\tan\frac{\pi\nu}{2}+1}{\tan\frac{\pi\nu}{2}-1}+\frac{\omega}{2\mathcal{J}}\frac{2\sin\pi\nu}{\sin\pi\nu-1}+\mathcal{O}\left(\frac{\omega^2}{\mathcal{J}^2}\right)
\end{equation}
The leading order piece agrees with $\xi_\nu$ found in \cite{Polchinski:2016xgd}.
From $\eqref{exactxi}$, it is easy to see
\begin{equation}
\xi_\omega(\nu)=\frac{1}{ \xi_\omega(-\nu)}
\label{xi relation}
\end{equation}
An alternative way to see that the exact eigenfunction should be a linear combination of Bessel functions is to notice that they are solutions to the following differential equation
\begin{equation} \label{exactDE}
\left[\omega^2 + \partial_z^2 -\frac{\nu^2-1/4}{\left(z+\frac{1}{2\mathcal{J}}\right)^2}\right] \chi^{\textrm{(ex)}}_{\omega,\nu}(z)=0
\end{equation}
If we make a shift $z\rightarrow z-\frac{1}{2\mathcal{J}}$, we can see that $\eqref{exactDE}$ turns into a Bessel equation. 
\subsection{Quantization Condition} \label{sapp:quantcond}
\begin{figure}[t!]
	\begin{center}
		\scalebox{0.6}{\hspace{10pt} \includegraphics{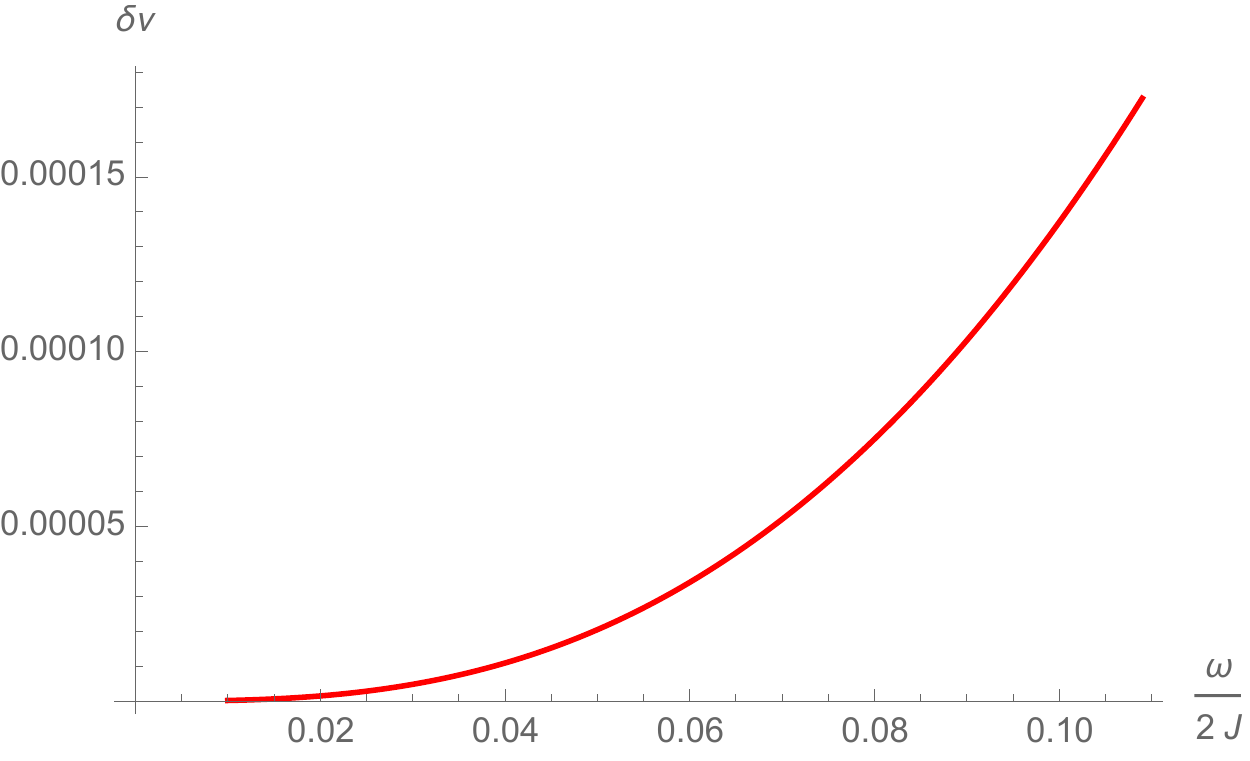}} 
	\end{center}
	\caption{Plot of the absolute difference between the allowed values of $\nu$, where we have defined $\delta\nu=\nu_{\text{MS}}-\nu_{\text{exact}}$. We see that the difference is almost zero for small values of $\frac{\omega}{2\mathcal{J}}$, which corresponds to the conformal limit.}
	\label{fig:MSexact}
\end{figure}

Let us impose the following boundary condition on the exact eigenfunctions
\begin{equation}
\chi^{(\text{ex})}_{\omega,\nu} (z)\Big | _{z=0}= 0
\end{equation}
Using equation $\eqref{eigenexact}$, imposing the above boundary condition requires us to solve the following equation in order to get the quantized values of $\nu$
\begin{equation} \label{quantcond}
J_\nu\left(\frac{\omega}{2\mathcal{J}}\right)+\xi_\omega(\nu)\ J_{-\nu}\left(\frac{\omega}{2\mathcal{J}}\right)=0
\end{equation}
We can solve this equation numerically and do a polynomial fit of order two for the curve given by Figure \ref{fig:MSexact} to get
\begin{equation}
\nu=\frac{3}{2}+a_1 \frac{\omega}{\mathcal{J}}+a_2 \left(\frac{\omega}{\mathcal{J}}\right)^2
\end{equation}
where $a_1=0.318048$ and $a_2=0.0046914$. This is in good agreement with the results obtained in \cite{Maldacena:2016hyu}. This can be seen as follows
\begin{equation} \label{nuMS}
\begin{split}
\nu_{\text{MS}}&=\frac{3}{2}+n\left(\frac{1-v}{v}\right)\\
&=\frac{3}{2}+\frac{\omega}{\pi \mathcal{J}}+\mathcal{O}\left(\frac{\omega}{\mathcal{J}}\right)^3
\end{split}
\end{equation}
We have used equation (2.31) of \cite{Maldacena:2016hyu}, to get from the first to second line of $\eqref{nuMS}$.

Another important observation is the following. The quantized values for $\nu$ one gets by solving the exact quantization equation $\eqref{quantcond}$ is in very good agreement with those obtained by solving the following equation 
\begin{equation} \label{xiroot}
\xi_\omega(\nu)=0 \implies \nu_n=2n+\frac{3}{2}+\frac{\omega}{\pi \mathcal{J}}
\end{equation}
for small values of $\frac{\omega}{2\mathcal{J}}$. It should be noted that the quantized values obtained by solving $\eqref{xiroot}$ hold true for any coupling. The discrepancy between the two methods arises only at order $\left(\frac{\omega}{2\mathcal{J}}\right)^{2\nu}$. This can be understood by recalling the asymptotic form of the Bessel function for small argument
\begin{equation} \label{Besselsmall}
J_\nu(z) \sim \left(\frac{z}{2}\right)^{\nu} \frac{1}{\Gamma(1+\nu)}
\end{equation}
Using $\eqref{Besselsmall}$ in $\eqref{quantcond}$, requires one to solve the following equation
\begin{equation}
\xi_\omega(\nu) \sim \left(\frac{\omega}{2\mathcal{J}}\right)^{2\nu}
\end{equation}

\begin{figure}[t!]
	\begin{center}
		\scalebox{0.6}{\hspace{10pt} \includegraphics{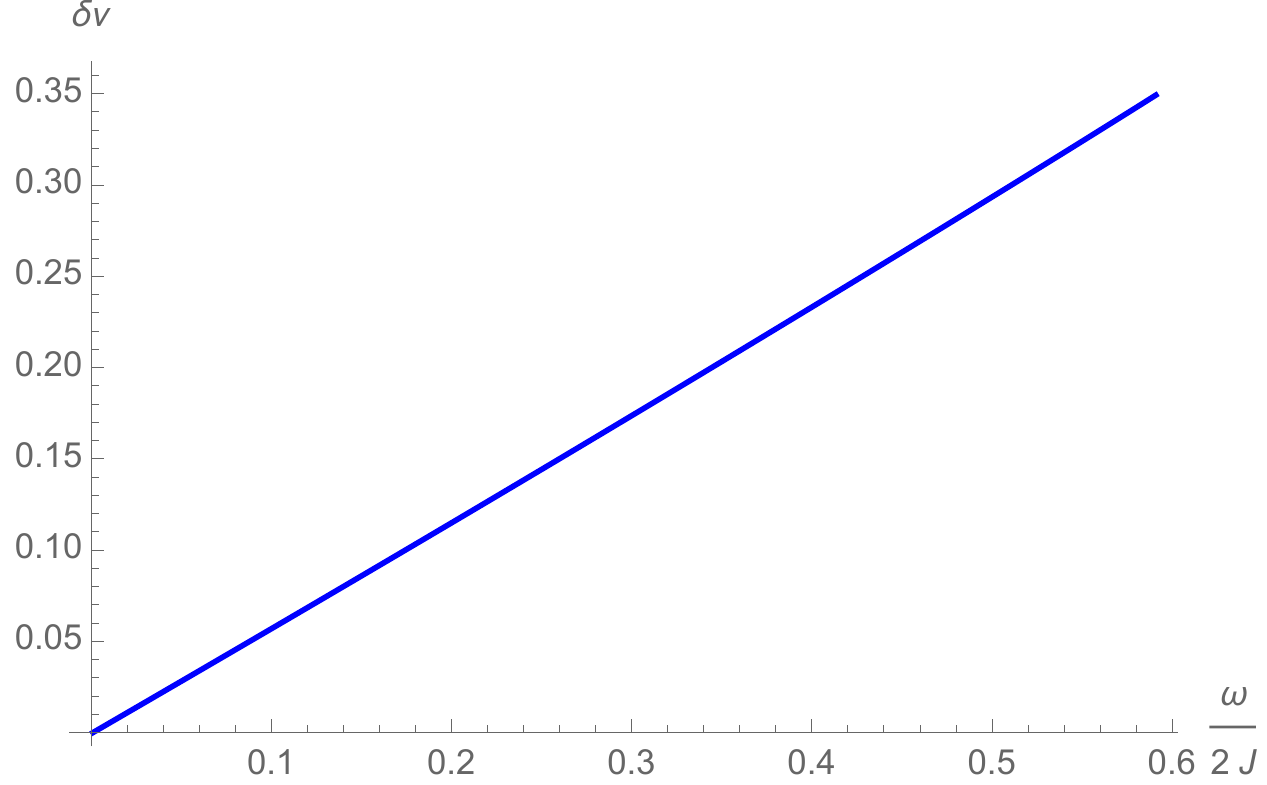}} 
	\end{center}
	\caption{Plot of the absolute difference between the allowed values of $\nu$ from two different methods, where we have defined $\delta\nu=\nu_{\text{approx}}-\nu_{\text{exact}}$ where $\nu_{\text{approx}}$ refers to the values of $\nu$ obtained by solving $\eqref{xiroot}$. We see that the difference grows with $\frac{\omega}{2\mathcal{J}}.$}
	\label{fig:exactapprox}
\end{figure}
The following Table \ref{numerical} gives the difference in quantized values for $\nu$ from the exact and approximate methods as a function of $\frac{\omega}{2\mathcal{J}}$. It provides numerical data in support of the above claim (see also Figure $\ref{fig:MSexact}$)
\begin{table}[h!]
\begin{center}
\begin{tabular}{| c | c |} \hline
$\frac{\omega}{2\mathcal{J}}$ & $\delta\nu$  \\ \hline
0.001 & 2.10064$\times 10^{-10}$\\ \hline
0.051 & 2.16037$\times 10^{-5}$\\ \hline
0.101 & 1.40694$\times 10^{-4}$\\ \hline
0.501 & 7.42506$\times 10^{-3}$\\ \hline
\end{tabular}
\end{center}
\caption{Difference in quantized values for $\nu$ from the exact and approximate methods as a function of $\frac{\omega}{2\mathcal{J}}$.}
\label{numerical}
\end{table}
\subsection{Perturbative determination of exact normalization} \label{sapp:pertnorm}
We derive $\eqref{exactnorm}$ here.
Let us consider $\nu=2n+\frac{3}{2}+\frac{\vert\omega\vert}{\pi\mathcal{J}}$ where $\xi_\omega(\nu)$ vanishes. So, the integral we need to perform is
\begin{equation} \label{exnorm}
\begin{split}
\int_0^\infty\frac{dz}{z+\frac{1}{2\mathcal{J}}} J_{\nu_1}(\vert \omega \vert \tilde{z}) J_{\nu_2}(\vert \omega \vert \tilde{z}) 
=\hat{N}_{\nu_1}\ \delta(\nu_1-\nu_2) 
\end{split}
\end{equation}
We want to determine the normalization in $\eqref{exnorm}$. In order to do so, we perform a $1/ \mathcal{J}$ expansion of the integrand. This gives
\begin{equation} 
\int_0^\infty\frac{dz}{z+\frac{1}{2\mathcal{J}}} J_{\nu_1}(\vert \omega \vert \tilde{z}) J_{\nu_2}(\vert \omega \vert \tilde{z}) = I_1+I_2+I_3
\end{equation}
where
\begin{equation} \label{I1}
\begin{split}
I_1&=\int_0^{\infty} \frac{dz}{z} J_{\nu_1}(\vert \omega \vert z) J_{\nu_2}(\vert \omega \vert z) =\frac{1}{2\nu_1} \delta(\nu_1-\nu_2)
\end{split}
\end{equation}

\begin{equation} \label{I2}
\begin{split}
I_2&=\frac{\omega}{2\mathcal{J}} \int_0^{\infty} \frac{dz}{z} \left[J_{\nu_1}(\vert \omega \vert z) J^\prime_{\nu_2}(\vert \omega \vert z) + J^\prime_{\nu_1}(\vert \omega \vert z) J_{\nu_2}(\vert \omega \vert z)-\frac{J_{\nu_1}(\vert \omega \vert z) J_{\nu_2}(\vert \omega \vert z)}{z}\right]\\
&=0\qquad\qquad\qquad (\text{for}\ \nu_1+\nu_2 >1)
\end{split}
\end{equation}
\begin{equation} \label{I3}
\begin{split}
I_3&=\left(\frac{\omega}{2\mathcal{J}}\right)^2 \int_0^{\infty} \frac{dz}{z} \Bigg[\frac{1}{2}\left(J^{\prime\prime}_{\nu_2}(\vert \omega \vert z)+2J^\prime_{\nu_1}(\vert \omega \vert z)J^\prime_{\nu_2}(\vert \omega \vert z)+J_{\nu_1}(\vert \omega \vert z)J^{\prime\prime}_{\nu_2}(\vert \omega \vert z)\right)
\\&-\frac{J_{\nu_1}(\vert \omega \vert z) J^\prime_{\nu_2}(\vert \omega \vert z) + J^\prime_{\nu_1}(\vert \omega \vert z) J_{\nu_2}(\vert \omega \vert z)}{z}+\frac{J_{\nu_1}(\vert \omega \vert z) J_{\nu_2}(\vert \omega \vert z)}{z^2}\Bigg]
\\&=0 \qquad\qquad\qquad (\text{for}\ \nu_1+\nu_2 >2)
\end{split}
\end{equation}
Based on the results of the integrals $\eqref{I1}-\eqref{I3}$, we conclude that
\begin{equation}
\hat{N}_{\nu_1}=\frac{1}{2\nu_1}+\mathcal{O}\left(\frac{\vert\omega\vert}{2\mathcal{J}}\right)^3
\end{equation}

\section{Detail Evaluation of Non-zero Mode Bi-local Propagator $\mathcal{D}_c$}
\label{app:nuint}
In this Appendix, we give details of the evaluation of $\mathcal{D}_c$ given by $\eqref{Dcfin}$. Since there is no enhancement for this part, we do not need to keep track of the $\left(|\omega |\mathcal{J}^{-1}\right)$ corrections and we can evaluate it in the conformal limit as usual.  Let us consider the continuous integral first in $\eqref{Dc}$ which we denote by $I_c$
\begin{equation} \label{Ic-main}
I_c \, = \, 4q^2 (z z^\prime)^{\frac{1}{2}} \int_{-\infty}^\infty \frac{d\omega}{2\pi}\ e^{i\omega(t-t^\prime)} \int \frac{d\nu}{N_\nu}\ \frac{Z^*_{\nu}(\vert \omega \vert z) Z_{\nu}(\vert \omega \vert z^\prime)}{\tilde{g}(\nu)-1} \, ,
\end{equation}
with $\nu = ir $.
The $\nu$ integral can be written as
\begin{equation} \label{Ic}
\begin{split}
&\quad \ \int_0^\infty dr\ \frac{ir}{2\sin\pi i r}\ \frac{Z_{-ir}(\vert \omega \vert z) Z_{ir}(\vert \omega \vert z^\prime)}{\tilde{g}(ir)-1}\\
&= \, \frac{1}{i} \int_{-i\infty}^{i\infty} d\nu\ \frac{\nu}{2\sin\pi \nu}\ \frac{Z_{-\nu}(\vert \omega \vert z) J_{\nu}(\vert \omega \vert z^\prime)}{\tilde{g}(\nu)-1}\\
&= \, (T_1+T_2) \, .
\end{split}
\end{equation}
We have extended the limits over the entire imaginary axis and defined $\nu=ir$ in going from the first to second line of $\eqref{Ic}$ using (\ref{xi relation}).
For $z>z^\prime$ we close the contour clockwise in the right half plane, which gives
\begin{equation} \label{T1}
T_1 \, = \, \frac{1}{i} \int_{-i\infty}^{i\infty} d\nu\ \frac{\nu}{2\sin\pi \nu}\ \frac{J_{-\nu}(\vert \omega \vert z) J_{\nu}(\vert \omega \vert z^\prime)}{\tilde{g}(\nu)-1} \, ,
\end{equation}
and 
\begin{equation} \label{T2}
T_2 \, = \, \frac{1}{i} \int_{-i\infty}^{i\infty} d\nu\ \frac{\nu}{2\sin\pi \nu}\ \frac{J_{\nu}(\vert \omega \vert z) J_{\nu}(\vert \omega \vert z^\prime)}{\tilde{g}(\nu)-1}\ \xi(-\nu) \, .
\end{equation}
$T_1$ only has a simple pole at $\nu=\frac{3}{2}$,where $\tilde{g}(3/2)=1$.
\footnote{
The poles at $\nu = n$ with $n$ being a positive integer will cancel with the same poles of $T_2$, so we don't write down these contributions here.
}
We can evaluate the residue to get
\begin{equation} \label{T11}
T_1 \, = \, \frac{3\pi}{2} \frac{J_{-\frac{3}{2}}(\vert \omega \vert z)J_{\frac{3}{2}}(\vert \omega \vert z^\prime)}{\tilde{g}^\prime\left(\frac{3}{2}\right)} \, .
\end{equation}
The $T_2$ term is a little more complicated. It has simple poles at $\nu=2n+\frac{3}{2}$ where $\xi_\nu=0$. We can evaluate the residue there to get
\begin{equation} \label{T22}
T_2 \, = \, - \, I_d \, + \, \left[\frac{1}{i} \int_{-i\infty}^{i\infty} d\nu\ \frac{\nu}{2\sin\pi \nu}\ \frac{J_{\nu}(\vert \omega \vert z) J_{\nu}(\vert \omega \vert z^\prime)}{\tilde{g}(\nu)-1}\ \xi(-\nu)\right]_{\nu=\frac{3}{2}} \, ,
\end{equation}
where the second term is only the $\nu=3/2$ contribution and
	\begin{equation}
		I_d \, = \, \sum_{n=1}^\infty (4n+3)\ \frac{J_{2n+\frac{3}{2}}(\vert \omega \vert z) J_{2n+\frac{3}{2}}(\vert \omega \vert z^\prime)}{\tilde{g}(2n+\frac{3}{2})-1} \, .
	\end{equation}
Let us consider the integrand in the second term of $\eqref{T22}$. It has singularities coming from $\xi(-\nu)$ as well as $\tilde{g}(\nu)-1$. Let us expand the following quantity around $\nu=\frac{3}{2}$
\begin{equation} \label{doublepole}
\frac{\xi(-\nu)}{\tilde{g}(\nu)-1} \, = \, \frac{1}{\left(\nu-\frac{3}{2}\right)^2 \tilde{g}^\prime\left(\frac{3}{2}\right) \xi^\prime\left(\frac{3}{2}\right)}\left[1-\frac{\left(\nu-\frac{3}{2}\right)}{2} \left(\frac{\xi^{\prime\prime}(\frac{3}{2})}{\xi^{\prime}(\frac{3}{2})}+\frac{\tilde{g}^{\prime\prime}\left(\frac{3}{2}\right)}{\tilde{g}^{\prime}\left(\frac{3}{2}\right)}\right)\right] \, .
\end{equation}
We have a double pole coming from the first term of $\eqref{doublepole}$ and a simple pole from the second term. Using $\eqref{doublepole}$ in $\eqref{T22}$, and evaluating the residues, we get
	\begin{align} 
	\label{Ic1}
		I_c \, = \, 4q^2 (z z^\prime)^{\frac{1}{2}} &\int_{-\infty}^\infty \frac{d\omega}{2\pi}\ e^{i\omega(t-t^\prime)}
		\Bigg[ \frac{3\pi}{2} \frac{J_{-\frac{3}{2}}(\vert \omega \vert z^>)J_{\frac{3}{2}}(\vert \omega \vert z^<)}{\tilde{g}^\prime\left(\frac{3}{2}\right)} \, - \, I_d \\
		&\ + \, \frac{\pi}{\tilde{g}^\prime\left(\frac{3}{2}\right) \xi^\prime\left(\frac{3}{2}\right)} \left( \frac{3}{2} \, \partial_{\nu} + 1 - \frac{3\xi''(\frac{3}{2})}{4\xi'(\frac{3}{2})} 
		- \frac{3\tilde{g}''(\frac{3}{2})}{4\tilde{g}'(\frac{3}{2})} \right) J_{\nu}(|\omega| z) J_{\nu}(|\omega| z') \bigg|_{\nu=\frac{3}{2}} \Bigg] \, . \nonumber
	\end{align}
The first term in $\eqref{Dc}$ exactly cancels with $I_d$ in $\eqref{Ic1}$.
Using $\tilde{g}'(\frac{3}{2}) =\frac{3}{2},  \tilde{g}''(\frac{3}{2}) = 1, \xi^\prime\left(\frac{3}{2}\right)=-\frac{\pi}{2}$ and $ \xi^{\prime\prime}\left(\frac{3}{2}\right)=0$ in $\eqref{Ic1}$, we get $\eqref{Dcfin}$ We have dropped the $\omega$ subscript while writing $\xi$ here because we are working in the conformal limit.

\section{Integrals of Products of Bessel Functions}
\label{app:Bessel product}
In this appendix, we consider the following integrals involving a product of two Bessel functions:
	\begin{align}
		&\quad \ \int_{-\infty}^{\infty} d\omega \, e^{-i\omega(t-t')} \, J_{\pm \nu}(|\omega| z^>) \, J_{\nu}(|\omega| z^<) \nonumber\\
		&= \, 2 \int_0^{\infty} d\omega \, \cos(\omega(t-t')) \, J_{\pm \nu}(\omega z^>) \, J_{\nu}(\omega z^<) \, .
	\label{eq:cos-integral}
	\end{align}
The plus sign case was already evaluated in Appendix D of \cite{Polchinski:2016xgd}.
The result is
	\begin{equation}
		2 \int_0^{\infty} d\omega \, \cos(\omega(t-t')) \, J_{\nu}(|\omega| z^>) \, J_{\nu}(|\omega| z^<) \, = \, \frac{1}{\pi \sqrt{z z'}} \, \mathcal{I}^{(1)}_{\nu}(\xi) \, , 
	\label{eq:plus-sign-integral}
	\end{equation}
with
	\begin{align}
		\mathcal{I}^{(1)}_{\nu}(\xi) \, \equiv \, \left\{
		\begin{array}{lc}
			2Q_{\nu - 1/2}(\xi) \qquad & (\xi>1) \\
			Q_{\nu - 1/2}(\xi + i \varepsilon) + Q_{\nu - 1/2}(\xi - i \varepsilon) \qquad & (-1<\xi<1) \\
			- 2\sin (\pi \nu)Q_{\nu - 1/2} (-\xi), & (\xi<-1)
		\end{array} \right.
	\end{align}
where 
	\begin{equation}
		\xi \, \equiv \, \frac{z^2+z'^2-(t-t')^2}{2zz'} \, .
	\label{xi}
	\end{equation}
This is related to the geodesic distance of AdS$_2$, which is given by $d(t, z; t', z') \, = \, \ln(\xi + \sqrt{\xi^2-1})$.

The minus sign integral is more involved. Let us start from Eq.(D.11) of \cite{Polchinski:2016xgd}, which reads
\footnote{Note that there is a typo in Eq.(D.11) of \cite{Polchinski:2016xgd} missing the $\overline{\omega}^{-2\nu}$ factor in the integrand.}
	\begin{align}
		&\quad \, \int_0^{\infty} dx \, e^{-\alpha x} \, J_{\nu}(\beta x) \, J_{-\nu}(\gamma x) \nonumber\\
		&= \, \frac{\left( - \frac{\beta \gamma}{\alpha^2} \right)^{\nu}}{\sqrt{\pi} \, \alpha \, \Gamma(\nu+\frac{1}{2})\Gamma(1-\nu)}
		\int_0^{\pi} d\phi \, (\sin\phi)^{2\nu} \, \left( - \frac{\overline{\omega}^2}{\alpha^2} \right)^{-\nu} {}_2F_1\left( \frac{1}{2}, 1; 1-\nu; - \frac{\overline{\omega}^2}{\alpha^2} \right) \, ,
	\label{negative-sign-integral}	
	\end{align}
with $\overline{\omega}=\sqrt{\beta^2+\gamma^2-2\beta \gamma \cos\phi}$.
In order to obtain the cosine integral (\ref{eq:cos-integral}), we need to evaluate the analytical continuation by $\alpha= a e^{i\theta}$ with $\theta=0 \to \pm\pi/2$
as explained in Appendix D of \cite{Polchinski:2016xgd}.
One has to be careful about this analytical continuation since the hypergeometric function ${}_2F_1(a,b;c;z)$ has a branch cut along the real $z$ axis through $1<z<\infty$.
Anyway, one should evaluate the $\phi$ integral before this analytical continuation.
For this integral it is convenient to rewrite the hypergeometric function in the integrand using the hypergeometric function identity (for example see 9.131.2 of \cite{Gradshteyn:1994}) as
	\begin{align}
		&\quad \, \left( - \frac{\overline{\omega}^2}{\alpha^2} \right)^{-\nu} {}_2F_1\left( \frac{1}{2}, 1; 1-\nu; - \frac{\overline{\omega}^2}{\alpha^2} \right) \nonumber\\
		&= \, \left( \frac{2\nu}{2\nu+1} \right) \, {}_2F_1\left( \nu+1, \nu+\frac{1}{2}; \nu+\frac{3}{2}; \frac{2\beta\gamma}{\alpha^2}(\xi-\cos\phi) \right) \nonumber\\
		&\qquad - \, \frac{1}{\sqrt{\pi}} \left( \frac{2\nu}{2\nu+1} \right) \, \Gamma(\nu+\tfrac{3}{2}) \Gamma(-\nu) \,
		{}_2F_1\left( \nu+1, \nu+\frac{1}{2}; \nu+1; - \frac{\overline{\omega}^2}{\alpha^2} \right) \nonumber\\
		&= \, \left( \frac{2\nu}{2\nu+1} \right) \, \sum_{n=0}^{\infty} \frac{(\nu+1)_n(\nu+\tfrac{1}{2})_n}{n! (\nu+\tfrac{3}{2})_n} 
		\left( \frac{2\beta\gamma}{\alpha^2} \right)^n (\xi-\cos\phi)^n \nonumber\\
		&\qquad - \, \frac{1}{\sqrt{\pi}} \left( \frac{2\nu}{2\nu+1} \right) \, \Gamma(\nu+\tfrac{3}{2}) \Gamma(-\nu)
		\left[ \frac{2\beta\gamma}{\alpha^2} (\xi - \cos\phi) \right]^{-\nu-\frac{1}{2}} \, , 
	\label{hypergeo-integrad}
	\end{align}
where we assumed $|2\beta \gamma \zeta / \alpha^2|<1$ and for the second equality,
we expanded the hypergeometric function in the power series with the rising Pochhammer symbol $(x)_n$.
We also defined 
	\begin{equation}
		\xi \, \equiv \, \frac{\alpha^2+\beta^2+\gamma^2}{2\beta\gamma} \, ,
	\end{equation}
which after the analytical continuation agrees with Eq.(\ref{xi}).
Therefore, now the $\phi$ integral can be performed as
	\begin{align}
		&\ \int_0^{\pi} d\phi \, (\sin \phi)^{2\nu} (\xi-\cos\phi)^s \\
		&= \left\{
		\begin{array}{l}
			4^{\nu} \, \pi \, \frac{(1+\xi)^s\Gamma(\nu+\frac{1}{2})}{\cos(\pi\nu) \Gamma(\frac{1}{2}-\nu)\Gamma(2\nu+1)} \,
			{}_2F_1\left( \nu+\frac{1}{2}, -s; 2\nu+1; \frac{2}{1+\xi} \right) \, , \hspace{70pt} (\xi>1) \\[8pt]
			2^{\nu} \, \Gamma(\nu+\tfrac{1}{2}) \Big[ \frac{(-1)^s \pi \, 2^{\nu+s}}{\Gamma(2\nu+s+1)\Gamma(\frac{1}{2}-\nu-s)\cos(\pi(\nu+s))} \,
			{}_2F_1\left(-2\nu-s, -s; \frac{1}{2}-\nu-s; \frac{1+\xi}{2} \right) \\[8pt]
			\hspace{30pt} + \frac{(1+\xi)^{\frac{1}{2}+\nu+s} \Gamma(1+s) [1-e^{i\pi s} \cos(\pi \nu)/\cos(\pi(\nu+s))]}{\sqrt{2} \Gamma(\frac{3}{2}+\nu+s)} 
			{}_2F_1\left(\frac{1}{2}-\nu, \frac{1}{2}+\nu; \frac{3}{2}+\nu+s; \frac{1+\xi}{2} \right) \Big] \, , \\[8pt]
			\hspace{320pt} \,  (-1<\xi<1) \nonumber\\[2pt]
			4^{\nu} (1-\xi)^s \, \frac{\Gamma^2(\nu+\frac{1}{2})}{\Gamma(2\nu+1)} \, {}_2F_1\left(\nu+\frac{1}{2}, -s; 2\nu+1; \frac{2}{1-\xi} \right) \, , \hspace{90pt} (\xi<-1)
		\end{array} \right.
	\end{align}
where $s$ is a general real number.
One can see that for any value of $\xi$ (and for any value of $s$),
the analytical continuation $\alpha= a e^{i\theta}$ with $\theta=0 \to \pm\pi/2$ does not hit the branch cut of the hypergeometric function.
Therefore, we can perform a naive analytical continuation for this $\phi$-integral part.

We note that for the cosine integral (\ref{eq:cos-integral}), the contribution from the first term in Eq.(\ref{hypergeo-integrad}) vanishes
due to the $\alpha^{-1}$ factor in Eq.(\ref{negative-sign-integral}) after the analytical continuation.
Therefore, for the cosine integral (\ref{eq:cos-integral}), the contribution solely comes from the second term in Eq.(\ref{hypergeo-integrad}), which is written as
	\begin{equation}
		2 \int_0^{\infty} dx \, \cos(a x) \, J_{\nu}(\beta x) \, J_{-\nu}(\gamma x) \, = \, \frac{1}{\pi \sqrt{\beta \gamma}} \ \mathcal{I}^{(2)}_{\nu}(\xi) \, ,
	\end{equation}
with
	\begin{align}
		\mathcal{I}^{(2)}_{\nu}(\xi) \, \equiv \, \left\{
		\begin{array}{lc}
			2\cos(\pi \nu) \, Q_{\nu-\frac{1}{2}}(\xi) \, , \qquad & (\xi>1) \\
			\pi \, P_{\nu-\frac{1}{2}}(-\xi) \, , \qquad & (-1<\xi<1) \\
			2\cos(\pi \nu) \, Q_{\nu-\frac{1}{2}}(-\xi) \, , \qquad & (\xi<-1)
		\end{array} \right.
	\end{align}

\section{Details of Evaluation of the Second Order Eigenvalue Shift} \label{app:lambda2}
In this Appendix, we explain how to get $\eqref{eig2}$ from $\eqref{l02}$, which we can rewrite as
\beq \label{l02ev}
\lambda_0^{(2)} \, =  \int \tilde{\chi}_k^{(0)} \cdot \widetilde{\mathcal{K}}^{(1)} \cdot \tilde{\chi}_0^{(1)}
		\, + \, \int \tilde{\chi}_0^{(0)} \cdot \widetilde{\mathcal{K}}^{(2)} \cdot \tilde{\chi}_0^{(0)} \,  
\eeq
The second term in $\eqref{l02ev}$ is given by
\beq \label{K2mat}
 \int \tilde{\chi}_0^{(0)} \cdot \widetilde{\mathcal{K}}^{(2)} \cdot \tilde{\chi}_0^{(0)} = \frac{1}{10}
\eeq
Using $\eqref{ef1}$, the first term in $\eqref{l02ev}$ can be written as
\beq
  \int \tilde{\chi}_k^{(0)} \cdot \widetilde{\mathcal{K}}^{(1)} \cdot \tilde{\chi}_0^{(1)} = I_1 +I_2
\eeq
where
\beq
\begin{split}
I_1 &=\int_{-\infty}^\infty dt \int_0^\infty dz\ \left(\frac{\sqrt{3}e^{-i\omega t}}{\sqrt{z}} J_{\frac{3}{2}}(|\omega| z)\right) \widetilde{\mathcal{K}}^{(1)}  \Bigg(\frac{\sqrt{3}e^{i\omega t}}{2\sqrt{z}} \left[J^\prime_{\frac{3}{2}} (\vert \omega \vert z)-\frac{1}{2z} J_{\frac{3}{2}} (\vert \omega \vert z)\right]\Bigg) \\
&= -\frac{1}{10}
\end{split}
\eeq
and 
\beq \label{I2}
\begin{split} 
I_2 &=\int_{-\infty}^\infty dt  \int_0^\infty dz\ \Bigg[\frac{2\sqrt{3}e^{-i\omega t}}{\pi \sqrt{z}} \left[ \frac{d}{d\nu}J_{\nu} ( \vert \omega \vert z) \right]_{\nu=\frac{3}{2}} + \frac{e^{-i\omega t}}{\pi \sqrt{3z}} J_{\frac{3}{2}} (\vert \omega \vert z)\Bigg] \widetilde{\mathcal{K}}^{(1)}    u^{(0)} _{\omega, \nu} (t,z)\\
&= \frac{1}{\pi^2}
\end{split}
\eeq
$\widetilde{\mathcal{K}}^{(1)}$ to be used above is given by $\eqref{LK1}$. Using $\eqref{K2mat}-\eqref{I2}$ in $\eqref{l02ev}$, we get $\eqref{eig2}$,up to a factor of two, which arises because of $\eqref{fseven}$.


\bibliographystyle{JHEP}
\bibliography{Refs} 

\providecommand{\href}[2]{#2}\begingroup\raggedright\begin{thebibliography}{10}

\bibitem{Sachdev:1992fk}
S.~Sachdev and J.~Ye, \emph{{Gapless spin fluid ground state in a random,
  quantum Heisenberg magnet}},
  \href{https://doi.org/10.1103/PhysRevLett.70.3339}{\emph{Phys. Rev. Lett.}
  {\bfseries 70} (1993) 3339}
  [\href{https://arxiv.org/abs/cond-mat/9212030}{{\ttfamily
  cond-mat/9212030}}].

\bibitem{Kitaev:2015}
A.~Kitaev, \emph{{Talks given at the Fundamental Physics Prize Symposium and
  KITP seminars}}, .

\bibitem{Sachdev:2015efa}
S.~Sachdev, \emph{{Bekenstein-Hawking Entropy and Strange Metals}},
  \href{https://doi.org/10.1103/PhysRevX.5.041025}{\emph{Phys. Rev. X}
  {\bfseries 5} (2015) 041025}
  [\href{https://arxiv.org/abs/1506.05111}{{\ttfamily 1506.05111}}].

\bibitem{Polchinski:2016xgd}
J.~Polchinski and V.~Rosenhaus, \emph{{The Spectrum in the Sachdev-Ye-Kitaev
  Model}}, \href{https://doi.org/10.1007/JHEP04(2016)001}{\emph{JHEP}
  {\bfseries 04} (2016) 001}
  [\href{https://arxiv.org/abs/1601.06768}{{\ttfamily 1601.06768}}].

\bibitem{Maldacena:2016hyu}
J.~Maldacena and D.~Stanford, \emph{{Remarks on the Sachdev-Ye-Kitaev model}},
  \href{https://doi.org/10.1103/PhysRevD.94.106002}{\emph{Phys. Rev. D}
  {\bfseries 94} (2016) 106002}
  [\href{https://arxiv.org/abs/1604.07818}{{\ttfamily 1604.07818}}].

\bibitem{Jevicki:2016bwu}
A.~Jevicki, K.~Suzuki and J.~Yoon, \emph{{Bi-Local Holography in the SYK
  Model}}, \href{https://doi.org/10.1007/JHEP07(2016)007}{\emph{JHEP}
  {\bfseries 07} (2016) 007}
  [\href{https://arxiv.org/abs/1603.06246}{{\ttfamily 1603.06246}}].

\bibitem{Jevicki:2016ito}
A.~Jevicki and K.~Suzuki, \emph{{Bi-Local Holography in the SYK Model:
  Perturbations}}, \href{https://doi.org/10.1007/JHEP11(2016)046}{\emph{JHEP}
  {\bfseries 11} (2016) 046}
  [\href{https://arxiv.org/abs/1608.07567}{{\ttfamily 1608.07567}}].

\bibitem{Davison:2016ngz}
R.~A. Davison, W.~Fu, A.~Georges, Y.~Gu, K.~Jensen and S.~Sachdev,
  \emph{{Thermoelectric transport in disordered metals without quasiparticles:
  The Sachdev-Ye-Kitaev models and holography}},
  \href{https://doi.org/10.1103/PhysRevB.95.155131}{\emph{Phys. Rev. B}
  {\bfseries 95} (2017) 155131}
  [\href{https://arxiv.org/abs/1612.00849}{{\ttfamily 1612.00849}}].

\bibitem{Gross:2017hcz}
D.~J. Gross and V.~Rosenhaus, \emph{{The Bulk Dual of SYK: Cubic Couplings}},
  \href{https://doi.org/10.1007/JHEP05(2017)092}{\emph{JHEP} {\bfseries 05}
  (2017) 092} [\href{https://arxiv.org/abs/1702.08016}{{\ttfamily
  1702.08016}}].

\bibitem{Gross:2017aos}
D.~J. Gross and V.~Rosenhaus, \emph{{All point correlation functions in SYK}},
  \href{https://doi.org/10.1007/JHEP12(2017)148}{\emph{JHEP} {\bfseries 12}
  (2017) 148} [\href{https://arxiv.org/abs/1710.08113}{{\ttfamily
  1710.08113}}].

\bibitem{Kitaev:2017awl}
A.~Kitaev and S.~J. Suh, \emph{{The soft mode in the Sachdev-Ye-Kitaev model
  and its gravity dual}},
  \href{https://doi.org/10.1007/JHEP05(2018)183}{\emph{JHEP} {\bfseries 05}
  (2018) 183} [\href{https://arxiv.org/abs/1711.08467}{{\ttfamily
  1711.08467}}].

\bibitem{Das:2017wae}
S.~R. Das, A.~Ghosh, A.~Jevicki and K.~Suzuki, \emph{{Space-Time in the SYK
  Model}}, \href{https://doi.org/10.1007/JHEP07(2018)184}{\emph{JHEP}
  {\bfseries 07} (2018) 184}
  [\href{https://arxiv.org/abs/1712.02725}{{\ttfamily 1712.02725}}].

\bibitem{Sarosi:2017ykf}
G.~Sárosi, \emph{{AdS$_{2}$ holography and the SYK model}},
  \href{https://doi.org/10.22323/1.323.0001}{\emph{PoS} {\bfseries Modave2017}
  (2018) 001} [\href{https://arxiv.org/abs/1711.08482}{{\ttfamily
  1711.08482}}].

\bibitem{Rosenhaus:2018dtp}
V.~Rosenhaus, \emph{{An introduction to the SYK model}},
  \href{https://doi.org/10.1088/1751-8121/ab2ce1}{\emph{J. Phys. A} {\bfseries
  52} (2019) 323001} [\href{https://arxiv.org/abs/1807.03334}{{\ttfamily
  1807.03334}}].

\bibitem{Gu:2016oyy}
Y.~Gu, X.-L. Qi and D.~Stanford, \emph{{Local criticality, diffusion and chaos
  in generalized Sachdev-Ye-Kitaev models}},
  \href{https://doi.org/10.1007/JHEP05(2017)125}{\emph{JHEP} {\bfseries 05}
  (2017) 125} [\href{https://arxiv.org/abs/1609.07832}{{\ttfamily
  1609.07832}}].

\bibitem{Gross:2016kjj}
D.~J. Gross and V.~Rosenhaus, \emph{{A Generalization of Sachdev-Ye-Kitaev}},
  \href{https://doi.org/10.1007/JHEP02(2017)093}{\emph{JHEP} {\bfseries 02}
  (2017) 093} [\href{https://arxiv.org/abs/1610.01569}{{\ttfamily
  1610.01569}}].

\bibitem{Berkooz:2016cvq}
M.~Berkooz, P.~Narayan, M.~Rozali and J.~Simón, \emph{{Higher Dimensional
  Generalizations of the SYK Model}},
  \href{https://doi.org/10.1007/JHEP01(2017)138}{\emph{JHEP} {\bfseries 01}
  (2017) 138} [\href{https://arxiv.org/abs/1610.02422}{{\ttfamily
  1610.02422}}].

\bibitem{Fu:2016vas}
W.~Fu, D.~Gaiotto, J.~Maldacena and S.~Sachdev, \emph{{Supersymmetric
  Sachdev-Ye-Kitaev models}},
  \href{https://doi.org/10.1103/PhysRevD.95.026009}{\emph{Phys. Rev. D}
  {\bfseries 95} (2017) 026009}
  [\href{https://arxiv.org/abs/1610.08917}{{\ttfamily 1610.08917}}].

\bibitem{Nishinaka:2016nxg}
T.~Nishinaka and S.~Terashima, \emph{{A note on Sachdev--Ye--Kitaev like model
  without random coupling}},
  \href{https://doi.org/10.1016/j.nuclphysb.2017.11.012}{\emph{Nucl. Phys. B}
  {\bfseries 926} (2018) 321}
  [\href{https://arxiv.org/abs/1611.10290}{{\ttfamily 1611.10290}}].

\bibitem{Erdmenger:2016jjg}
J.~Erdmenger, C.~Hoyos, A.~O'Bannon, I.~Papadimitriou, J.~Probst and J.~M. Wu,
  \emph{{Two-point Functions in a Holographic Kondo Model}},
  \href{https://doi.org/10.1007/JHEP03(2017)039}{\emph{JHEP} {\bfseries 03}
  (2017) 039} [\href{https://arxiv.org/abs/1612.02005}{{\ttfamily
  1612.02005}}].

\bibitem{Turiaci:2017zwd}
G.~Turiaci and H.~Verlinde, \emph{{Towards a 2d QFT Analog of the SYK Model}},
  \href{https://doi.org/10.1007/JHEP10(2017)167}{\emph{JHEP} {\bfseries 10}
  (2017) 167} [\href{https://arxiv.org/abs/1701.00528}{{\ttfamily
  1701.00528}}].

\bibitem{Peng:2017kro}
C.~Peng, \emph{{Vector models and generalized SYK models}},
  \href{https://doi.org/10.1007/JHEP05(2017)129}{\emph{JHEP} {\bfseries 05}
  (2017) 129} [\href{https://arxiv.org/abs/1704.04223}{{\ttfamily
  1704.04223}}].

\bibitem{Dartois:2017xoe}
S.~Dartois, H.~Erbin and S.~Mondal, \emph{{Conformality of $1/N$ corrections in
  Sachdev-Ye-Kitaev-like models}},
  \href{https://doi.org/10.1103/PhysRevD.100.125005}{\emph{Phys. Rev. D}
  {\bfseries 100} (2019) 125005}
  [\href{https://arxiv.org/abs/1706.00412}{{\ttfamily 1706.00412}}].

\bibitem{Yoon:2017nig}
J.~Yoon, \emph{{SYK Models and SYK-like Tensor Models with Global Symmetry}},
  \href{https://doi.org/10.1007/JHEP10(2017)183}{\emph{JHEP} {\bfseries 10}
  (2017) 183} [\href{https://arxiv.org/abs/1707.01740}{{\ttfamily
  1707.01740}}].

\bibitem{a:2018kvh}
A.~M. García-García, Y.~Jia and J.~J. Verbaarschot, \emph{{Exact moments of
  the Sachdev-Ye-Kitaev model up to order $1/N^2$}},
  \href{https://doi.org/10.1007/JHEP04(2018)146}{\emph{JHEP} {\bfseries 04}
  (2018) 146} [\href{https://arxiv.org/abs/1801.02696}{{\ttfamily
  1801.02696}}].

\bibitem{Nosaka:2018iat}
T.~Nosaka, D.~Rosa and J.~Yoon, \emph{{The Thouless time for mass-deformed
  SYK}}, \href{https://doi.org/10.1007/JHEP09(2018)041}{\emph{JHEP} {\bfseries
  09} (2018) 041} [\href{https://arxiv.org/abs/1804.09934}{{\ttfamily
  1804.09934}}].

\bibitem{Peng:2018zap}
C.~Peng, \emph{{$\mathcal{N}=(0,2)$ SYK, Chaos and Higher-Spins}},
  \href{https://doi.org/10.1007/JHEP12(2018)065}{\emph{JHEP} {\bfseries 12}
  (2018) 065} [\href{https://arxiv.org/abs/1805.09325}{{\ttfamily
  1805.09325}}].

\bibitem{Jia:2018ccl}
Y.~Jia and J.~J. Verbaarschot, \emph{{Large $N$ expansion of the moments and
  free energy of Sachdev-Ye-Kitaev model, and the enumeration of intersection
  graphs}}, \href{https://doi.org/10.1007/JHEP11(2018)031}{\emph{JHEP}
  {\bfseries 11} (2018) 031}
  [\href{https://arxiv.org/abs/1806.03271}{{\ttfamily 1806.03271}}].

\bibitem{Nayak:2019khe}
P.~Nayak, J.~Sonner and M.~Vielma, \emph{{Eigenstate Thermalisation in the
  conformal Sachdev-Ye-Kitaev model: an analytic approach}},
  \href{https://doi.org/10.1007/JHEP10(2019)019}{\emph{JHEP} {\bfseries 10}
  (2019) 019} [\href{https://arxiv.org/abs/1903.00478}{{\ttfamily
  1903.00478}}].

\bibitem{Qi:2020ian}
X.-L. Qi and P.~Zhang, \emph{{The Coupled SYK model at Finite Temperature}},
  \href{https://doi.org/10.1007/JHEP05(2020)129}{\emph{JHEP} {\bfseries 05}
  (2020) 129} [\href{https://arxiv.org/abs/2003.03916}{{\ttfamily
  2003.03916}}].

\bibitem{Klebanov:2020kck}
I.~R. Klebanov, A.~Milekhin, G.~Tarnopolsky and W.~Zhao, \emph{{Spontaneous
  Breaking of $U(1)$ Symmetry in Coupled Complex SYK Models}},
  \href{https://arxiv.org/abs/2006.07317}{{\ttfamily 2006.07317}}.

\bibitem{Gurau:2016lzk}
R.~Gurau, \emph{{The complete $1/N$ expansion of a SYK--like tensor model}},
  \href{https://doi.org/10.1016/j.nuclphysb.2017.01.015}{\emph{Nucl. Phys. B}
  {\bfseries 916} (2017) 386}
  [\href{https://arxiv.org/abs/1611.04032}{{\ttfamily 1611.04032}}].

\bibitem{Witten:2016iux}
E.~Witten, \emph{{An SYK-Like Model Without Disorder}},
  \href{https://doi.org/10.1088/1751-8121/ab3752}{\emph{J. Phys. A} {\bfseries
  52} (2019) 474002} [\href{https://arxiv.org/abs/1610.09758}{{\ttfamily
  1610.09758}}].

\bibitem{Klebanov:2016xxf}
I.~R. Klebanov and G.~Tarnopolsky, \emph{{Uncolored random tensors, melon
  diagrams, and the Sachdev-Ye-Kitaev models}},
  \href{https://doi.org/10.1103/PhysRevD.95.046004}{\emph{Phys. Rev. D}
  {\bfseries 95} (2017) 046004}
  [\href{https://arxiv.org/abs/1611.08915}{{\ttfamily 1611.08915}}].

\bibitem{Peng:2016mxj}
C.~Peng, M.~Spradlin and A.~Volovich, \emph{{A Supersymmetric SYK-like Tensor
  Model}}, \href{https://doi.org/10.1007/JHEP05(2017)062}{\emph{JHEP}
  {\bfseries 05} (2017) 062}
  [\href{https://arxiv.org/abs/1612.03851}{{\ttfamily 1612.03851}}].

\bibitem{Krishnan:2016bvg}
C.~Krishnan, S.~Sanyal and P.~N. Bala~Subramanian, \emph{{Quantum Chaos and
  Holographic Tensor Models}},
  \href{https://doi.org/10.1007/JHEP03(2017)056}{\emph{JHEP} {\bfseries 03}
  (2017) 056} [\href{https://arxiv.org/abs/1612.06330}{{\ttfamily
  1612.06330}}].

\bibitem{Li:2017hdt}
T.~Li, J.~Liu, Y.~Xin and Y.~Zhou, \emph{{Supersymmetric SYK model and random
  matrix theory}}, \href{https://doi.org/10.1007/JHEP06(2017)111}{\emph{JHEP}
  {\bfseries 06} (2017) 111}
  [\href{https://arxiv.org/abs/1702.01738}{{\ttfamily 1702.01738}}].

\bibitem{Gurau:2017xhf}
R.~Gurau, \emph{{Quenched equals annealed at leading order in the colored SYK
  model}}, \href{https://doi.org/10.1209/0295-5075/119/30003}{\emph{EPL}
  {\bfseries 119} (2017) 30003}
  [\href{https://arxiv.org/abs/1702.04228}{{\ttfamily 1702.04228}}].

\bibitem{Itoyama:2017emp}
H.~Itoyama, A.~Mironov and A.~Morozov, \emph{{Rainbow tensor model with
  enhanced symmetry and extreme melonic dominance}},
  \href{https://doi.org/10.1016/j.physletb.2017.05.043}{\emph{Phys. Lett. B}
  {\bfseries 771} (2017) 180}
  [\href{https://arxiv.org/abs/1703.04983}{{\ttfamily 1703.04983}}].

\bibitem{Krishnan:2017ztz}
C.~Krishnan, K.~V.~P. Kumar and S.~Sanyal, \emph{{Random Matrices and
  Holographic Tensor Models}},
  \href{https://doi.org/10.1007/JHEP06(2017)036}{\emph{JHEP} {\bfseries 06}
  (2017) 036} [\href{https://arxiv.org/abs/1703.08155}{{\ttfamily
  1703.08155}}].

\bibitem{Narayan:2017qtw}
P.~Narayan and J.~Yoon, \emph{{SYK-like Tensor Models on the Lattice}},
  \href{https://doi.org/10.1007/JHEP08(2017)083}{\emph{JHEP} {\bfseries 08}
  (2017) 083} [\href{https://arxiv.org/abs/1705.01554}{{\ttfamily
  1705.01554}}].

\bibitem{deMelloKoch:2017bvv}
R.~de~Mello~Koch, D.~Gossman and L.~Tribelhorn, \emph{{Gauge Invariants,
  Correlators and Holography in Bosonic and Fermionic Tensor Models}},
  \href{https://doi.org/10.1007/JHEP09(2017)011}{\emph{JHEP} {\bfseries 09}
  (2017) 011} [\href{https://arxiv.org/abs/1707.01455}{{\ttfamily
  1707.01455}}].

\bibitem{Azeyanagi:2017drg}
T.~Azeyanagi, F.~Ferrari and F.~I. Schaposnik~Massolo, \emph{{Phase Diagram of
  Planar Matrix Quantum Mechanics, Tensor, and Sachdev-Ye-Kitaev Models}},
  \href{https://doi.org/10.1103/PhysRevLett.120.061602}{\emph{Phys. Rev. Lett.}
  {\bfseries 120} (2018) 061602}
  [\href{https://arxiv.org/abs/1707.03431}{{\ttfamily 1707.03431}}].

\bibitem{Giombi:2017dtl}
S.~Giombi, I.~R. Klebanov and G.~Tarnopolsky, \emph{{Bosonic tensor models at
  large $N$ and small $\epsilon$}},
  \href{https://doi.org/10.1103/PhysRevD.96.106014}{\emph{Phys. Rev. D}
  {\bfseries 96} (2017) 106014}
  [\href{https://arxiv.org/abs/1707.03866}{{\ttfamily 1707.03866}}].

\bibitem{Ferrari:2017jgw}
F.~Ferrari, V.~Rivasseau and G.~Valette, \emph{{A New Large $N$ Expansion for
  General Matrix--Tensor Models}},
  \href{https://doi.org/10.1007/s00220-019-03511-7}{\emph{Commun. Math. Phys.}
  {\bfseries 370} (2019) 403}
  [\href{https://arxiv.org/abs/1709.07366}{{\ttfamily 1709.07366}}].

\bibitem{Benedetti:2018goh}
D.~Benedetti and R.~Gurau, \emph{{2PI effective action for the SYK model and
  tensor field theories}},
  \href{https://doi.org/10.1007/JHEP05(2018)156}{\emph{JHEP} {\bfseries 05}
  (2018) 156} [\href{https://arxiv.org/abs/1802.05500}{{\ttfamily
  1802.05500}}].

\bibitem{Krishnan:2018jsp}
C.~Krishnan and K.~Pavan~Kumar, \emph{{Complete Solution of a Gauged Tensor
  Model}},  \href{https://arxiv.org/abs/1804.10103}{{\ttfamily 1804.10103}}.

\bibitem{Delporte:2018iyf}
N.~Delporte and V.~Rivasseau, \emph{{The Tensor Track V: Holographic Tensors}},
   in \emph{{17th Hellenic School and Workshops on Elementary Particle Physics
  and Gravity}}, 4, 2018, \href{https://arxiv.org/abs/1804.11101}{{\ttfamily
  1804.11101}}.

\bibitem{Diaz:2018eik}
P.~Diaz and J.~Rosabal, \emph{{Spontaneous Symmetry Breaking in Tensor
  Theories}}, \href{https://doi.org/10.1007/JHEP01(2019)094}{\emph{JHEP}
  {\bfseries 01} (2019) 094}
  [\href{https://arxiv.org/abs/1809.10153}{{\ttfamily 1809.10153}}].

\bibitem{deMelloKoch:2019lsx}
R.~De~Mello~Koch, D.~Gossman, N.~Hasina~Tahiridimbisoa and A.~L. Mahu,
  \emph{{Holography for Tensor models}},
  \href{https://doi.org/10.1103/PhysRevD.101.046004}{\emph{Phys. Rev. D}
  {\bfseries 101} (2020) 046004}
  [\href{https://arxiv.org/abs/1910.13982}{{\ttfamily 1910.13982}}].

\bibitem{Benedetti:2019sop}
D.~Benedetti and I.~Costa, \emph{{$SO(3)$-invariant phase of the $O(N)^3$
  tensor model}},
  \href{https://doi.org/10.1103/PhysRevD.101.086021}{\emph{Phys. Rev. D}
  {\bfseries 101} (2020) 086021}
  [\href{https://arxiv.org/abs/1912.07311}{{\ttfamily 1912.07311}}].

\bibitem{Cotler:2016fpe}
J.~S. Cotler, G.~Gur-Ari, M.~Hanada, J.~Polchinski, P.~Saad, S.~H. Shenker
  et~al., \emph{{Black Holes and Random Matrices}},
  \href{https://doi.org/10.1007/JHEP05(2017)118}{\emph{JHEP} {\bfseries 05}
  (2017) 118} [\href{https://arxiv.org/abs/1611.04650}{{\ttfamily
  1611.04650}}].

\bibitem{Stanford:2017thb}
D.~Stanford and E.~Witten, \emph{{Fermionic Localization of the Schwarzian
  Theory}}, \href{https://doi.org/10.1007/JHEP10(2017)008}{\emph{JHEP}
  {\bfseries 10} (2017) 008}
  [\href{https://arxiv.org/abs/1703.04612}{{\ttfamily 1703.04612}}].

\bibitem{Mertens:2017mtv}
T.~G. Mertens, G.~J. Turiaci and H.~L. Verlinde, \emph{{Solving the Schwarzian
  via the Conformal Bootstrap}},
  \href{https://doi.org/10.1007/JHEP08(2017)136}{\emph{JHEP} {\bfseries 08}
  (2017) 136} [\href{https://arxiv.org/abs/1705.08408}{{\ttfamily
  1705.08408}}].

\bibitem{Raben:2018sjl}
T.~G. Raben and C.-I. Tan, \emph{{Minkowski conformal blocks and the Regge
  limit for Sachdev-Ye-Kitaev-like models}},
  \href{https://doi.org/10.1103/PhysRevD.98.086009}{\emph{Phys. Rev. D}
  {\bfseries 98} (2018) 086009}
  [\href{https://arxiv.org/abs/1801.04208}{{\ttfamily 1801.04208}}].

\bibitem{Mertens:2018fds}
T.~G. Mertens, \emph{{The Schwarzian theory --- origins}},
  \href{https://doi.org/10.1007/JHEP05(2018)036}{\emph{JHEP} {\bfseries 05}
  (2018) 036} [\href{https://arxiv.org/abs/1801.09605}{{\ttfamily
  1801.09605}}].

\bibitem{Maldacena:2018lmt}
J.~Maldacena and X.-L. Qi, \emph{{Eternal traversable wormhole}},
  \href{https://arxiv.org/abs/1804.00491}{{\ttfamily 1804.00491}}.

\bibitem{Lam:2018pvp}
H.~T. Lam, T.~G. Mertens, G.~J. Turiaci and H.~Verlinde, \emph{{Shockwave
  S-matrix from Schwarzian Quantum Mechanics}},
  \href{https://doi.org/10.1007/JHEP11(2018)182}{\emph{JHEP} {\bfseries 11}
  (2018) 182} [\href{https://arxiv.org/abs/1804.09834}{{\ttfamily
  1804.09834}}].

\bibitem{Blommaert:2018oro}
A.~Blommaert, T.~G. Mertens and H.~Verschelde, \emph{{The Schwarzian Theory - A
  Wilson Line Perspective}},
  \href{https://doi.org/10.1007/JHEP12(2018)022}{\emph{JHEP} {\bfseries 12}
  (2018) 022} [\href{https://arxiv.org/abs/1806.07765}{{\ttfamily
  1806.07765}}].

\bibitem{Chen:2019qqe}
Y.~Chen and P.~Zhang, \emph{{Entanglement Entropy of Two Coupled SYK Models and
  Eternal Traversable Wormhole}},
  \href{https://doi.org/10.1007/JHEP07(2019)033}{\emph{JHEP} {\bfseries 07}
  (2019) 033} [\href{https://arxiv.org/abs/1903.10532}{{\ttfamily
  1903.10532}}].

\bibitem{Belokurov:2018fnn}
V.~V. Belokurov and E.~T. Shavgulidze, \emph{{Simple rules of functional
  integration in the Schwarzian theory: SYK correlators}},
  \href{https://arxiv.org/abs/1811.11863}{{\ttfamily 1811.11863}}.

\bibitem{Belokurov:2018aol}
V.~V. Belokurov and E.~T. Shavgulidze, \emph{{Correlation functions in the
  Schwarzian theory}},
  \href{https://doi.org/10.1007/JHEP11(2018)036}{\emph{JHEP} {\bfseries 11}
  (2018) 036} [\href{https://arxiv.org/abs/1804.00424}{{\ttfamily
  1804.00424}}].

\bibitem{Khveshchenko:2017mvj}
D.~Khveshchenko, \emph{{Thickening and sickening the SYK model}},
  \href{https://doi.org/10.21468/SciPostPhys.5.1.012}{\emph{SciPost Phys.}
  {\bfseries 5} (2018) 012} [\href{https://arxiv.org/abs/1705.03956}{{\ttfamily
  1705.03956}}].

\bibitem{Cardella:2019kum}
M.~A. Cardella, \emph{{Derivation of the two Schwarzians effective action for
  the Sachdev-Ye-Kitaev spectral form factor}},
  \href{https://arxiv.org/abs/1907.09570}{{\ttfamily 1907.09570}}.

\bibitem{Ghosh:2019rcj}
A.~Ghosh, H.~Maxfield and G.~J. Turiaci, \emph{{A universal Schwarzian sector
  in two-dimensional conformal field theories}},
  \href{https://doi.org/10.1007/JHEP05(2020)104}{\emph{JHEP} {\bfseries 05}
  (2020) 104} [\href{https://arxiv.org/abs/1912.07654}{{\ttfamily
  1912.07654}}].

\bibitem{Iliesiu:2020qvm}
L.~V. Iliesiu and G.~J. Turiaci, \emph{{The statistical mechanics of
  near-extremal black holes}},
  \href{https://arxiv.org/abs/2003.02860}{{\ttfamily 2003.02860}}.

\bibitem{Gervais:1975pa}
J.-L. Gervais, A.~Jevicki and B.~Sakita, \emph{{Perturbation Expansion Around
  Extended Particle States in Quantum Field Theory. 1.}},
  \href{https://doi.org/10.1103/PhysRevD.12.1038}{\emph{Phys. Rev. D}
  {\bfseries 12} (1975) 1038}.

\bibitem{Gervais:1975yg}
J.-L. Gervais, A.~Jevicki and B.~Sakita, \emph{{Collective Coordinate Method
  for Quantization of Extended Systems}},
  \href{https://doi.org/10.1016/0370-1573(76)90049-1}{\emph{Phys. Rept.}
  {\bfseries 23} (1976) 281}.

\bibitem{Teitelboim:1983ux}
C.~Teitelboim, \emph{{Gravitation and Hamiltonian Structure in Two Space-Time
  Dimensions}}, \href{https://doi.org/10.1016/0370-2693(83)90012-6}{\emph{Phys.
  Lett. B} {\bfseries 126} (1983) 41}.

\bibitem{Jackiw:1984je}
R.~Jackiw, \emph{{Lower Dimensional Gravity}},
  \href{https://doi.org/10.1016/0550-3213(85)90448-1}{\emph{Nucl. Phys. B}
  {\bfseries 252} (1985) 343}.

\bibitem{Almheiri:2014cka}
A.~Almheiri and J.~Polchinski, \emph{{Models of AdS$_{2}$ backreaction and
  holography}}, \href{https://doi.org/10.1007/JHEP11(2015)014}{\emph{JHEP}
  {\bfseries 11} (2015) 014} [\href{https://arxiv.org/abs/1402.6334}{{\ttfamily
  1402.6334}}].

\bibitem{Jensen:2016pah}
K.~Jensen, \emph{{Chaos in AdS$_2$ Holography}},
  \href{https://doi.org/10.1103/PhysRevLett.117.111601}{\emph{Phys. Rev. Lett.}
  {\bfseries 117} (2016) 111601}
  [\href{https://arxiv.org/abs/1605.06098}{{\ttfamily 1605.06098}}].

\bibitem{Maldacena:2016upp}
J.~Maldacena, D.~Stanford and Z.~Yang, \emph{{Conformal symmetry and its
  breaking in two dimensional Nearly Anti-de-Sitter space}},
  \href{https://doi.org/10.1093/ptep/ptw124}{\emph{PTEP} {\bfseries 2016}
  (2016) 12C104} [\href{https://arxiv.org/abs/1606.01857}{{\ttfamily
  1606.01857}}].

\bibitem{Engelsoy:2016xyb}
J.~Engelsöy, T.~G. Mertens and H.~Verlinde, \emph{{An investigation of
  AdS$_{2}$ backreaction and holography}},
  \href{https://doi.org/10.1007/JHEP07(2016)139}{\emph{JHEP} {\bfseries 07}
  (2016) 139} [\href{https://arxiv.org/abs/1606.03438}{{\ttfamily
  1606.03438}}].

\bibitem{Mandal:2017thl}
G.~Mandal, P.~Nayak and S.~R. Wadia, \emph{{Coadjoint orbit action of Virasoro
  group and two-dimensional quantum gravity dual to SYK/tensor models}},
  \href{https://doi.org/10.1007/JHEP11(2017)046}{\emph{JHEP} {\bfseries 11}
  (2017) 046} [\href{https://arxiv.org/abs/1702.04266}{{\ttfamily
  1702.04266}}].

\bibitem{Das:2017pif}
S.~R. Das, A.~Jevicki and K.~Suzuki, \emph{{Three Dimensional View of the
  SYK/AdS Duality}}, \href{https://doi.org/10.1007/JHEP09(2017)017}{\emph{JHEP}
  {\bfseries 09} (2017) 017}
  [\href{https://arxiv.org/abs/1704.07208}{{\ttfamily 1704.07208}}].

\bibitem{Taylor:2017dly}
M.~Taylor, \emph{{Generalized conformal structure, dilaton gravity and SYK}},
  \href{https://doi.org/10.1007/JHEP01(2018)010}{\emph{JHEP} {\bfseries 01}
  (2018) 010} [\href{https://arxiv.org/abs/1706.07812}{{\ttfamily
  1706.07812}}].

\bibitem{Grumiller:2017qao}
D.~Grumiller, R.~McNees, J.~Salzer, C.~Valcárcel and D.~Vassilevich,
  \emph{{Menagerie of AdS$_{2}$ boundary conditions}},
  \href{https://doi.org/10.1007/JHEP10(2017)203}{\emph{JHEP} {\bfseries 10}
  (2017) 203} [\href{https://arxiv.org/abs/1708.08471}{{\ttfamily
  1708.08471}}].

\bibitem{Das:2017hrt}
S.~R. Das, A.~Ghosh, A.~Jevicki and K.~Suzuki, \emph{{Three Dimensional View of
  Arbitrary $q$ SYK models}},
  \href{https://doi.org/10.1007/JHEP02(2018)162}{\emph{JHEP} {\bfseries 02}
  (2018) 162} [\href{https://arxiv.org/abs/1711.09839}{{\ttfamily
  1711.09839}}].

\bibitem{Nayak:2018qej}
P.~Nayak, A.~Shukla, R.~M. Soni, S.~P. Trivedi and V.~Vishal, \emph{{On the
  Dynamics of Near-Extremal Black Holes}},
  \href{https://doi.org/10.1007/JHEP09(2018)048}{\emph{JHEP} {\bfseries 09}
  (2018) 048} [\href{https://arxiv.org/abs/1802.09547}{{\ttfamily
  1802.09547}}].

\bibitem{Gaikwad:2018dfc}
A.~Gaikwad, L.~K. Joshi, G.~Mandal and S.~R. Wadia, \emph{{Holographic dual to
  charged SYK from 3D Gravity and Chern-Simons}},
  \href{https://doi.org/10.1007/JHEP02(2020)033}{\emph{JHEP} {\bfseries 02}
  (2020) 033} [\href{https://arxiv.org/abs/1802.07746}{{\ttfamily
  1802.07746}}].

\bibitem{Lala:2018yib}
A.~Lala and D.~Roychowdhury, \emph{{SYK/AdS duality with Yang-Baxter
  deformations}}, \href{https://doi.org/10.1007/JHEP12(2018)073}{\emph{JHEP}
  {\bfseries 12} (2018) 073}
  [\href{https://arxiv.org/abs/1808.08380}{{\ttfamily 1808.08380}}].

\bibitem{Gonzalez:2018enk}
H.~A. González, D.~Grumiller and J.~Salzer, \emph{{Towards a bulk description
  of higher spin SYK}},
  \href{https://doi.org/10.1007/JHEP05(2018)083}{\emph{JHEP} {\bfseries 05}
  (2018) 083} [\href{https://arxiv.org/abs/1802.01562}{{\ttfamily
  1802.01562}}].

\bibitem{Moitra:2019bub}
U.~Moitra, S.~K. Sake, S.~P. Trivedi and V.~Vishal, \emph{{Jackiw-Teitelboim
  Gravity and Rotating Black Holes}},
  \href{https://doi.org/10.1007/JHEP11(2019)047}{\emph{JHEP} {\bfseries 11}
  (2019) 047} [\href{https://arxiv.org/abs/1905.10378}{{\ttfamily
  1905.10378}}].

\bibitem{Moitra:2019xoj}
U.~Moitra, S.~K. Sake, S.~P. Trivedi and V.~Vishal, \emph{{Jackiw-Teitelboim
  Model Coupled to Conformal Matter in the Semi-Classical Limit}},
  \href{https://doi.org/10.1007/JHEP04(2020)199}{\emph{JHEP} {\bfseries 04}
  (2020) 199} [\href{https://arxiv.org/abs/1908.08523}{{\ttfamily
  1908.08523}}].

\bibitem{Hirano:2019iwt}
S.~Hirano and Y.~Lei, \emph{{Nearly AdS$_2$ holography in quantum CGHS model}},
  \href{https://doi.org/10.1007/JHEP01(2020)178}{\emph{JHEP} {\bfseries 01}
  (2020) 178} [\href{https://arxiv.org/abs/1910.12542}{{\ttfamily
  1910.12542}}].

\bibitem{Afshar:2019axx}
H.~Afshar, H.~A. González, D.~Grumiller and D.~Vassilevich, \emph{{Flat space
  holography and the complex Sachdev-Ye-Kitaev model}},
  \href{https://doi.org/10.1103/PhysRevD.101.086024}{\emph{Phys. Rev. D}
  {\bfseries 101} (2020) 086024}
  [\href{https://arxiv.org/abs/1911.05739}{{\ttfamily 1911.05739}}].

\bibitem{Alkalaev:2019xuv}
K.~Alkalaev and X.~Bekaert, \emph{{Towards higher-spin AdS$_2$/CFT$_1$
  holography}}, \href{https://doi.org/10.1007/JHEP04(2020)206}{\emph{JHEP}
  {\bfseries 04} (2020) 206}
  [\href{https://arxiv.org/abs/1911.13212}{{\ttfamily 1911.13212}}].

\bibitem{Tarnopolsky:2018env}
G.~Tarnopolsky, \emph{{Large $q$ expansion in the Sachdev-Ye-Kitaev model}},
  \href{https://doi.org/10.1103/PhysRevD.99.026010}{\emph{Phys. Rev. D}
  {\bfseries 99} (2019) 026010}
  [\href{https://arxiv.org/abs/1801.06871}{{\ttfamily 1801.06871}}].

\bibitem{Streicher:2019wek}
A.~Streicher, \emph{{SYK Correlators for All Energies}},
  \href{https://doi.org/10.1007/JHEP02(2020)048}{\emph{JHEP} {\bfseries 02}
  (2020) 048} [\href{https://arxiv.org/abs/1911.10171}{{\ttfamily
  1911.10171}}].

\bibitem{Choi:2019bmd}
C.~Choi, M.~Mezei and G.~Sárosi, \emph{{Exact four point function for large
  $q$ SYK from Regge theory}},
  \href{https://arxiv.org/abs/1912.00004}{{\ttfamily 1912.00004}}.

\bibitem{Jevicki:2014mfa}
A.~Jevicki, K.~Jin and J.~Yoon, \emph{{1/N and loop corrections in higher spin
  AdS$_4$/CFT$_3$ duality}},
  \href{https://doi.org/10.1103/PhysRevD.89.085039}{\emph{Phys. Rev. D}
  {\bfseries 89} (2014) 085039}
  [\href{https://arxiv.org/abs/1401.3318}{{\ttfamily 1401.3318}}].

\bibitem{Gu:2019jub}
Y.~Gu, A.~Kitaev, S.~Sachdev and G.~Tarnopolsky, \emph{{Notes on the complex
  Sachdev-Ye-Kitaev model}},
  \href{https://doi.org/10.1007/JHEP02(2020)157}{\emph{JHEP} {\bfseries 02}
  (2020) 157} [\href{https://arxiv.org/abs/1910.14099}{{\ttfamily
  1910.14099}}].

\bibitem{Das2020:wip}
S.~R. Das, A.~Ghosh, A.~Jevicki and K.~Suzuki, \emph{{Work in progress}}, .

\bibitem{Gradshteyn:1994}
I.~S. Gradshteyn and I.~M. Ryzhik, \emph{{Tables of Integrals, Series, and
  Products}}, vol.~7. 1994.

\end{thebibliography}\endgroup



\providecommand{\href}[2]{#2}\begingroup\raggedright\endgroup

\addcontentsline{toc}{section}{References}


\end{document}